\documentclass[11pt,onecolumn,journal]{IEEEtran}

\usepackage[T1]{fontenc}  
\usepackage[utf8]{inputenc}

\usepackage[english]{babel}

\usepackage{amsmath}
\usepackage{stmaryrd}

\usepackage{graphicx}
\usepackage[numbers,sort&compress]{natbib}

\usepackage[]{times}
\usepackage{amsmath}
\usepackage{amsfonts}
\usepackage{amssymb}
\usepackage{wasysym}

\usepackage{physics}
\usepackage{tabularx}
\usepackage{multirow}
\usepackage{enumerate}
\usepackage{subfigure}
\usepackage{bm}
\usepackage{bbm}
\usepackage{multirow}
\usepackage{xcolor}
\usepackage{url}
\usepackage{breakurl} 
\usepackage{wrapfig}
\usepackage{sidecap}
\usepackage[footnotesize]{caption}
\usepackage{float}
\usepackage{newproof}
\usepackage{mathrsfs}
\usepackage{algorithmic}
\usepackage{algorithm}
\usepackage{stmaryrd}
\usepackage{upgreek}
\usepackage{tcolorbox}
\usepackage{tikz}
\usepackage{quantikz}
\usepackage{fontawesome5}
\usetikzlibrary{quantikz2}
\usetikzlibrary{positioning,arrows.meta,calc}
\definecolor{genA}{RGB}{34,139,34}
\definecolor{genB}{RGB}{230,97,1} 
\usepackage{graphicx}
\graphicspath{{./figures/}}
\DeclareGraphicsExtensions{.pdf,.png,.jpg}
\usepackage{standalone}

\definecolor{darkgreen}{RGB}{0,170,0} 
\newcommand{\darkgreen}[1]{\textcolor{darkgreen}{#1}}
\newcommand{\red}[1]{\textcolor{red}{#1}}

\newcommand{\blue}[1]{\textcolor{blue}{#1}}
\definecolor{UniBlue}{RGB}{83,121,170}
\definecolor{DarkBlue}{RGB}{0,0,165}
\definecolor{MyOrange}{RGB}{228,108,10}
\definecolor{DarkGray}{RGB}{127,127,127}
\definecolor{MyGray}{RGB}{217,217,217}
\definecolor{mygreen}{RGB}{0,176,80}

\newcommand{\orange}[1]{\textcolor{MyOrange}{#1}}

\usepackage[colorlinks=true, allcolors=DarkBlue]{hyperref}

\usepackage{soul}
\setstcolor{red} 
\setul{0.3ex}{0.7pt} 
\definecolor{darkgray}{rgb}{0.66, 0.66, 0.66}
\setulcolor{darkgray}

\usepackage{cancel}

\def\mathcolor#1#{\@mathcolor{#1}}
\def\@mathcolor#1#2#3{%
  \protect\leavevmode
  \begingroup
    \color#1{#2}#3%
  \endgroup
}
\newcommand{\bDelta}{\textcolor{blue}{\Delta}}
\newcommand{\rDelta}{\textcolor{red}{\Delta}}
\newcommand{\beps}{\boldsymbol \epsilon}

\newcommand{\mathred}[1]{\mathcolor{red}{#1}}
\newcommand{\mathblue}[1]{\mathcolor{blue}{#1}}

\newcommand{\rX}{\textcolor{red}{X}}
\newcommand{\bZ}{\textcolor{blue}{Z}}
\newcommand{\gY}{\textcolor{darkgreen}{Y}}

\newcommand{\mycite}[1]{[\nobreak\hspace{-1pt}\cite{#1}\nobreak\hspace{-1pt}]} 

\usepackage{etoolbox}
\newrobustcmd{\outline}[1]{\iftoggle{draft}{\noindent \orange{ {\bf Section Outline:} #1 \\}}}

\newcommand{\pmatrixstretch}{\renewcommand{\arraystretch}{0.8}\setlength{\arraycolsep}{3pt}}

\tikzset{
    noise_gate/.style={
        draw,
        thick,
        minimum width=2cm,
        minimum height=0.8cm,
        align=center
    }
}

\newcommand{\noisegate}[2]{%
    \gate[style=noise_gate]{%
        \mathcolor{red}{\mathrm{#1_{#2}}}\;%
        \text{\textcolor{orange}{\faBolt}}\;%
        \mathcolor{blue}{\mathrm{#1_{#2}}}%
    }%
    \push{\phantom{\rule{0.4cm}{0pt}}}%
}

\newcommand{\noisesyndrome}[2]{%
    \gate[style=noise_gate]{%
        \text{\textcolor{orange}{\faBolt}}\;%
        \mathcolor{#1}{\mathrm{ \tilde{s}_{#2}}}%
    }%
    \push{\phantom{\rule{0.4cm}{0pt}}}%
}

\newcommand{\deltasyndrome}[2]{%
    \gate[style=noise_gate]{%
        \text{\textcolor{orange}{\faBolt}}\;%
        \mathcolor{#1}{\mathrm{\Delta s_{#2}}}%
    }%
    \push{\phantom{\rule{0.4cm}{0pt}}}%
}

\newcommand{\noisethunder}[0]{
    \push{\text{\textcolor{orange}{\Large\faBolt}}}
}

\providetoggle{draft}
\settoggle{draft}{true}

\begin{document}
\title{\vspace*{-8mm}Quantum Low-Density Parity-Check Codes}

\author{Bane~Vasi\'c$^{(1)}$, Valentin~Savin$^{(2)}$,
Michele Pacenti$^{(1)}$, Shantom Borah$^{(1)}$, and Nithin Raveendran$^{(1)}$
\\[2mm]%
{\normalsize ${}^{(1)}$\, University of Arizona, Department of ECE, Center for Quantum Networks, Tucson, Arizona, USA 
}\\
{\normalsize ${}^{(2)}$\,Universit\'e Grenoble Alpes, CEA-L\'eti, Grenoble, France  
}
\vspace*{-5mm}
}

\maketitle

\begin{abstract}
Quantum error correction (QEC) is a cornerstone of quantum computing, enabling reliable information processing in the presence of noise. Sparse stabilizer codes -- referred to generally as quantum low-density parity-check (QLDPC) codes -- have risen to the forefront of QEC research in recent years. This can be attributed to several key factors. First, classical LDPC codes admit low-complexity belief propagation iterative decoding and near-capacity performance,  which contributed to the early interest in QLDPC codes. Then, the result promising constant overhead fault tolerance using QLDPC codes led to the search for code families that go beyond the long-holding $\sqrt{n}$ scaling barrier of minimum distance for codelength $n$. This resulted in recent breakthroughs in the construction of  QLDPC codes, which, combined with efficient decoding algorithms and the development of fault-tolerant protocols operating on QLDPC-encoded quantum information, provide a promising pathway to low-overhead, fault-tolerant quantum computation.
However, despite their potential,  challenges remain, particularly in constructing and decoding finite-length codes that account for, or efficiently leverage, specific characteristics of quantum hardware, such as connectivity, topology, native gate sets, and noise models. This article provides an in-depth examination of QLDPC codes and their iterative decoders, catering to an information theory audience\footnote{We make minimal use of quantum jargon.} with no or limited background in quantum mechanics. We discuss the theoretical underpinnings, explore unique characteristics of quantum channels, and delineate key code constructions and decoding algorithms, ultimately highlighting the impact and future prospects of QLDPC codes in quantum information science.
\end{abstract}

\section{Introduction}
\label{sec:intro}
In classical systems, LDPC codes are widely used and have been standardized for many applications, including wireless networks, data storage, optical links, etc. The authors of this article have been involved in research and development of LDPC codes since the beginning, from attending the very first presentation on the topic by David MacKay at Bell Labs in 1998 (a few weeks after, Peter Shor gave the first talk on quantum codes, organized by Robert Calderbank), and very early interactions with Tom Richardson and Rudiger Urbanke on LDPC code constructions, to the present day. We collaborated with other pioneers of LDPC codes,  Shu Lin, Dariush Divsalar, and David Declercq, on topics ranging from theoretical analysis and code and decoder constructions to hardware implementation and commercialization. We also worked on classical fault tolerance using LDPC codes, a topic not widely known, and introduced to us by Alexander Kuznetsov from the Russian Academy of Sciences, who moved to the USA, and has collaborated with us for many years. Thus, the reader should not be surprised that we start this article on quantum LDPC (QLDPC) codes by viewing QLDPC codes and iterative decoding through the lenses and aesthetics of classical fault tolerance and classical \emph{modern coding theory} (LDPC codes). And who knows, perhaps this article becomes a grain of sand for a motivation for information theorists and coding theorists to look into this set of problems.

Interestingly, MacKay~\emph{et al.}~\cite{mackay_quantum} proposed using LDPC codes for \emph{quantum error correction} (QEC) in 2004, but it took about twenty years for QLDPC to become mainstream. The first spark came from the quantum physics community, Gottesman~\cite{Gottesman97,gottesman2013fault} was the first to observe the potential of LDPC codes for QEC. His motivation to consider LDPC codes was not any of the features classical LDPC codes are celebrated for: capacity achieving, ability to correct a linear fraction of errors in code length, low-complexity decoding, nothing of it. It was fault tolerance. It seems that at the time he wrote his dissertation, Gottesman was not familiar with the classical theory of fault tolerance, but he demonstrated excellent intuition, rediscovering key concepts in the quantum setting. He, of course, did much more, including introducing the stabilizer formalism for quantum codes. So let us begin by approaching the QLDPC coding problem through the lens of classical fault tolerance. For a start, we consider only faulty memory, we leave the computation by faulty quantum gates for later.

\subsection{Classical Fault Tolerance Setting}
\label{sec:intro:classical_ft}
The memory registers in which the bits are stored are assumed to be unreliable, and each bit is flipped independently of others with known probability $\alpha$. These registers are connected to a \emph{correction circuit}, which periodically updates the values in the registers. For a correction circuit to be able to correct these bit flips, the information must be in the coded form. This is the celebrated Taylor-Kuznetsov (TK) classical fault-tolerant scheme~\cite{taylor1,kuznetsov}.

Let $\mathbf{m}$ be a \emph{message} vector composed of $k$ bits. We encode $\mathbf{m}$ using an $(n,k)$ linear block code $\mathcal{C}$ with generator matrix $G$. The resulting $n$-bit codeword $\mathbf{x}=\mathbf{m} G$, is then stored in a memory in $n$ memory locations. For now, let us assume that encoding is perfect, i.e., there are no errors during multiplication with $G$  -- we will revisit this assumption later. There is no reason for much worry, since the rows of $G$ are themselves codewords of the same code, and the same method we are crafting to protect an arbitrary codeword can be applied to protect them, thus to 
protect encoding. The coded bits can be accessed from the memory by a correction circuit in discrete ``clock cycles'', and for simplicity, we assume that every step of processing of bits takes the same unit \emph{memory} time $\Delta \tau=1$ 
(we also ignore the fact that in order to be processed, bits must be moved from their actual physical location to the physical location of the (input of the) circuit that processes them). In between that discrete unit time instants, $1 \le \tau \le T$, each of $n$ bits may get flipped (even if idling and not being processed) with probability $\alpha$. The result is the sequence of $n$-bit error vectors $\beps^{(1)},\beps^{(2)},\ldots \beps^{(T)}$,
where the elements of the random error vectors $\beps^{(\ell)}$ are i.i.d. and $\text{Ber}(\alpha)$, the Bernoulli distribution with  parameter $\alpha$.

If no correction was applied, the content of the memory during these $T$ time steps during which the information is stored in the memory would be $\mathbf{y}^{(1)},\mathbf{y}^{(2)},\ldots \mathbf{y}^{(T)}$, where
$\mathbf{y}^{(0)}=\mathbf{x}$, $\mathbf{y}^{(\tau)} = \mathcal{M}(\mathbf{y}^{(\tau-1)})=\mathbf{y}^{(\tau-1)}+\beps^{(\tau)}$,
$\mathcal{M}$ denotes the operation of the memory in one time step. Clearly, $\mathbf{y}^{(\tau)}=\mathbf{x}+ \sum_{\ell \le \tau} \beps^{(\ell)}$. We refer to $\mathbf{e}^{(\tau)}= \sum_{\ell \le \tau} \beps^{(\ell)}$ as the \emph{accumulated} error vector at time $\tau$. The elements of $\mathbf{e}^{(\tau)}$ are i.i.d. and $\text{Ber}(\frac{1-(1-\alpha)^\tau}{2})$. For large $\tau$ (and $T$), the content of the memory would ``diffuse'' from $\mathbf{x}$, and become uniformly random with probability $\frac{1}{2}$ and independent of $\mathbf{x}$, which means that the information stored in $\mathbf{x}$ would be lost almost certainly. Thus, the time $T$ for which the information can be stored in memory cannot be arbitrarily large.

\subsection{Coding for Faulty Memories}
\label{sec:intro:faulty_memory}
If we apply error correction at time $\tau=T$, and if the accumulated vector $\mathbf{e}^{(T)}$ is  correctable with high probability by the decoding algorithm operating on the code $\mathcal{C}$, 
then the memory is said to be \emph{stable}\footnote{Technically, it is not the memory itself that is said to be stable, but rather a family of memories $\{\mathcal{M}_k\}$, where each memory $\mathcal{M}_k$ can store $k$ bits of information. We require that, for any $T>0$ and $\delta > 0$, there exists $k>0$ such that $p_k(T) < \delta$, where $p_k(T)$ is the probability that decoding the $\mathcal{M}_k$ memory at time $T$ fails~\cite{taylor1}.}. 
The above statement is informal, but the operation of the memory is equivalent to that of a binary symmetric channel (BSC), so the information theory results for coded transmission over the BSC apply directly. We may also not want to wait for $T$ time steps to apply error correction, and if we have fast technology to implement the correction circuit,  we can perform error correction at every time step $\tau$. This requires that all operations in the correction circuit be executed at a faster clock rate, to ensure that errors are corrected by the time $\tau+1$.

And we hope we maintain the information in the memory for $T$ correcting rounds by means of correcting corrupt codewords. Let $\mathbf{y}^{(\tau)}=\mathbf{x} +\mathbf{e}^{(\tau)}$ denote the corrupt codeword at the beginning of the $\tau$-th correction round. The vector $\mathbf{e}^{(\tau)}$ is the accumulated error vector at the beginning of $\tau$-th round. Ideally, if the correction circuit can correct all errors in prior rounds, then the only remaining error before $\tau$ round of correction would be  $\mathbf{e}^{(\tau)}=\beps^{(\tau)}$. But would this be possible when the correction circuit is made of unreliable components? Let us first see what happens in the \emph{perfect}, reliable, correction circuit.

The correction circuit consists of three sub-circuits for the following operations:
\begin{eqnarray}
  \label{eq: S}
  \mathbf{s}            &\leftarrow& \mathcal{S}(\mathbf{y}) \quad \enspace \enspace \text{(\emph{syndrome} computation)} \\
  \label{eq: D}
  \widehat{\mathbf{e}}  &\leftarrow& \mathcal{D}(\mathbf{s}) \quad \enspace \enspace \text{(\emph{decoding})}\\
  \label{eq: R}
  \mathbf{y}'            &\leftarrow& \mathcal{R}(\mathbf{y},\widehat{\mathbf{e}}) \quad \text{(\emph{recovery})}
\end{eqnarray}
where their respective functions are $\mathcal{S}$, $\mathcal{D}$, and $\mathcal{R}$. The function of $\mathcal{S}$ is $\mathbf{s}=\mathbf{y} H^\text{T}$, where $H$ is the parity check matrix of the error correcting code $\mathcal{C}$, the operation $\mathcal{D}$ is an iterative decoding algorithm, or a decoder, and the recovery $\mathcal{R}$ is $\mathbf{y}' = \mathbf{y} +\widehat{\mathbf{e}}$. Note that the recovery operation overwrites $\mathbf{y}$ with $\mathbf{y}'$, so that $\mathbf{y}'$ becomes the memory content at the end of the correction round (alternatively, we could have written $\mathbf{y} \leftarrow \mathbf{y} +\widehat{\mathbf{e}}$, but we use the prime notation to avoid later ambiguity).

If for some error pattern $\mathbf{e}$, the decoder $\mathcal{D}$ produces the output $\widehat{\mathbf{e}}=\mathcal{D}(\mathbf{e}H^\text{T})=\mathbf{e}$, then $\mathbf{e}$ is said to be correctable by the decoder $\mathcal{D}$, otherwise it is uncorrectable. The number of bits in which $\widehat{\mathbf{e}}$ and $\mathbf{e}$ differ, i.e., the Hamming weight of the \emph{residual error} $\Delta {\mathbf{e}} = \mathbf{e}+ \widehat{\mathbf{e}}$, tells us how many errors remain in the codeword after  applying the recovery operation $\mathcal{R}$. Thus, the residual error at the end of $(\tau-1)$-th correction is $\Delta {\mathbf{e}}^{(\tau-1)}$, so at the beginning of the $\tau$-th correction round $\mathbf{y}^{(\tau)}=\mathbf{x} +\Delta {\mathbf{e}}^{(\tau-1)}+\beps^{(\tau)}$, and the error that the decoder needs to correct is
$\Delta {\mathbf{e}}^{(\tau-1)} + \beps^{(\tau)}$.

\subsection{Need for Speed in a Correction Circuit}
\label{sec:intro:need_for_speed}
As mentioned earlier, the clock cycle of the correction circuit, denoted by $\Delta \lambda$, must be much shorter than that of the memory, $\Delta \tau$, to ensure all the computational operations are completed within one memory clock cycle. Each elementary operation in $\mathcal{S}$, $\mathcal{D}$, and $\mathcal{R}$, i.e., each single-input or two-input \emph{Boolean} gate, takes a unit \emph{computational} time to be completed.
By the nature of correction circuits, some operations in it are sequential; thus, we introduce the parameter $D$ that we call computational \emph{circuit depth}. Since all computations in the correction circuits must be completed in one memory clock cycle, $D$ shows how much the computation clock is faster than the memory clock, $\Delta \tau= D \Delta \lambda$.

\subsection{Perfect and Faulty Gates, Additive Perturbations}
\label{sec:intro:perturbations}
So far, the operation of $\mathcal{S}$, $\mathcal{D}$, and $\mathcal{R}$ were assumed to be perfect, and only errors occurred due to $\mathcal{M}$, the memory errors. In classical fault tolerance theory, the assumptions are harsh: \emph{all} Boolean gates used in the correction circuit, in \emph{all} sub-circuits $\mathcal{S}$, $\mathcal{D}$, and $\mathcal{R}$, including the registers used to store intermediate results are assumed to be faulty. Faulty means that they are subject to independent and transient faults, and for simplicity, we assume the faults have the same and known probability $\beta$. 
We denote these faulty operations by $\widetilde{\mathcal{S}}$, $\widetilde{\mathcal{D}}$, and $\widetilde{\mathcal{R}}$, producing faulty results
\begin{eqnarray}
   \label{eq: faulty_operations_S_tilde}   
   \mathbf{s}+\Delta\mathbf{s}
   &\leftarrow& \widetilde{\mathcal{S}}(\mathbf{y})     \\
   \label{eq: faulty_operations_D_tilde}
   \widehat{\mathbf{e}}+\Delta \widehat{\mathbf{e}}
   &\leftarrow& \widetilde{\mathcal{D}}(\mathbf{s}+\Delta\mathbf{s}) \\
   \label{eq: faulty_operations_R_tilde}  
   \mathbf{y}' + \Delta \mathbf{y}'
   &\leftarrow& \widetilde{\mathcal{R}}(\mathbf{y}, \widehat{\mathbf{e}}+\Delta \widehat{\mathbf{e}}),
\end{eqnarray}
where $\mathbf{s}$, $\widehat{\mathbf{e}}$, $\mathbf{y}'$ are the results of the fault-free operations $\mathcal{S}$, $\mathcal{D}$, and $\mathcal{R}$ (with the same inputs as the faulty ones), 
and the additive perturbations $\Delta\mathbf{s}$, $\Delta \widehat{\mathbf{e}}$, $\Delta \mathbf{y}$  arise from the faulty nature of $\widetilde{\mathcal{S}}$, $\widetilde{\mathcal{D}}$, and $\widetilde{\mathcal{R}}$.

\subsection{Fundamental Results on Coded Classical Fault Tolerant Memories}
\label{sec:intro:coded_classical_memories}
The remarkable result by Taylor~\cite{taylor1} and Kuznetsov~\cite{kuznetsov} is that, despite all components being unreliable, it is still possible -- under some assumptions -- to make the entire memory stable. 
The code $\mathcal{C}$ used is an LDPC code with some properties (for more details see our work~\mycite{07_VC_J_33,07_CV_J_133,06_IMCSVB_T_140, elsa_faulty_faid_density_evolution_tcomm_2015,ngassa2015density}\,\footnote{Throughout this paper, [\hspace{-1pt}[\hspace{2pt}]\hspace{-1pt}] denotes work of the authors. We are not original. Based on the authors' best knowledge, the true inventor of this referencing convention is Professor Rick Wesel.}). 
There is much more to be said about these conditions and code properties, but the fundamental reason why LDPC codes are conducive to fault tolerance is that the syndrome computation ($\mathcal{S}$) and decoding ($\mathcal{D}$) procedures can be implemented using constant depth Boolean circuits (the same holds for the recovery procedure $\mathcal{R}$, but it is a general property valid irrespective of the code). For syndrome computation, this follows directly from the low-density property of the parity-check matrix $H$, while for decoding, it results from the use of iterative decoding algorithms, such as Bit-Flipping of Gallager-B (initially considered by Taylor and Kuznetsov, though more general message-passing decoders can also be used~\mycite{elsa_faulty_faid_density_evolution_tcomm_2015,ngassa2015density}). To be more precise, we note that the depth of these circuits depends only on the weights of the rows (corresponding to parity-checks) and columns (corresponding to bits) of the parity check matrix $H$, rather than its size. Moreover, it can be shown that it is sufficient to use so-called \emph{regular} LDPC codes, with constant row-weight $\rho$ and column-weight $\gamma$. The key idea is that the LDPC class allows to keep $\gamma$ and $\rho$ constant and small, thereby keeping the depth of the correction circuit constant and, consequently, maintaining a bounded probability of errors during the correction process. If the computation and memory error rates ($\beta, \alpha$) are sufficiently small, then by adjusting the coding rate and increasing the code length to obtain a sufficiently strong code, there is a hope that the faulty correction circuit corrects more errors than it introduces in the computational process of correction. 
This deserves a few explanations, as the faulty syndrome computation $\widetilde{\mathcal{S}}$ may produce a wrong syndrome $\widetilde{\mathbf{s}} = \mathbf{s}+\Delta \mathbf{s}$ for the decoder $\widetilde{\mathcal{D}}$, which is itself faulty. Even if we had a perfect decoder, how could we possibly correct all these errors when the syndrome itself is wrong? 
Of course, with memories built entirely from unreliable components, it is unrealistic to expect that the content of the memory will be exactly restored; nevertheless, we can hope that the underlying information is preserved.
The key idea is that we will tolerate residual errors, provided they occur at a sufficiently low rate, so that a perfect, fault-free decoder can still correct them.
Remarkably, Taylor and Kuznetsov's results show that this is indeed the case: even if error correction is performed all along the way with a faulty syndrome and a faulty decoder, we still retain sufficient correction capability to prevent the stored data word from drifting away from the original codeword, so that a perfect decoder can faithfully recover the latter.

\subsection{Building Quantum Fault-Tolerant Model from Classical Bricks}
\label{sec:intro:quantum_ft_models}
\begin{figure}[H]
\centering
\subfigure[]{\label{fig: gate_perfect}\includegraphics[width=0.3\linewidth]{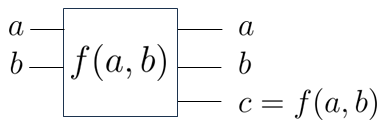}}
\centering
\subfigure[]{\label{fig: gate_faulty_classical}\includegraphics[width=0.3\linewidth]{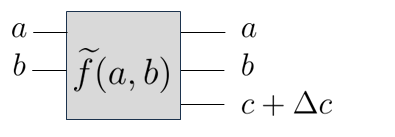}}
\centering
\subfigure[]{\label{fig: gate_faulty_quantum}\includegraphics[width=0.3\linewidth]{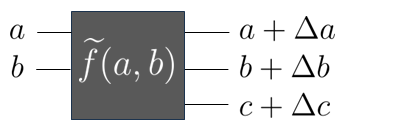}}
\vspace{-6 pt}
\caption{\protect\subref{fig: gate_perfect} Perfect gate,
         \protect\subref{fig: gate_faulty_classical} Faulty classical Boolean gate
         \protect\subref{fig: gate_faulty_quantum} Faulty ``quantum'' Boolean gate
}
\label{fig: faulty_gates}
\end{figure}

Now, we start introducing spices and flavors dictated by quantumness of memories and correction circuits, but we still stay in the classical world -- we explain the quantum origins of these assumptions later. First, the major problem is that as opposed to classical faulty gates, which alter only their output, quantum gates not only alter the output, but the input arguments as well. This is depicted in Fig.~\ref{fig: faulty_gates}.
Fig.~\ref{fig: gate_perfect} shows a perfect two-input logic gate performing Boolean function $f(a,b)$ on input arguments $a$ and $b$, and producing output $c=f(a,b)$. Fig.~\ref{fig: gate_faulty_classical} shows the \emph{faulty} gate $\widetilde{f}$ in classical fault-tolerance literature. The ideal output $c=f(a,b)$ is altered by (adding a binary) perturbation $\Delta c$, but  $a$ and $b$ are kept intact. Fig.~\ref{fig: gate_faulty_quantum} is a classical equivalent of a faulty quantum gate (technically, this is valid only for quantum Clifford gates, which are sufficient for implementing quantum error correction). After performing the operation of the gate $\widetilde{f}$, the inputs $a$ and $b$ become $a + \Delta a$ and $b + \Delta b$, together with the output being $c + \Delta c$.

Fortunately, decoding is a classical algorithm that can be performed by classical hardware. Since the work of Taylor and Kuznetsov there have been decades of progress in classical hardware -- for example, highly reliable Complementary Metal-Oxide-Semiconductor (CMOS) technology -- and, as a result, we can assume the decoder to be perfectly reliable. Thus $\widetilde{\mathcal{D}} = \mathcal{D}$, and the only faulty gates are encountered in $\widetilde{\mathcal{S}}$ and $\widetilde{\mathcal{R}}$; the situation seems even better than in the classical setting! Yet, the quantum setting comes with its own complications. These stem from the nature of the errors to be corrected and from the fact that classical information about those errors (i.e., the classical syndrome needed by the decoder) can only be obtained through measurement operations, which in turn may destroy the encoded quantum information. To prevent this, syndrome measurement is performed indirectly:  ancilla (auxiliary) qubits interact with the encoded qubits and are then measured. This will be discussed later; here, our goal is merely to introduce a quantum flavor into the classical setting, without delving too deeply into quantum aspects.  But bringing quantum measurement into the discussion allows us to highlight a fundamental point:  the ultimate purpose of any computation, even quantum, is to serve classical humans, and as such they need only classical answers. Thus, in the end, quantumness is sidestepped by allowing a destructive measurement of the encoded qubits after the last correction round (more explanation will be given later). The classical information resulting from this measurement forms a codeword of a classical code, possibly corrupted by residual errors. If these errors are correctable (by our perfect, fault-free decoder), the quantum memory is said to be stable.
This closely mirrors the classical fault-tolerant memory case, showing that classical and quantum settings are not fundamentally different. In both cases, the underlying principle is to prevent residual errors from accumulating beyond the maximum correctable rate of the perfect decoder.

Note that from the above quantum faulty gate operation shown in Fig.~\ref{fig: gate_faulty_quantum}, it follows that we need to modify the definition of  $\widetilde{\mathcal{S}}$ from that given in Eq.~\eqref{eq: faulty_operations_S_tilde}, so that the faulty operations become
\begin{eqnarray}
   \label{eq: faulty_operations_S_Q}
   (\mathbf{y} + \Delta \mathbf{y}, \mathbf{s}+\Delta\mathbf{s}) &\leftarrow& \widetilde{\mathcal{S}}(\mathbf{y}) \\
   \label{eq: faulty_operations_D_Q}
   \widehat{\mathbf{e}}+\Delta \widehat{\mathbf{e}}    &\leftarrow& \widetilde{\mathcal{D}}(\mathbf{s}+\Delta\mathbf{s}) \\
   \label{eq: faulty_operations_R_Q}
   \mathbf{y}' + \Delta \mathbf{y}'                       &\leftarrow& \widetilde{\mathcal{R}}(\mathbf{y}+\Delta \mathbf{y},\widehat{\mathbf{e}}+\Delta \widehat{\mathbf{e}})
\end{eqnarray} 
Thus, the data vector $\mathbf{y}$ gets modified twice in the correction circuit, once in syndrome computation~(\ref{eq: faulty_operations_S_Q}), once in memory content recovery~(\ref{eq: faulty_operations_R_Q}) (recall that the prime notation indicates an update of the memory content).

\subsection{My Name is \red{Red}, My Name is \blue{Blue} -- Additional Quantum Constraints}
\label{sec:intro:quantum_constraints}
The reader is by now probably already tired of our obsession with classical fault-tolerance and our promise that the quantum part is coming next, but we need one more classical digression. Let suppose that there are two messages, red $\mathcolor{red}{\mathbf{m}}$ and blue $\mathcolor{blue}{\mathbf{m}}$, two (very specially and jointly designed) codes $\mathcolor{red}{\mathcal{C}}$, and $\mathcolor{blue}{\mathcal{C}}$, producing two codewords $\mathcolor{red}{\mathbf{x}}$ and $\mathcolor{blue}{\mathbf{x}}$ that are stored at $2n$ memory locations. The corresponding corrupted codewords are $\mathcolor{red}{\mathbf{y}}$ and $\mathcolor{blue}{\mathbf{y}}$, and the syndromes are $\mathcolor{red}{\mathbf{s}}$ and $\mathcolor{blue}{\mathbf{s}}$. For now, let us ignore the question about why $\mathcolor{red}{\mathcal{C}}$ and $\mathcolor{blue}{\mathcal{C}}$ are so special and how they are related, an answer will come naturally in the next section when we introduce the stabilizer formalism. 

For our discussion, let us assume that the memory has the additional constraint that only one of these stored codewords can be accessed at a time, i.e., implying that the correction of red and blue codewords must be done in a sequential fashion, first red, then blue, red, blue, red, blue, etc. The corresponding correction circuits are $\mathcolor{red}{\widetilde{\mathcal{S}}}$, $\mathcolor{red}{\widetilde{\mathcal{D}}}$, and $\mathcolor{red}{\widetilde{\mathcal{R}}}$, and
$\mathcolor{blue}{\widetilde{\mathcal{S}}}$, $\mathcolor{blue}{\widetilde{\mathcal{D}}}$, and $\mathcolor{blue}{\widetilde{\mathcal{R}}}$.
The mechanisms of blue and red correction circuits is the same, and different color are used to indicate that different parity check matrices $\mathcolor{red}{H}$ and $\mathcolor{blue}{H}$ are used.

But there is another complication - correcting $\mathcolor{red}{\mathbf{y}}$ perturbs $\mathcolor{blue}{\mathbf{y}}$, and vice versa, more specifically, computing the syndrome $\mathcolor{red}{\mathbf{s}}$ in $\mathcolor{red}{\widetilde{\mathcal{S}}}$, alters both $\mathcolor{red}{\mathbf{y}}$ and  $\mathcolor{blue}{\mathbf{y}}$, and the same holds for $\mathcolor{blue}{\widetilde{\mathcal{S}}}$. Therefore, assuming that during one correction round the red correction is done before the blue, we have (note the slight change in the use of single, double, or triple primes in the notation):
\begin{eqnarray}
   \nonumber
   (
    \mathcolor{red}{\widetilde{\mathbf{y}}'},
    \mathcolor{blue}{\widetilde{\mathbf{y}}'},
    \mathcolor{red}{\widetilde{\mathbf{s}}}
    ) &\leftarrow&
    \mathcolor{red}{\widetilde{\mathcal{S}}}(
     \mathcolor{red}{ \mathbf{y}},
     \mathcolor{blue}{ \mathbf{y}}
     ) \\
   \nonumber
   \mathcolor{red}{\widetilde{\widehat{\mathbf{e}}}}
   &\leftarrow&
   \mathcolor{red}{\widetilde{\mathcal{D}}}(\mathcolor{red}{\widetilde{\mathbf{s}}}) \\
   \nonumber
   \mathcolor{red}{\widetilde{\mathbf{y}}''}
   &\leftarrow&
   \mathcolor{red}{\widetilde{\mathcal{R}}}(\mathcolor{red}{\widetilde{\mathbf{y}}'} ,\mathcolor{red}{\widetilde{\widehat{\mathbf{e}}}}),
\end{eqnarray}
for the red correction, and
\begin{eqnarray}
   \nonumber
   (
    \mathcolor{blue}{\widetilde{\mathbf{y}}''},
    \mathcolor{red}{\widetilde{\mathbf{y}}'''},
    \mathcolor{blue}{\widetilde{\mathbf{s}}}
    ) &\leftarrow&
    \mathcolor{blue}{\widetilde{\mathcal{S}}}
    (
    \mathcolor{blue}{\widetilde{\mathbf{y}}'},
    \mathcolor{red}{\widetilde{\mathbf{y}}''}
    ) \\
   \nonumber
   \mathcolor{blue}{\widetilde{\widehat{\mathbf{e}}}}
   &\leftarrow&
   \mathcolor{blue}{\widetilde{\mathcal{D}}}(\mathcolor{blue}{\widetilde{\mathbf{s}}})\\
   \nonumber
   \mathcolor{blue}{\widetilde{\mathbf{y}}'''}
   &\leftarrow&
   \mathcolor{blue}{\widetilde{\mathcal{R}}}(\mathcolor{blue}{\widetilde{\mathbf{y}}''} ,\mathcolor{blue}{\widetilde{\widehat{\mathbf{e}}}} ).
\end{eqnarray}
for blue. 
Each ``tilde'' red and blue intermediate data vectors obtained by faulty sub-processors can be expressed through their perfect, fault-free versions and the corresponding additive perturbations. That is: 
 $\mathcolor{red}{\widetilde{\mathbf{y}}'}=\mathcolor{red}{\mathbf{y}'}+ \rDelta \mathcolor{red}{\mathbf{y}'}$, $\mathcolor{red}{\widetilde{\mathbf{y}}''}=\mathcolor{red}{\mathbf{y}''}+ \rDelta \mathcolor{red}{\mathbf{y}''}$,
$\mathcolor{red}{\widetilde{\mathbf{y}}'''}=\mathcolor{red}{\mathbf{y}'''}+ \rDelta \mathcolor{red}{\mathbf{y}'''}$,
$\mathcolor{blue}{\widetilde{\mathbf{y}}'}=\mathcolor{blue}{\mathbf{y}'}+ \bDelta \mathcolor{blue}{\mathbf{y}'}$, $\mathcolor{blue}{\widetilde{\mathbf{y}}''}=\mathcolor{blue}{\mathbf{y}''}+ \bDelta \mathcolor{blue}{\mathbf{y}''}$, and
$\mathcolor{blue}{\widetilde{\mathbf{y}}'''}=\mathcolor{blue}{\mathbf{y}'''}+ \bDelta \mathcolor{blue}{\mathbf{y}'''}$ (note that fault-free syndrome computation does not modify the input data vector, thus in the above notation $\mathcolor{red}{\mathbf{y}'} = \mathcolor{red}{\mathbf{y}}$, $\mathcolor{red}{\mathbf{y}'''} = \mathcolor{red}{\widetilde{\mathbf{y}}''}$,
$\mathcolor{blue}{\mathbf{y}'} = \mathcolor{blue}{\mathbf{y}}$, and $\mathcolor{blue}{\mathbf{y}''} = \mathcolor{blue}{\widetilde{\mathbf{y}}'}$). 
Perturbation vectors arising from the same faulty sub-processor can be correlated, resulting in errors that affect both red and blue data vectors. Properly characterizing and accounting for these correlations is critical for maximizing decoding accuracy.

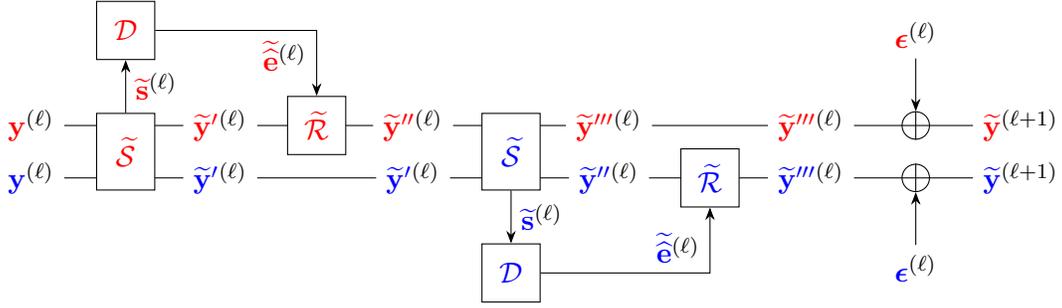
\begin{figure}[H]
  \centering
  \begin{tikzpicture}[
     box/.style={draw,fill=white,  minimum width=22pt, minimum height=22pt, align=center},
     sbox/.style={draw,fill=white,  minimum width=22pt, minimum height=29pt, align=center},
    >=Stealth
]

\def\vsp{0pt}
\def\vvsp{20pt}
\def\hsp{45pt}

\node (yb0) {$\mathblue{\mathbf{y}}^{(\ell)}$};
\node[right=\hsp of yb0] (yb1) {$\mathblue{\widetilde{\mathbf{y}}'}^{(\ell)}$};
\node[right=\hsp of yb1] (yb2) {$\mathblue{\widetilde{\mathbf{y}}'}^{(\ell)}$};
\node[right=\hsp of yb2] (yb3) {$\mathblue{\widetilde{\mathbf{y}}''}^{(\ell)}$};
\node[right=\hsp of yb3] (yb4) {$\mathblue{\widetilde{\mathbf{y}}'''}^{(\ell)}$};
\node[right=\hsp of yb4] (yb5) {$\mathblue{\widetilde{\mathbf{y}}}^{(\ell+1)}$};

\draw (yb0) -- (yb1) -- (yb2) -- (yb3) -- (yb4) -- (yb5);

\node[above=\vsp of yb0] (yr0) {$\mathred{\mathbf{y}}^{(\ell)}$};
\node[above=\vsp of yb1] (yr1) {$\mathred{\widetilde{\mathbf{y}}'}^{(\ell)}$};
\node[above=\vsp of yb2] (yr2) {$\mathred{\widetilde{\mathbf{y}}''}^{(\ell)}$};
\node[above=\vsp of yb3] (yr3) {$\mathred{\widetilde{\mathbf{y}}'''}^{(\ell)}$};
\node[above=\vsp of yb4] (yr4) {$\mathred{\widetilde{\mathbf{y}}'''}^{(\ell)}$};
\node[above=\vsp of yb5] (yr5) {$\mathred{\widetilde{\mathbf{y}}}^{(\ell+1)}$};

\draw (yr0) -- (yr1) -- (yr2) -- (yr3) -- (yr4) -- (yr5);

\node[sbox] (Sr) at ($(yb0)!0.5!(yr1)$) {$\mathred{\widetilde{\mathcal{S}}}$};

\node[box, above=\vvsp of Sr] (Dr) {$\mathred{\mathcal{D}}$};

\node[box] (Rr) at ($(yr1)!0.5!(yr2)$) {$\mathred{\widetilde{\mathcal{R}}}$};

\draw[->] (Sr) -- (Dr) node[midway,right] {$\mathred{\widetilde{\mathbf{s}}}^{(\ell)}$};

\draw[->] (Dr) -| (Rr) node[midway,below left] {$\mathred{\widetilde{\widehat{\mathbf{e}}}}^{(\ell)}$};

\node[sbox] (Sb) at ($(yb2)!0.5!(yr3)$) {$\mathblue{\widetilde{\mathcal{S}}}$};

\node[box, below=\vvsp of Sb] (Db) {$\mathblue{\mathcal{D}}$};

\node[box] (Rb) at ($(yb3)!0.5!(yb4)$) {$\mathblue{\widetilde{\mathcal{R}}}$};

\draw[->] (Sb) -- (Db) node[midway,right] {$\mathblue{\widetilde{\mathbf{s}}}^{(\ell)}$};

\draw[->] (Db) -| (Rb) node[midway,above left] {$\mathblue{\widetilde{\widehat{\mathbf{e}}}}^{(\ell)}$};

\node[draw, circle, minimum size=5pt] (or) at ($(yr4)!0.5!(yr5)$) {};
\draw (or.north) -- (or.south);

\node[above=\vvsp of or] (er) {$\mathred{\beps}^{(\ell)}$};
\draw[->] (er) -- (or);

\node[draw, circle, minimum size=5pt] (ob) at ($(yb4)!0.5!(yb5)$) {};
\draw (ob.north) -- (ob.south);

\node[below=\vvsp of ob] (eb) {$\mathblue{\beps}^{(\ell)}$};
\draw[->] (eb) -- (ob);

\end{tikzpicture}
  \caption{Faulty correction circuit}\label{fig: faulty_correction_circuits_rb}
\end{figure}

Thus, along the correction rounds, we have the sequence of  red and blue corrupted codewords,
$\mathcolor{red}{\mathbf{y}}^{(1)},\mathcolor{red}{\mathbf{y}}^{(2)},\ldots \mathcolor{red}{\mathbf{y}}^{(T)}$ and
$\mathcolor{blue}{\mathbf{y}}^{(1)},\mathcolor{blue}{\mathbf{y}}^{(2)},\ldots \mathcolor{blue}{\mathbf{y}}^{(T)}$, wherein in each correction round we have
$\mathcolor{red}{\mathbf{y}}^{(\ell)}=\mathcolor{red}{\widetilde{\mathbf{y}}'''}^{(\ell-1)} + \mathcolor{red}{\beps}^{(\ell)}$
and
$\mathcolor{blue}{\mathbf{y}}^{(\ell)}=\mathcolor{blue}{\widetilde{\mathbf{y}}'''}^{(\ell-1)} + \mathcolor{blue}{\beps}^{(\ell)}$, with $\mathcolor{red}{\beps}^{(\ell)}$ and $\mathcolor{blue}{\beps}^{(\ell)}$ denoting the red and blue memory errors,
and the initialization is as follows
$\mathcolor{red}{\mathbf{y}}^{(0)}=\mathcolor{red}{\mathbf{x}}$, and
$\mathcolor{blue}{\mathbf{y}}^{(0)}=\mathcolor{blue}{\mathbf{x}}$. This is illustrated in Fig.~\ref{fig: faulty_correction_circuits_rb}. 
The residual red error before the correction round is $\mathcolor{red}{\widetilde{\mathbf{y}}'''}^{(\ell-1)} + \mathcolor{red}{\mathbf{x}}$. The accumulated red error is  $\mathcolor{red}{\mathbf{e}}^{(\ell)} = \mathcolor{red}{\widetilde{\mathbf{y}}'''}^{(\ell-1)} + \mathcolor{red}{\mathbf{x}} + \mathcolor{red}{\beps}^{(\ell)} = \mathcolor{red}{\mathbf{y}}^{(\ell)} + \mathcolor{red}{\mathbf{x}}$,  and its estimate is $\mathcolor{red}{\widetilde{\widehat{\mathbf{e}}}}^{(\ell)}$. The same applies to blue errors.

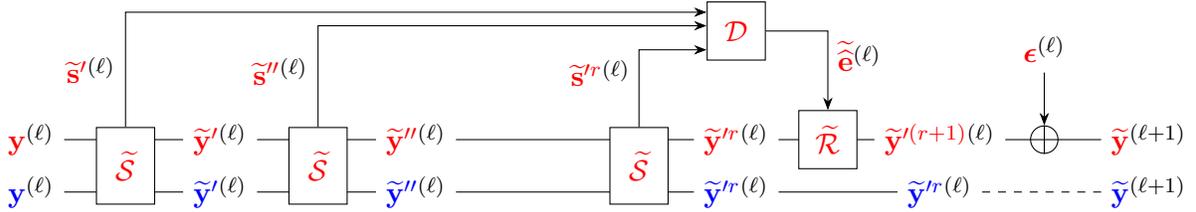
\begin{figure}[H]
  \centering
\begin{tikzpicture}[
     box/.style={draw,fill=white,  minimum width=22pt, minimum height=22pt, align=center},
     sbox/.style={draw,fill=white,  minimum width=22pt, minimum height=29pt, align=center},
    >=Stealth
]

\def\vsp{0pt}
\def\vvsp{20pt}
\def\hsp{45pt}

\node (yb0) {$\mathblue{\mathbf{y}}^{(\ell)}$};
\node[right=\hsp of yb0] (yb1) {$\mathblue{\widetilde{\mathbf{y}}'}^{(\ell)}$};
\node[right=\hsp of yb1] (yb2) {$\mathblue{\widetilde{\mathbf{y}}''}^{(\ell)}$};
\node[right=2*\hsp of yb2] (yb3) {$\mathblue{\widetilde{\mathbf{y}}'^{r}}^{(\ell)}$};
\node[right=\hsp of yb3] (yb4) {$\mathblue{\widetilde{\mathbf{y}}'^{r}}^{(\ell)}$};
\node[right=\hsp of yb4] (yb5) {$\mathblue{\widetilde{\mathbf{y}}}^{(\ell+1)}$};

\draw (yb0) -- (yb1) -- (yb2) -- (yb3) -- (yb4);
\draw[dashed] (yb4) -- (yb5);

\node[above=\vsp of yb0] (yr0) {$\mathred{\mathbf{y}}^{(\ell)}$};
\node[above=\vsp of yb1] (yr1) {$\mathred{\widetilde{\mathbf{y}}'}^{(\ell)}$};
\node[above=\vsp of yb2] (yr2) {$\mathred{\widetilde{\mathbf{y}}''}^{(\ell)}$};
\node[above=\vsp of yb3] (yr3) {$\mathred{\widetilde{\mathbf{y}}'^{r}}^{(\ell)}$};
\node[above=\vsp of yb4] (yr4) {$\mathred{\widetilde{\mathbf{y}}'^{(r+1)}}^{(\ell)}$};
\node[above=\vsp of yb5] (yr5) {$\mathred{\widetilde{\mathbf{y}}}^{(\ell+1)}$};

\draw (yr0) -- (yr1) -- (yr2) -- (yr3) -- (yr4) -- (yr5);

\node[sbox] (Sr1) at ($(yb0)!0.5!(yr1)$) {$\mathred{\widetilde{\mathcal{S}}}$};

\node[sbox] (Sr2) at ($(yb1)!0.5!(yr2)$) {$\mathred{\widetilde{\mathcal{S}}}$};

\node[sbox, right=2.2*\hsp of Sr2] (Sr3) {$\mathred{\widetilde{\mathcal{S}}}$};

\node[box, above=\vvsp of yr3] (Dr) {$\mathred{\mathcal{D}}$};

\node[box] (Rr) at ($(yr3)!0.45!(yr4)$) {$\mathred{\widetilde{\mathcal{R}}}$};

\draw[->] (Sr1) |- ($(Dr.west)+(0,0.25)$) node[pos=0.35,below left] {$\mathred{\widetilde{\mathbf{s}}'}^{(\ell)}$};

\draw[->] (Sr2) |- ($(Dr.west)+(0,0.07)$) node[pos=0.39,below left] {$\mathred{\widetilde{\mathbf{s}}''}^{(\ell)}$};

\draw[->] (Sr3) |- ($(Dr.west)-(0,0.25)$) node[pos=0.51,below left] {$\mathred{\widetilde{\mathbf{s}}'^{r}}^{(\ell)}$};

\draw[->] (Dr) -| (Rr) node[midway,below right] {$\mathred{\widetilde{\widehat{\mathbf{e}}}}^{(\ell)}$};

\node[draw, circle, minimum size=5pt] (or) at ($(yr4)!0.5!(yr5)$) {};
\draw (or.north) -- (or.south);

\node[above=\vvsp of or] (er) {$\mathred{\beps}^{(\ell)}$};
\draw[->] (er) -- (or);

\end{tikzpicture}
  \caption{Faulty correction circuit with repeated syndrome measurements}\label{fig: repeated_syndrome_measurement_decoding}
\end{figure}

Finally, we note that there are alternative ways of decoding and recovery operations. One such alternative (for correcting red errors, the blue errors are corrected similarly) is shown in Fig.~\ref{fig: repeated_syndrome_measurement_decoding}. 
The main difference is that multiple syndromes are collected and decoded jointly, which may help handle errors occurring during faulty syndrome computation (which may not be detected by the computed syndrome, but can be detected by subsequent syndromes). This gives rise to a decoding window that extends in time, an approach known as space-time decoding. Decoding windows can also overlap, allowing errors introduced by the last faulty syndrome computation in each window to be handled. 

\subsection{Outline of the Rest of the Article}
\label{sec:intro:outline}
In the remainder of the article, we proceed to highlight the major innovations and contributions to  QLDPC codes, starting from (a) quantum noise models, continuing with (b) quantum specific constraints on structure of LDPC codes and recent
advancements in QLDPC code construction, and (c) basic principles of iterative decoding and advanced decoding techniques, and finish with (d) remarks and open problems. We discuss the critical challenges of QEC, which quantum systems must overcome to reach full-scale, fault-tolerant quantum computing. In particular, we stress how major technology hurdles, cognizant of hardware constraints, have to be addressed while meeting the reliability demands. The authors hope that the paper gives a holistic perspective on this topic, bridging theory and practice to motivate the Information Theory Society members and wider audience to explore a broad spectrum of challenges associated with QEC and fault-tolerance.

\section{Basics}
\label{sec:basics}
In this section, we provide basic definitions and concepts of quantum mechanics that will be needed in subsequent sections. The only originality in it comes from possible uncorrected typos, even this disclaimer is not original. We start by introducing a channel model for quantum memory, discussing commuting operators and symplectic inner product. Then, we introduce stabilizer codes and relate them to classical error correction codes, and finally give details of faulty quantum gates used to compute the syndrome. For a more general and detailed introduction, the reader may refer to~\cite{nielsen2010quantum}, in particular Chapters~2, 4, 8, and 10. Readers already familiar with the fundamentals of quantum computation may skip ahead to Section~\ref{sec:noise}.

\subsection{Qubits, Unitary Evolution, and Projective Measurements}
\label{sec:basics:qubits}
A \emph{pure} quantum state $\ket{\psi}$ is a unit-norm column vector (ket) in a two-dimensional complex vector space, that is $\ket{\psi} = a \ket{0} + b \ket{1}$, where $a,b \in \mathbb{C}$, and vectors  $\ket{0}=\pmatrixstretch\begin{pmatrix} 1 \\ 0 \end{pmatrix}$ and $\ket{1}=\pmatrixstretch\begin{pmatrix} 0 \\ 1 \end{pmatrix}$ form the so-called \emph{computational} basis. Therefore, unlike classical bit having ``state'' either 0 or 1, a qubit $\ket{\psi}$ is a linear superposition of basis states $\ket{0}$ and $\ket{1}$.
If $\bra{\psi}$ denotes the row vector (bra) that is the complex conjugate of $\ket{\psi}$, then from the unity of the norm it follows that $\braket{\psi}=|a|^2 + |b|^2 =1$.

Similarly, an $n$-qubit state can be written as $\ket{\psi} = a_0\ket{00 \ldots 0} + a_1\ket{00 \ldots 1} + \ldots + a_{2^n-1}\ket{11 \ldots 1}$, where the basis vectors are $\ket{x_1x_2 \ldots x_n} \triangleq \ket{x_1}\otimes \ket{x_2} \ldots \otimes \ket{x_n}, \ket{x_i} \in \{\ket{0},\ket{1} \}$, where $\otimes$ is Kronecker (tensor) product, and the coefficients $a_i \in \mathbb{C}$, with $\sum_i|a_i|^2 =1$. Each computational basis vector has only a single nonzero (one) element at the position whose natural binary code is
$x_1x_2 \ldots x_n$. For $n=3$, they are shown below with the positions indicated on the right.

\begin{equation*}
  \ket{000}=\left(
  \begin{array}{c}
  1 \\
  0 \\
  0 \\
    0 \\
  \vdots \\
  0
  \end{array}
  \right),
 \ket{001}=\left(
  \begin{array}{c}
  0 \\
  1 \\
  0 \\
    0 \\
  \vdots \\
  0
  \end{array}
  \right),
   \ket{010}=\left(
  \begin{array}{c}
  0 \\
  0 \\
  1 \\
    0 \\
  \vdots \\
  0
  \end{array}
  \right),
     \ket{011}=\left(
  \begin{array}{c}
  0 \\
  0 \\
  0 \\
  1 \\
  \vdots \\
  0
  \end{array}
  \right), \ldots
       \ket{111}=\left(
  \begin{array}{c}
  0 \\
  0 \\
  0 \\
  0 \\
  \vdots \\
  1
  \end{array}
  \right)
\quad
\begin{array}{c}
  0 \\
  1 \\
  2 \\
  3 \\
  \vdots \\
  7
  \end{array}
\end{equation*}

Quantum states evolve through unitary operators, which are linear transformations that preserve inner products. Specifically, an $n$-qubit unitary operator $U$ is a linear operator that satisfies $UU^\dagger = U^\dagger U = I$, where $U^\dagger$ is the Hermitian adjoint (conjugate transpose) of $U$. If $\ket{\psi} \in \mathbb{C}^{2^n}$ is a unit-norm vector (a quantum state), so is  $U\ket{\psi}$.

Hermitian operators, that is, linear operators $O$ satisfying $O = O^\dagger$, form another important class of operators in quantum mechanics. They are also known as \emph{observables} and are associated with the concept of projective measurements. Each observable admits a spectral decomposition into real eigenvalues and corresponding eigenprojectors. When a quantum state $\ket{\psi}$ is measured using an observable $O$, the state is projected, probabilistically, onto one of the eigenspaces of $O$. The probability of projecting into an eigenspace associated with eigenvalue $o_i$ is given by $\bra{\psi}P_i\ket{\psi} = ||P_i\ket{\psi}||^2$, where $P_i$ is the projector onto that eigenspace. The post-measurement state is the normalized projected state $\frac{P_i\ket{\psi}}{||P_i\ket{\psi}||}$ and the outcome of the measurement is the corresponding eigenvalue $o_i$.

\subsection{Pauli Operators and Errors}
\label{sec:basics:errors}
Since nature prevents us from observing quantum states, we better stop right away talking about them. After all, the goal of quantum error correction is not to observe quantum states but only to protect them. From now on, we will merely talk about errors, which we will model as undesired actions of Pauli operators on a quantum state (see Section~\ref{sec:noise} for a more detailed discussion of noise models).

Let us begin by introducing the single-qubit Pauli matrices $X= \pmatrixstretch\begin{pmatrix} 0 & 1 \\ 1 & 0 \end{pmatrix}$, $Y= i\pmatrixstretch\begin{pmatrix} 0 & -1 \\ 1 & 0 \end{pmatrix}$, and $Z= \pmatrixstretch\begin{pmatrix} 1 & 0 \\ 0 & -1 \end{pmatrix}$. These matrices are both unitary and Hermitian, hence self-inverse (i.e., $X^2=Y^2=Z^2=I$, the identity matrix), with eigenvalues $\pm1$. 
The Pauli $Z$ matrix has eigenstates $\ket{0}$ and $\ket{1}$, which form the computational, or $Z$-basis. The Pauli $X$ matrix has eigenstates $\ket{+} = (\ket{0} + \ket{1})/\sqrt{2}$ and $\ket{-} = (\ket{0} - \ket{1})/\sqrt{2}$, forming the $X$-basis, also called the phase basis. The Pauli $Y$ matrix has eigenstates $\ket{+i} = (\ket{0} + i \ket{1})/\sqrt{2}$ and $\ket{-i} = (\ket{0} - i \ket{1})/\sqrt{2}$, forming the $Y$-basis. The Pauli matrices anticommute, meaning $\{X,Y\} = \{X,Z\} = \{Y,Z\} = 0$, and they satisfy the relation $Y = i X Z$.

\begin{figure}[!t]
  \centering
  \includegraphics[width=0.4\linewidth]{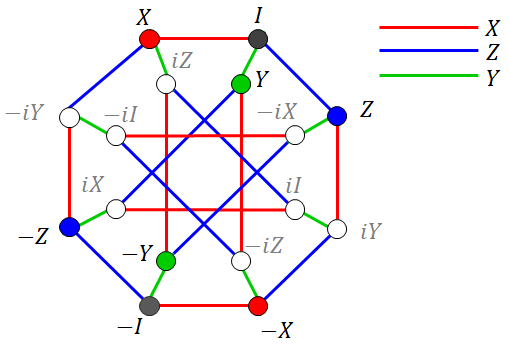}
  \caption{Cayley diagram of Pauli group $\mathcal{P}_1$}\label{fig: pauli_group}
\end{figure}

The  single-qubit Pauli group, denoted $\mathcal{P}_1$, is the group generated by $I$, $X$, $Y$, and $Z$, and consists of all products of these matrices with coefficients $\pm 1, \pm i$. Its Cayley diagram is shown in Fig.~\ref{fig: pauli_group}, where Pauli matrices (with $\pm1$ coefficients) are depicted by colored filled circles, and colored arrows indicate right multiplication by the corresponding Pauli matrices. The $n$-qubit Pauli group $\mathcal{P}_n$  is the $n$-fold tensor product of $\mathcal{P}_1$. Thus, $\mathcal{P}_n=\{\pm,\pm i \} \times \{I,X,Y,Z\}^{\otimes n}$, and $|\mathcal{P}_n|=4\times4^{n}$ since there are $4^n$ possible tensor products, and another factor of 4 is for the possible product by $\pm 1$ and $\pm i$.

We use Pauli matrices to model errors in a quantum channel.
Let us first examine their action on a single-qubit state: $X(a \ket{0} + b \ket{1}) = a \ket{1} + b \ket{0}$, so it acts as a \emph{bit-flip}, exchanging $Z$-basis states  $\ket{0}$ and $\ket{1}$; $Z(a \ket{0} + b \ket{1}) = a \ket{0} - b \ket{1}$, so it acts as a \emph{phase-flip}, exchanging $X$-basis states  $\ket{+}$ and $\ket{-}$; finally, since $Y = iXZ$, it acts as both a bit- and a phase-flip (up to a global phase $i$ that has no physical significance in quantum mechanics).

A \emph{single-qubit Pauli channel} acts on a single-qubit state $\ket{\psi}$ by randomly applying the operators $I$, $X$, $Y$, or $Z$ with probabilities $p_I$, $p_X$, $p_Y$, $p_Z$,  which satisfy $p_I + p_X + p_Y + p_Z = 1$. One of the most common examples is the \emph{single-qubit depolarizing channel}  with error probability $p$, for which $p_I=1-p$, and $p_X = p_Y = p_Z = p/3$. 
For an $n$-qubit system, if we consider a single-qubit depolarizing channel acting independently on each qubit, the resulting total error $E = E_1\otimes\cdots\otimes E_n\in \{I, X, Y, Z\}^{\otimes n}$ has probability $\Pr(E) = (1-\alpha)^{(n-|E|)} (\alpha/3)^{|E|}$ that depends only on the error \emph{weight} $|E|$, i.e., the number of non-identity\footnote{By a common abuse of language, we include the identity among the Pauli matrices and refer to $X$, $Y$, and $Z$ as the non-identity (or non-trivial) Pauli matrices.} Pauli matrices among the $E_i$.  Alternatively, we may consider that any non-trivial error $E \neq I^{\otimes n}$ occurs with equal probability (similarly to the single-qubit case), which is known as the \emph{$n$-qubit depolarizing channel}. The precise probability of an error will not be important for our purposes here; for now, it is only important to note that errors are modeled as single- or multi-qubit Hermitian Pauli operators (the reader noticed that we ignore the coefficients $\pm i$, since they correspond to a global phase with no physical significance).

\subsection{From Classical to Stabilizer Codes}
\label{sec:basics:stabilizers}

Classical bits are protected against errors by enforcing linear dependencies -- parity checks -- between them. Formally, we consider a parity-check matrix $H$ and define the code space as its kernel, that is, the set of vectors  $\mathbf{x}$ (the codewords) satisfying $\mathbf{x} H^{\text{T}} = 0$. If $H$ has $n$ columns and $n-k$ independent rows, then the code space has dimension $k$, allowing us to encode $k$ \emph{logical} (uncoded) bits into $n$ \emph{physical} (coded) bits. If a codeword $\mathbf{x}$ is corrupted by an error $\mathbf{e}$ producing a non-zero syndrome $\mathbf{s} = (\mathbf{x}+\mathbf{e}) H^{\text{T}} = \mathbf{e}H^{\text{T}}$, then we can detect the error, and possibly correct it. The minimum distance of the code is defined as the minimum weight of a nonzero codeword, or, by analogy with the quantum stabilizer code case below, the minimum weight of an undetectable (nonzero) error.  A code with minimum distance $d$ can correct any $\left\lfloor\frac{d-1}{2}\right\rfloor$ bit errors.
 
Protecting quantum states against Pauli errors follows a similar principle, exploiting the concept of stabilizer codes.  Let $\mathcal{S}=\langle h_1,h_2,\ldots, h_{n-k} \rangle$ be an Abelian subgroup of $\mathcal{P}_n$, where the notation $\langle \cdot \rangle$ indicates that the subgroup is generated by its elements $h_i$, $1 \le i \le n-k$, which we assume to be independent for simplicity. Every element of $\mathcal{S}$ can be written as a product of its (independent and commuting) generators. We also require that $\mathcal{S}$ does not contain $-I$ (to ensure that the code space, defined below, is nontrivial), which in particular implies that all elements of $\mathcal{S}$ are Hermitian (not containing $\pm i$ coefficients). The stabilizer code defined by $\mathcal{S}$ is the subspace of quantum states left invariant by the elements of $\mathcal{S}$, that is, quantum states $\ket{\psi}$ satisfying  $h\ket{\psi} = \ket{\psi}$, $\forall h\in \mathcal{S}$. It is of course sufficient that this condition be satisfied for generators $h_i$, $1\leq i \leq n-k$. Elements of $\mathcal{S}$ are called \emph{stabilizers} and $\mathcal{S}$ itself is called the \emph{stabilizer group} (we shall also refer to stabilizers as parity checks, and the reason for this will become clear in the next section). It can be seen that each independent stabilizer halves the dimension of the code space, thus, for a stabilizer group $\mathcal{S}$ with $n-k$ independent generators, the code space has dimension  $2^k$, and allows encoding $k$ \emph{logical} (uncoded) qubits into $n$ \emph{physical} (coded) qubits. 

Let $E$ be an $n$-qubit Pauli error acting on a code state $\ket{\psi}$, and let $\widetilde{\ket{\psi}} = E\ket{\psi}$ denote the error corrupted state. Note that $E$ either commutes or anticommutes with the stabilizer generators. If $E$  commutes with all the generators (and thus, all the stabilizers), we have $h_i\widetilde{\ket{\psi}} = h_i E\ket{\psi} = Eh_i\ket{\psi} = E\ket{\psi} = \widetilde{\ket{\psi}}$. Hence, $\widetilde{\ket{\psi}}$ is also a code state, and we have no way to detect that an error has happened. Such an undetectable error lives in the \emph{centralizer group}\footnote{The standard notation for the centralizer group is $Z(\cdot)$, not to be confused with the Pauli-$Z$ operator; the intended meaning should be clear from the context.} $Z(\mathcal{S}) = \{E \in \mathcal{P}_n \mid Eh = hE, \forall h\in \mathcal{S}\}$ (it can be shown that the centralizer of a stabilizer group is equal to its normalizer, though we will not need this fact here). The centralizer $Z(\mathcal{S})$ contains the stabilizer group $\mathcal{S}$ itself, but errors $E\in \mathcal{S}$ are not harmful, as they act trivially on the code space, that is, $\widetilde{\ket{\psi}} = \ket{\psi}$. Errors $E\in Z(\mathcal{S}) \setminus \mathcal{S}$ are problematic, as they cannot be detected and act non-trivially on the code space. Such errors correspond to so-called \emph{logical Pauli operators}: Pauli operators that leave the code space invariant but act non-trivially on it. Similarly to the classical case, the minimum distance of a stabilizer code is defined as the minimum weight of an undetectable error that acts non-trivially on the code space: $d = \min \{ |E| \mid E \in  Z(\mathcal{S}) \setminus \mathcal{S}\}$. A stabilizer code with minimum distance $d$ can correct any $\left\lfloor\frac{d-1}{2}\right\rfloor$ Pauli errors. A stabilizer code encoding $k$ qubits into $n$ qubits with distance $d$ is denoted as an $[[n,k,d]]$ stabilizer code.

So far, we have only discussed undetectable errors, which leave the code space invariant. Fortunately, errors can also be detected and corrected. If the error $E$ anticommutes with at least one generator, say $h_i$, then $h_i\widetilde{\ket{\psi}} = -\widetilde{\ket{\psi}}$, and thus $\widetilde{\ket{\psi}}$ lies outside the code space. Although it lies outside the code space, $\widetilde{\ket{\psi}}$ remains an eigenstate of $h_i$,  with eigenvalue $-1$. This is where quantum measurement comes into play. Since stabilizers are Hermitian Pauli operators, they can be measured following the principle of projective measurement discussed in Section~\ref{sec:basics:qubits}. Because $\widetilde{\ket{\psi}}$ lies in the $(\pm 1)$-eigenspaces of the generators (with the sign depending on their commutation or anticommutation with the error), measuring a generator leaves $\widetilde{\ket{\psi}}$ unchanged, while revealing the corresponding eigenvalue. The collection of these measurement outcomes defines the \emph{syndrome} $\mathbf{s} = (s_1,s_2,\dots,s_{n-k})$.  Viewing $\mathbf{s}$ as a binary vector (rather than $\pm 1$), we have  $h_i\widetilde{\ket{\psi}} = (-1)^{s_i} \widetilde{\ket{\psi}} \Leftrightarrow (-1)^{s_i} E h_i = h_i E$. This allows us to detect errors, and the next section will explain how to correct them by translating commutation and anticommutation properties revealed by syndrome measurements into linear binary equations. We will say more about how stabilizer generators are measured in Section~\ref{sec:basics:cnot-gate}.

\subsection{Commutativity as Parity Check Verification and Syndrome Decoding}
\label{sec:basics:commutativity}

Note that Pauli operators play a dual role: they represent both the errors affecting qubits and the stabilizers used to detect and correct these errors. In the previous section we used different notation for errors ($E$) and stabilizers ($h_i$).  Here, we shall use $P$ to denote Pauli operators in general, whenever we wish to avoid committing to a specific role as an error or stabilizer. 

To illustrate the commutation and anticommutation of two Pauli operators, we begin with a simple example. Let $P_1=X \otimes X \otimes I \otimes Z \otimes I \otimes Y$ be an $n=6$-qubit Hermitian Pauli operator  (we often write it for short as a string $P_1=XXIZIY$ or by listing only nontrivial operators and their positions $P_1=X_1X_2Z_4Y_6$), and let $P_2=I \otimes Z \otimes I \otimes I \otimes I \otimes X$ ($P_2=IZIIIX$ or $P_2=Z_2X_6$) be another such operator. The answer to whether these two operators commute can be obtained by counting the number of places in which the corresponding strings contain distinct (and thus anticommuting) non-trivial operators. In our example, these positions are $2$ and $6$ - two positions. Since this number is even, $P_1$ and $P_2$ commute, and this is true in general. If it were odd, they would anticommute.
To facilitate verification of commutativity by an algebraic condition, we introduce a binary representation of Pauli operators.

A Pauli error on  $n$ qubits can be expressed as a binary error vector of length $2n$ by mapping the Pauli operators to binary tuples as follows: $I \rightarrow (\mathcolor{red}{0},\mathcolor{blue}{0}),X \rightarrow (\mathcolor{red}{1},\mathcolor{blue}{0}),Z \rightarrow (\mathcolor{red}{0},\mathcolor{blue}{1}),Y \rightarrow (\mathcolor{red}{1},\mathcolor{blue}{1})$.  
Thus, if $P$ is an $n$-qubit Pauli error, its binary representation is
$\mathbf{p}=(\mathcolor{red}{\mathbf{p}} \enspace \mathcolor{blue}{\mathbf{p}})$.
The binary error vector corresponding to $P_1=\rX \rX I \bZ I \gY$
is $\mathbf{p}_1=(\mathcolor{red}{\mathbf{p}}_1 \enspace  \mathcolor{blue}{\mathbf{p}}_1)=(\mathcolor{red}{110001}  \enspace \mathcolor{blue}{000101})$, and the one corresponding to $P_2=I \bZ III \rX$ is $\mathbf{p}_2=(\mathcolor{red}{\mathbf{p}}_2 \enspace  \mathcolor{blue}{\mathbf{p}}_2)=(\mathcolor{red}{000001}  \enspace \mathcolor{blue}{010000})$.
where for visual clarity, we use a color code in which red and blue denote the left (`{\red X}') and right (`{\blue Z}') parts of the vector, respectively. The $\gY$ at the last position in $P_1$, leads to 1 at the last positions of both $\mathcolor{red}{\mathbf{p}}_1$ and $\mathcolor{blue}{\mathbf{p}}_1$. We use green for \darkgreen{$Y$}, although it would make more sense to use purple, which is a mixture of blue and red, but we have an aesthetic excuse\footnote{Some other RGB mixtures once created live on their own. Green for example, orange too. For orange ask Almodovar or ask painters of facades in Cura\c{c}ao. Purple alone is visually incomplete, it needs another color, contrast. Yellow, for example. A field of lavender just before the sunset.}. 
The verification whether $P_1$ and $P_2$ commute reduces to counting the number of positions in which $\mathcolor{red}{\mathbf{p}}_1$ and $\mathcolor{blue}{\mathbf{p}}_2$ both have binary ones, and the number of positions in which
$\mathcolor{red}{\mathbf{p}}_2$ and $\mathcolor{blue}{\mathbf{p}}_1$ both have binary ones. This can be expressed in terms of the \emph{symplectic product} $\mathbf{p}_1 \odot \mathbf{p}_2 = (\mathcolor{red}{\mathbf{p}}_1 \enspace  \mathcolor{blue}{\mathbf{p}}_1) \odot (\mathcolor{red}{\mathbf{p}}_2 \enspace  \mathcolor{blue}{\mathbf{p}}_2) =\mathcolor{red}{\mathbf{p}}_1 \mathcolor{blue}{\mathbf{p}}_2^{\text{T}}+\mathcolor{blue}{\mathbf{p}}_1\mathcolor{red}{\mathbf{p}}_2 ^{\text{T}}$ (we shall tacitly assume binary, $\text{mod}\ 2$, operations). Thus, $P_1$ and $P_2$ commute if $\mathbf{p}_1 \odot \mathbf{p}_2 = 0$, and they anticommute if $\mathbf{p}_1 \odot \mathbf{p}_2 = 1$. 

In what follows, we shall refer to the binary representation 
$\mathbf{p}=(\mathcolor{red}{\mathbf{p}} \enspace \mathcolor{blue}{\mathbf{p}})$ as the \emph{symplectic representation} of  Pauli operator $P$. Depending on the context, we may sometimes drop the use of red and blue coloring in favor of the more standard notation $\mathbf{p}=(\mathbf{p}_X \enspace \mathbf{p}_Z)$ commonly used in the literature. While we retain the \textcolor{red}{red} and \textcolor{blue}{blue} notation throughout the rest of this section, we also introduce the $X$ and $Z$ subscripts below, to help the reader become familiar with both conventions.

We now return to stabilizer codes and consider a stabilizer group $\mathcal{S}=\langle h_1,h_2,\ldots, h_{n-k} \rangle$. The \emph{check matrix} of the corresponding stabilizer code is defined as the matrix whose rows are given by the symplectic representations of the generators $h_i$. We can write $H = (\mathred{H_X} \enspace \mathblue{H_Z})$, with its $i$th row given by $(\mathred{\mathbf{h}_{i,X}} \enspace \mathblue{\mathbf{h}_{i,Z}})$, the symplectic representation of Pauli operator $h_i$. The fact that the stabilizer group is Abelian, i.e., that its generators mutually commute, can be expressed as $\mathred{H_X} \mathblue{H_Z}^{\text{T}} + \mathblue{H_Z} \mathred{H_X}^{\text{T}} = 0$. An example of an $n=5$-qubit stabilizer group with four generators $h_1, h_2, h_3, h_4$ and corresponding check matrix $H$ is given below.
\begin{equation}
\label{eq:generators-and-H-matrix}
\begin{array}{ c@{\ =\ }c@{\ }c@{\ }c@{\ }c@{\ }c}
 h_1 & X & Z & Z & X & I  \\
 h_2 & I & X & Z & Z & X  \\
 h_3 & X & I & X & Z & Z  \\
 h_4 & Z & X & I & X & Z  
\end{array}\qquad
  H=\left(
       \begin{array}{ccccc@{\ \ }|@{\ \ }ccccc}
1 & 0 & 0 & 1 & 0 & 0 & 1 & 1 & 0 & 0 \\
0 & 1 & 0 & 0 & 1 & 0 & 0 & 1 & 1 & 0 \\
1 & 0 & 1 & 0 & 0 & 0 & 0 & 0 & 1 & 1 \\
0 & 1 & 0 & 1 & 0 & 1 & 0 & 0 & 0 & 1
       \end{array}
     \right)
\end{equation}
Moreover, the syndrome of a Pauli error $E$ (acting on any given code state) can be expressed as 
\begin{equation*}
    \mathbf{s} = \mathred{\mathbf{e}_X} \mathblue{H_Z}^{\text{T}} + \mathblue{\mathbf{e}_Z} \mathred{H_X}^{\text{T}},
\end{equation*} where $\mathbf{e} = (\mathred{\mathbf{e}_X} \enspace \mathblue{\mathbf{e}_Z})$ is the symplectic representation of $E$. 
Once the syndrome is obtained through measurements of the generators, the role of a \emph{syndrome decoder} is to find an error $\hat{E}$ -- optimally, the most likely one -- matching the measured syndrome. That is, whose symplectic representation $\hat{\mathbf{e}} = (\mathred{\hat{\mathbf{e}}_X} \enspace \mathblue{\hat{\mathbf{e}}_Z})$ satisfies the above equation.  
In the case where each qubit is affected by i.i.d. depolarizing noise (Section~\ref{sec:basics:errors}), the most likely error $\hat{E}$ corresponds to one with the minimum weight (note that in the symplectic representation, a 1 in the same position in both $\mathred{\hat{\mathbf{e}}_X}$ and $\mathblue{\hat{\mathbf{e}}_Z}$ corresponds to a single $Y$ error, so it contributes only once to the error weight).  

In contrast to the classical case, successful decoding does not require the decoded error to be identical to the one that has happened. It is enough that the residual error $E\hat{E}$ acts trivially on the code space, that is,  $E\hat{E} \in \mathcal{S}$\,\footnote{Note that multiplication of Pauli operators corresponds to addition (mod 2) in the symplectic representation.}. In terms of symplectic representation, this means that $\mathbf{e}+\hat{\mathbf{e}} \in \text{rowspace}(H)$, i.e., it is spanned by the rows of $H$. 
 This multiplicity of valid possibilities for a decoder may look like an advantage, as something that would make decoding easier, but it actually creates a difficulty for QLDPC codes and iterative decoders. An iterative decoder is, in essence, an inference algorithm -- a constraint satisfaction algorithm or a Bethe-free energy minimization algorithm --  and these algorithms can encounter convergence issues in the presence of multiple solutions. The decoding difficulty will manifest itself as ambiguous or oscillating messages within the iterative process, potentially preventing the algorithm from converging to a consistent solution. This will be discussed in Section~\ref{sec:decoder}.

\subsection{Encoding and Logical Operators}
Before going further, let us pause briefly to discuss how quantum information can be encoded in a stabilizer code. To begin, we first step back to classical codes to recall how they encode information. To perform encoding with a classical code, we use a generator matrix $G$, whose rows $\mathbf{g}_i$, $1\leq i \leq k$, form a basis of the code space. The $i$th logical bit -- i.e., the $k$-length vector $(0,\dots,1,\dots,0)$ with a single $1$ in position $i$ -- is encoded as $\mathbf{g}_i$. By linearity, an arbitrary message $(m_1,\dots,m_k)$ is then encoded as the linear combination $\sum_{i=1}^k m_i\mathbf{g}_i$. 

This gives us an intuition about how to proceed in the case of a stabilizer code. With a subtlety the reader may have already noticed: in the transition from classical linear codes to quantum stabilizer codes, the code space of the classical code retains the role of describing undetectable nontrivial errors. As such, it is identified in the quantum case with the space of logical Pauli operators $Z(\mathcal{S}) \setminus \mathcal{S}$, rather than the code (state) space of the stabilizer code. 

Since $ \mathcal{S}$ is generated by  $n-k$ independent, mutually commuting generators $h_1,\dots, h_{n-k}$, we can extend the set by $k$ independent,  mutually commuting logical operators $\bar{Z}_1,...,\bar{Z}_k \in Z(\mathcal{S}) \setminus \mathcal{S}$, which are also independent of the stabilizer generators. The notation $\bar{Z}_i$ is not used to highlight their membership in the centralizer group, but rather to indicate that they will serve as \emph{logical Pauli $Z$ operators}. 

Consider an uncoded computational basis state $\ket{m_1\dots m_k}$ of a $k$-qubit system, and let $Z_i$ be the Pauli $Z$ operator acting on the $i$th qubit ($Z_i = I\otimes \cdots  Z \cdots \otimes I$, with the single $Z$ acting in position $i$). We have $Z_i \ket{m_1\dots m_k} = (-1)^{m_i} \ket{m_1\dots m_k}$. Hence, $\ket{m_1\dots m_k}$ lies in the $+1$ eigenspace of $Z_i$ (i.e., is stabilized) if $m_i=0$, and in the $-1$ eigenspace (i.e., is anti-stabilized) if $m_i=1$. We  encode the $k$-qubit state $\ket{m_1\dots m_k}$ into the $n$-qubit state $\ket{\psi_{m_1\dots m_k}}$ that satisfies analogous stabilization conditions. Specifically, we require $\ket{\psi_{m_1\dots m_k}}$ be stabilized by the stabilized generators $h_1,\dots, h_{n-k}$ (so that it lies in the code space), stabilized by the logical operators $\bar{Z}_i$ whenever $m_i = 0$, and anti-stabilized by  $\bar{Z}_i$ whenever $m_i = 1$. These conditions uniquely determine the encoded state $\ket{\psi_{m_1\dots m_k}}$. So,  encoding maps a computational basis state $\ket{m_1\dots m_k}$ to $\ket{\psi_{m_1\dots m_k}}$, and by linearity, any  superposition of computational basis states is mapped to the corresponding superposition of  $\ket{\psi_{m_1\dots m_k}}$ states. 

The reader may notice that we have not provided an explicit expression for the encoded states $\ket{\psi_{m_1\dots m_k}}$ in the computational basis of the $n$-qubit system.  While this is possible, though more arduous, it is actually not necessary. The beauty of the stabilizer formalism is that it allows a compact description of these states in terms of stabilization or anti-stabilization by Pauli operators. Moreover, we can manipulate these states without requiring their explicit description in the computational basis. Let us explain how to apply logical Pauli $X$ operators on the encoded states. For a given set of logical operators $\bar{Z}_1,...,\bar{Z}_k \in Z(\mathcal{S}) \setminus \mathcal{S}$ as above (which specifies a choice of encoding), there exists a unique set of independent, mutually commuting logical operators $\bar{X}_1,...,\bar{X}_k \in Z(\mathcal{S}) \setminus \mathcal{S}$, such that $\bar{X}_i$ and $\bar{Z}_j$ commute if and only if $i=j$. These operators will serve as \emph{logical Pauli $X$ operators}. If $\bar{X}_i$ is applied on the encoded state $\ket{\psi_{m_1\dots m_i\dots m_k}}$, the resulting state will satisfy the same stabilization of anti-stabilization conditions for  Pauli operators $h_1,\dots,h_{n-k}$ and $\bar{Z_1},\dots,\bar{Z}_k$, except for $\bar{Z}_i$, because it anti-commutes with $\bar{X}_i$. Hence, we necessarily have $\bar{X}_i \ket{\psi_{m_1\dots m_i\dots m_k}} = \ket{\psi_{m_1\dots (1-m_i)\dots m_k}}$,  so that $\bar{X}_i$ acts as a Pauli $X$ operator on the $i$th logical qubit.

As a concrete example, for the stabilizer code defined in~\eqref{eq:generators-and-H-matrix}, which encodes $k=1$ logical qubit, we can choose the logical operators as $\bar{Z} = ZZZZZ$ and $\bar{X} = XXXXX$.

\subsection{Calderbank-Shor-Steane Codes}
\label{sec:basics:css_codes}
CSS codes, named after their inventors -- Rob Calderbank, Peter Shor~\cite{calderbank1996quantum_exists} and Andrew Steane~\cite{steane_multiple_1996} -- are a class of stabilizer codes whose generators can be divided into $X$-type and $Z$-type Pauli operators, meaning that each generator is either a tensor product of only Pauli $X$ and identity operators ($X$-type) or a tensor product of only Pauli $Z$ and identity operators ($Z$-type). Accordingly, in the symplectic representation, each stabilizer generator $h_i$ takes the form $(\mathred{\mathbf{h}_{i,X}} \enspace \mathblue{\mathbf{0}})$ for an $X$-type generator or $(\mathred{\mathbf{0}} \enspace \mathblue{\mathbf{h}_{i,Z}})$ for a $Z$-type generator. Consequently, the corresponding check matrix can be written in block form as\footnote{Note the slightly different definition of $\mathred{H_X}$ and $\mathblue{H_Z}$ compared to the general case of stabilizer codes, as they exclude the all-zero blocs below and above, respectively.}
 \begin{equation*}
       H = \left(\begin{array}{cc}
           \mathred{H_X} & \mathblue{0} \\
           \mathred{0} & \mathblue{H_Z}
       \end{array} \right),
   \end{equation*}
and the commutativity between $X$-type and $Z$-type generators rewrites as $\mathred{H_X} \mathblue{H_Z}^{\text{T}} = 0$. Matrices $H_X$ and $H_Z$ define two classical linear codes\footnote{Strictly speaking, $\ker(H_X)$ corresponds to $Z$-type Pauli errors commuting with the $X$-type generators of the stabilizer group. As such, it could more accurately be denoted  $\mathcal{C}_Z$ (and colored in blue); however, this is not the standard convention. We drop the use of red and blue colors in what follows, both to reduce visual discomfort and to prevent possible hardness in interpretation.} $\mathcal{C}_X = \ker(H_X)$ and  $\mathcal{C}_Z = \ker(H_Z)$, satisfying $\mathcal{C}_X^\perp  = \Im(H_X) \subseteq \mathcal{C}_Z$ and similarly $\mathcal{C}_Z^\perp = \Im(H_Z) \subseteq \mathcal{C}_X$. In this paper, we focus on \emph{CSS QLDPC codes}, defined as CSS codes for which  $\mathcal{C}_X$ and $\mathcal{C}_Z$ are classical LDPC codes.

In the binary representation,  $X$-type and $Z$-type stabilizers correspond to vector spaces $\mathcal{C}_X^\perp$ and $\mathcal{C}_Z^\perp$,  generated by the rows of  $H_X$ and $H_Z$, respectively (here, we discard the $\mathbf{0}$ part of the symplectic representation and refer to the remaining part simply as the binary representation). Likewise, the logical operators correspond to the set  $\left(\mathcal{C}_X \setminus \mathcal{C}_Z^{\perp}\right) \cup \left(\mathcal{C}_Z \setminus \mathcal{C}_X^{\perp}\right)$, and the quantum minimum distance of the code is given as
\begin{equation*}
    d_X = \min\{|\mathbf{e}| \mid \mathbf{e} \in \mathcal{C}_X \setminus \mathcal{C}_Z^{\perp}\}, \quad
     d_Z = \min\{|\mathbf{e}| \mid \mathbf{e} \in \mathcal{C}_Z \setminus \mathcal{C}_X^{\perp}\}, \quad
     d = \min (d_X, d_Z)
\end{equation*}
Note that distances $d_X$ and $d_Z$ of the CSS code can be larger than the classical minimum distances of $\mathcal{C}_X$ and $\mathcal{C}_Z$, since we expurgate the codewords that belong to the dual of the opposite code.  

For a Pauli error $E$ acting on a CSS code state, the syndrome can be expressed in terms of its symplectic representation $\mathbf{e} = (\mathbf{e}_X \enspace \mathbf{e}_Z)$, as $\mathbf{s} = (\mathbf{s}_X, \mathbf{s}_Z)$, with
\begin{equation*}
    \mathbf{s}_X = \mathbf{e}_Z H_X^{\text{T}} 
    \quad \text{and \quad }
    \mathbf{s}_Z = \mathbf{e}_X H_Z^{\text{T}}
\end{equation*}
Here, $\mathbf{s}_X$ collects the measurement outcomes of the $X$-type generators, which detect $Z$ and $Y$ errors, while $\mathbf{s}_Z$ collects the measurement outcomes of the $Z$-type generators, which detect $X$ and $Y$ errors. Since a $Y$ error can be viewed as a combination of $X$ and $Z$ errors, in the sequel we will simply refer to $X$ and $Z$ errors, implicitly including the corresponding components of any $Y$ errors. The separation of the syndrome into $X$-type and $Z$-type generators allows us to decode $X$ and $Z$ errors independently. While independent decoding is convenient, as it reduces the problem to decoding two classical (LDPC) codes, it is suboptimal because it ignores the correlations introduced by $Y$ errors between  $\mathbf{e}_X$ and $\mathbf{e}_Z$. To properly account for these correlations, quaternary decoding can be performed. Yet, in the case of LDPC codes with iterative message-passing decoding, the binary nature of the parity-checks renders quaternary decoding effectively a coordinated exchange of information, at each iteration, between two binary decoders~\cite{kuo_refined_2020,kuo_exploitingdegeneracy_2022}. This exchange of information can also be delayed until one of the decoders has converged, providing assistance to the other decoder if it struggles, resulting in a turbo-like decoding approach~\mycite{CorrelatedBF_decodingNithin_2023}.

\subsection{The CNOT Gate, Pauli Propagation, and Stabilizer Measurements}
\label{sec:basics:cnot-gate}

We have managed to reach the definition of stabilizer and CSS codes without ultimately introducing anything beyond Pauli operators acting on qubits and projective measurements. However, to explain how these measurements are actually implemented, we need to consider the circuits that realize them, with the minimal requirement being the introduction of the controlled-NOT (CNOT) operation. 

\begin{figure}[H]
  \centering
  \includegraphics[width=0.5\linewidth]{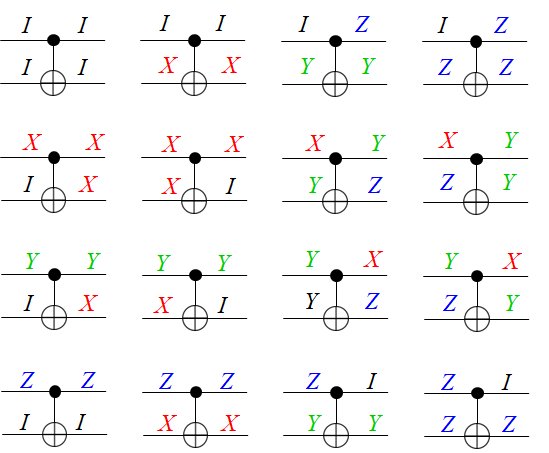}
  \caption{Propagation of Pauli operators through a CNOT gate - table of possibilities.}
  \label{fig: cnot_propagation_table}
\end{figure}

The CNOT operation is a two-qubit gate that flips the state of the target qubit if the control qubit is in the state $\ket{1}$, and does nothing if the control qubit is in $\ket{0}$. Assuming the first qubit is the control and the second the target, its action in the computational basis $\{ \ket{00}, \ket{01}, \ket{10},\ket{11}\}$ writes as $\text{CNOT}\ket{uv} = \ket{u,u\oplus v}$,  where $\oplus$ is the sum modulo 2,  and thus is represented by the matrix:
$$\text{CNOT}= \begin{pmatrix}
        1 & 0 & 0 & 0  \\
        0 & 1 & 0 & 0  \\
        0 & 0 & 0 & 1 \\
        0 & 0 & 1 & 0
        \end{pmatrix}
$$
The property of interest for us is that any two-qubit Pauli operator propagates through the CNOT gate to another two-qubit Pauli operator. Precisely, for any two-qubit Pauli operator $P_1\otimes P_2\in \mathcal{P}_2$, there exists $P'_1\otimes P'_2\in \mathcal{P}_2$ such that $\text{CNOT}(P_1\otimes P_2) = (P'_1\otimes P'_2)\text{CNOT}$. Fig.~\ref{fig: cnot_propagation_table} shows all 16 possibilities for the propagation of two-qubit Pauli operators. Qubits are represented by horizontal wires, while the CNOT gate is depicted as a ``$\bullet$'' on the control wire (first qubit) connected by a vertical line to a ``$\oplus$'' on the target wire (second qubit). Pauli operators $P_1$ and $P_2$ are placed above the corresponding wires to the left of the CNOT gate, and they propagate to Pauli operators  $P'_1$, $P'_2$  shown to the right of the CNOT gate.

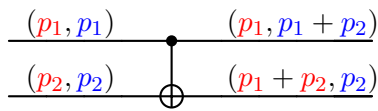
\begin{figure}[H]
  \centering
  \begin{quantikz}
\lstick{\raisebox{1.5ex}{\smash{$(\mathcolor{red}{p_1}, \mathcolor{blue}{p_1})$}}\hspace*{-10ex}} &&&& \ctrl{1} &&&&& \rstick{\hspace*{-12ex}\raisebox{1.5ex}{\smash{$(\mathcolor{red}{p_1}, \mathcolor{blue}{p_1}+\mathcolor{blue}{p_2})$}}} \qw \\
\lstick{\raisebox{1.5ex}{\smash{$(\mathcolor{red}{p_2}, \mathcolor{blue}{p_2})$}}\hspace*{-10ex}} &&&& \targ{}  &&&&& \rstick{\hspace*{-12ex}\raisebox{1.5ex}{\smash{$(\mathcolor{red}{p_1}+\mathcolor{red}{p_2}, \mathcolor{blue}{p_2})$}}} \qw
\end{quantikz}
  \caption{Propagation of Pauli operators - table of possibilities (binary representation)}
  \label{fig: cnot_propagation_binary}
\end{figure}

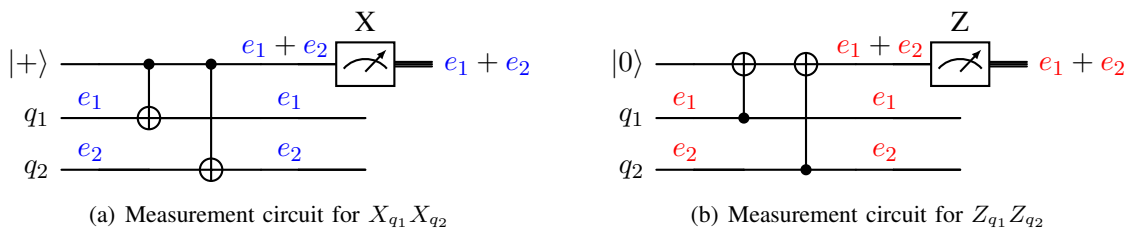
\begin{figure}[htbp]
    \centering
    \subfigure[Measurement circuit for $X_{q_1}X_{q_2}$]{
        \begin{quantikz}[row sep=0pt]
\lstick{$\ket{+}$} && \ctrl{1} & \ctrl{2} &\push{\raisebox{3ex}{\hbox to 0pt{\!\!\!\!$\mathcolor{blue}{e_1}+\mathcolor{blue}{e_2}$\hss}}}&&  \meter{$X$} \qw & \rstick{$\mathcolor{blue}{e_1}+\mathcolor{blue}{e_2}$} \cw \\
\lstick{$q_1$} \push{\raisebox{3ex}{\hbox to 0pt{\;\;$\mathcolor{blue}{e_1}$\hss}}} && \targ{} & \qw      &\push{\raisebox{3ex}{\hbox to 0pt{\;\;$\mathcolor{blue}{e_1}$\hss}}}&& \qw \\
\lstick{$q_2$} \push{\raisebox{3ex}{\hbox to 0pt{\;\;$\mathcolor{blue}{e_2}$\hss}}} && \qw     & \targ{} &\push{\raisebox{3ex}{\hbox to 0pt{\;\;$\mathcolor{blue}{e_2}$\hss}}}&& \qw
\end{quantikz}
        \label{fig:measurement-XX}
    }
    \quad
    \subfigure[Measurement circuit for $Z_{q_1}Z_{q_2}$]{
        \begin{quantikz}[row sep=0pt]
\lstick{$\ket{0}$} && \targ{} & \targ{} &\push{\raisebox{3ex}{\hbox to 0pt{\!\!\!\!$\mathcolor{red}{e_1}+\mathcolor{red}{e_2}$\hss}}}&&  \meter{$Z$} \qw & \rstick{$\mathcolor{red}{e_1}+\mathcolor{red}{e_2}$} \cw \\
\lstick{$q_1$} \push{\raisebox{3ex}{\hbox to 0pt{\;\;$\mathcolor{red}{e_1}$\hss}}} && \ctrl{-1} & \qw      &\push{\raisebox{3ex}{\hbox to 0pt{\;\;$\mathcolor{red}{e_1}$\hss}}}&& \qw \\
\lstick{$q_2$} \push{\raisebox{3ex}{\hbox to 0pt{\;\;$\mathcolor{red}{e_2}$\hss}}} && \qw     & \ctrl{-2} &\push{\raisebox{3ex}{\hbox to 0pt{\;\;$\mathcolor{red}{e_2}$\hss}}}&& \qw
\end{quantikz}
        \label{fig:measurement-ZZ}
    }
    \caption{Measurement circuits for $X$-type and $Z$-type stabilizer generators. For simplicity, we assume that each generator acts on only two qubits $q_1$ and $q_2$.}
    \label{fig:measurement-circuits}
\end{figure}

Using the binary representation of a Pauli operator introduced in Section~\ref{sec:basics:commutativity}, namely $P \rightarrow (\mathcolor{red}{p}, \mathcolor{blue}{p})$, the $16$ possibilities in  Fig.~\ref{fig: cnot_propagation_table} can be summarized as shown in Fig.~\ref{fig: cnot_propagation_binary}. In words, the \textcolor{red}{red} ($X$) component of the Pauli operator propagates from the control to the target qubit, while the \textcolor{blue}{blue} ($Z$) component propagates from the target to the control qubit. We can exploit this property to ``copy'' $X$ or $Z$ errors from the encoded qubits onto ancilla qubits, which are then measured in an appropriate basis to extract the syndrome information.  We illustrate this in Fig.~\ref{fig:measurement-circuits}. Consider an $X$-type stabilizer generator, which, for simplicity, we assume acts on only two qubits, $q_1$ and $q_2$. Recall that $X$-type generators are used to detect $Z$ errors. In Fig.~\ref{fig:measurement-XX} we consider an ancilla qubit that is initialized in state $\ket{+}$, the $(+1)$-eigenstate of $X$, controls two CNOT gates targeting qubits $q_1$ and $q_2$, and is then measured in the $X$ basis. The $Z$-components of the errors affecting qubits $q_1$ and $q_2$, shown in blue, propagate to the ancilla qubit. Measuring the ancilla qubit in the $X$-basis reveals the parity value $\mathcolor{blue}{e_1}+\mathcolor{blue}{e_2}$, since the ancilla state flips from $\ket{+}$ to $\ket{-}$ whenever this parity is odd. However, note that if $\mathcolor{blue}{e_1}$ occurs on qubit $q_1$ after the first CNOT gate -- e.g., due to noise in the gate -- it will no longer be detected by the measurement. Such situations can occur in faulty measurement circuits. Fig.~\ref{fig:measurement-ZZ} shows a similar measurement of a $Z$-type generator. The ancilla qubit is initialized this time in state $\ket{0}$, the $(+1)$-eigenstate of $Z$, it is targeted by two CNOT gates controlled by qubits $q_1$ and $q_2$, and is then measured in the $Z$ basis. The $X$-components of the errors affecting qubits $q_1$ and $q_2$ propagate to the ancilla qubit, whose measurement in the $Z$-basis reveals the parity value $\mathcolor{red}{e_1}+\mathcolor{red}{e_2}$.

\section{The Noise Model Hierarchy}
\label{sec:noise}
We shall now proceed to the task of outlining the noise models under which our quantum error correction architectures are designed to operate. From a systems perspective, error correction systems are often designed around a specific noise model, which implies that the ability of said noise model to accurately reflect physical phenomena greatly influences the overall effectiveness of the system. Yet, it is often the case that the most realistic noise models are also the ones that are the most difficult to mathematically analyze in a tractable manner. This tradeoff between mathematical tractability and physical accuracy often results in the establishment of a hierarchy of noise models of increasing accuracy but also increasing complexity. 

In this sense, quantum hardware is no different. Quantum computing hardware can be realized in various forms, including superconducting qubits~\cite{blais2021circuit, krantz2019quantum, rasmussen2021superconducting}, trapped ions~\cite{leibfried2003quantum, singer2010colloquium, castillo2023electronic}, photonic qubits~\cite{luo2023recent, chen2021quantum}, neutral atoms~\cite{wintersperger2023neutral, graham2022multi}, and several other modalities~\cite{de2021materials}. These hardware platforms vary greatly in form and function, yielding radically different hierarchies of noise models with varying levels of complexity. An elaborate discussion of all these noise models is well beyond the scope of this work. And so, we shall instead content ourselves with a set of sweeping assumptions that will allow us to treat all of these diverse technologies on equal footing.

The first assumption that we make is the assumption of two-level systems, i.e., qubits. One of the fundamental postulates of quantum mechanics is that the state space of a quantum system is described by a complex Hilbert space.\footnote{For the purposes of this work, we shall mostly deal with finite-dimensional Hilbert spaces, in which case a Hilbert space is essentially a vector space armed with an inner product.} In principle, the dimensionality of this Hilbert space could be arbitrarily large, or even infinite, as is the case for superconducting and photonic quantum systems. The assumption of a two-level system involves selecting two specific orthonormal vectors from this larger Hilbert space and assigning them as the two basis states of a qubit. This is akin to the classical concept of modulation, wherein a finite set of points is selected from a continuous space, so that the resulting space can be treated as a digital system. This assumption enables us to visualize the state of our qubit as a 2-dimensional complex vector space, as discussed in Section~\ref{sec:basics}.

The second assumption is that our qubit is initialized in a quantum state that is not entangled with the environment. Under this assumption, there exists a well-established result stating that any noisy channel acting on the state space of a qubit can be described as a completely positive trace preserving (CPTP) map~\cite{Nielsen, Wilde_2017}. From a computing perspective, this assumption is always true, and the qubit is almost always initialized in a fixed quantum state that is decoupled from the environment.

For the purposes of this article, the formal structure of CPTP maps will not be important to us, for we shall use a third approximation, known as the Pauli twirling approximation (PTA), which allows us to approximate an arbitrary CPTP map as a discrete Pauli channel~\cite{geller2013efficient}. A Pauli channel acting on a single qubit simply involves choosing a matrix at random, from the set of Pauli matrices ${I, X, Y, Z}$ (as was defined in Section~\ref{sec:basics}) with probabilities $p_I, p_X, p_Y, p_Z$ respectively, and applying this operation to the 2-dimensional vector representing the state of our qubit. Hence, we are left with the relatively simplified scenario of a two-state quantum system (qubit) where the noise acting upon this system is a stochastic mixture of the discrete set of Pauli operators $\{I, X, Y, Z\}$. 

It is worth noting that the first two assumptions are relatively weak in the sense that they do not usually have significant effect on the physical accuracy of the constructed noise model. However, there are situations in which the PTA turns out not to be a very good approximation (see~\cite{geller2013efficient} for details). Nevertheless, the PTA is usually a consistent staple of the quantum error correction literature due to the mathematical tractability of Pauli operators acting on quantum systems. This mathematical tractability follows largely from the stabilizer formalism~\cite{Gottesman97} and the Gottesman-Knill theorem~\cite{Nielsen, aaronson2004improved}, which states that the effect of Pauli errors on Clifford circuits -- a specific class of quantum circuits used to perform syndrome measurement in stabilizer codes  -- can be efficiently simulated on classical computers.

At this stage, it is important to clarify the difference between the notion of a \emph{noise model} and that of a \emph{noise channel}. In general, the quantum computers that we shall aim to stabilize will consist of a number of qubits, where each qubit is approximately modeled in accordance with the assumptions outlined in the preceding paragraphs. A noise model is a specification of the locations at which noise affects the set of qubits under consideration, whereas a noise channel is a specification of the structure of this noise. Under the PTA, the noise channel is simply a specification of the probability distribution $\{p_I, p_X, p_Y, p_Z\}$.

With these assumptions at hand, we may now begin building our hierarchy of noise models. In particular, we focus on three noise models for qubits, namely, (a) the code-capacity noise model, (b) the phenomenological noise model, and (c) the circuit-level noise model. These noise models are listed in an increasing order of accuracy/complexity, with the code-capacity model being the simplest / least-realistic and the circuit-level noise model being the most complex and realistic.
\begin{figure}

    \centering
    \renewcommand{\arraystretch}{1.5} 
    \setlength{\tabcolsep}{10pt} 
    
    \begin{tabular}{|>{\centering\arraybackslash}m{0.28\linewidth}|
                    >{\centering\arraybackslash}m{0.28\linewidth}|
                    >{\centering\arraybackslash}m{0.28\linewidth}|}
    \hline
    \multicolumn{3}{|c|}{\textbf{Noise Model Hierarchy}} \\ \hline
    \textbf{Code-Capacity Noise} & \textbf{Phenomenological Noise} & \textbf{Circuit-Level Noise} \\ \hline
    
    
    \vspace{1em}\scalebox{0.6}{\begin{quantikz}
    D \quad &
        \noisethunder & \gate[wires=3, style={minimum width=3cm, minimum height=3cm}]{\hspace{0.1cm}\text{\Large{SM Circuit}}\hspace{0.1cm}} & & &  \\
    \mathcolor{blue}{A} \quad & & &
        \meter[style={draw=blue, fill=white}]{} &
        \setwiretype{c} & \\
    \mathcolor{red}{A} \quad & & &
        \meter[style={draw=red, fill=white}]{} & 
        \setwiretype{c} & 
\end{quantikz}}\vspace{1em} &
    \vspace{1em}\scalebox{0.6}{\begin{quantikz}
    D \quad &
        \noisethunder & \gate[wires=3, style={minimum width=3cm, minimum height=3cm}]{\hspace{0.1cm}\text{\Large{SM Circuit}}\hspace{0.1cm}} & & &  \\
    \mathcolor{blue}{A} \quad & & &
        \meter[style={draw=blue, fill=white}]{} &
        \setwiretype{c} \noisethunder & \\
    \mathcolor{red}{A} \quad & & &
        \meter[style={draw=red, fill=white}]{} & 
        \setwiretype{c} \noisethunder & 
\end{quantikz}}\vspace{1em} &
    \vspace{1em}\scalebox{0.6}{\begin{quantikz}
    D \quad &
        \noisethunder & \gate[wires=3, style={minimum width=3cm, minimum height=3cm}]{\text{\textcolor{orange}{\Large\faBolt} \Large{SM Circuit}}} & & &  \\
    \mathcolor{blue}{A} \quad & & &
        \meter[style={draw=blue, fill=white}]{} &
        \setwiretype{c} \noisethunder & \\
    \mathcolor{red}{A} \quad & & &
        \meter[style={draw=red, fill=white}]{} & 
        \setwiretype{c} \noisethunder & 
\end{quantikz}}\vspace{1em} \\ \hline
    
    \vspace{0.5em}$\begin{gathered}
        \mathcolor{red}{\mathbf{e}}, \mathcolor{blue}{\mathbf{e}} \sim \text{\textcolor{orange}{\faBolt}} \\
        \mathcolor{blue}{\mathbf{\tilde{s}}} = \mathcolor{red}{\mathbf{e}} \mathcolor{blue}{{H^{\text{\color{black}T}}}} \\
        \mathcolor{red}{\mathbf{\tilde{s}}} = \mathcolor{blue}{\mathbf{e}} \mathcolor{red}{{H^{\text{\color{black}T}}}}
    \end{gathered}$\vspace{0.5em} &
    \vspace{0.5em}$\begin{gathered}
        \mathcolor{red}{\mathbf{e}}, \mathcolor{blue}{\mathbf{e}},
        \mathcolor{red}{\mathbf{\Delta s}}, \mathcolor{blue}{\mathbf{\Delta s}} 
        \sim \text{\textcolor{orange}{\faBolt}} \\
        \mathcolor{blue}{\mathbf{\tilde{s}}} = \mathcolor{red}{\mathbf{e}} \mathcolor{blue}{{H^{\text{\color{black}T}}}} + \mathcolor{blue}{\mathbf{\Delta s}} \\
        \mathcolor{red}{\mathbf{\tilde{s}}} = \mathcolor{blue}{\mathbf{e}} \mathcolor{red}{{H^{\text{\color{black}T}}}}+ \mathcolor{red}{\mathbf{\Delta s}}
    \end{gathered}$\vspace{0.5em} &
    \vspace{0.5em}$\begin{gathered}
        \mathcolor{red}{\mathbf{f}}, \mathcolor{blue}{\mathbf{f}} \sim \text{\textcolor{orange}{\faBolt}} \\
        \mathcolor{red}{\mathbf{e}} = \mathcolor{red}{\mathbf{f} F_D^{\text{\color{black}T}}}, \;\; 
        \mathcolor{blue}{\mathbf{\tilde{s}}} = \mathcolor{red}{\mathbf{f}} \mathcolor{blue}{F_S^{\text{\color{black}T}}}, \\
        \mathcolor{blue}{\mathbf{e}} = \mathcolor{blue}{\mathbf{f} F_D^{\text{\color{black}T}}}, \;\;
        \mathcolor{red}{\mathbf{\tilde{s}}} = \mathcolor{blue}{\mathbf{f}} \mathcolor{red}{F_S^{\text{\color{black}T}}}, \\
    \end{gathered}$\vspace{0.5em} \\ \hline
    
    \end{tabular}
    \caption{The noise model hierarchy. Noise channels are indicated as \textcolor{orange}{\faBolt}. The code-capacity model has noise channels only on the data qubits. The phenomenological model introduces syndrome noise. The circuit-level noise model introduces correlations between all these noise sources via the circuit faults in a noisy Syndrome Measurement (SM) circuit.}
    \label{fig:noise_model}

\end{figure}
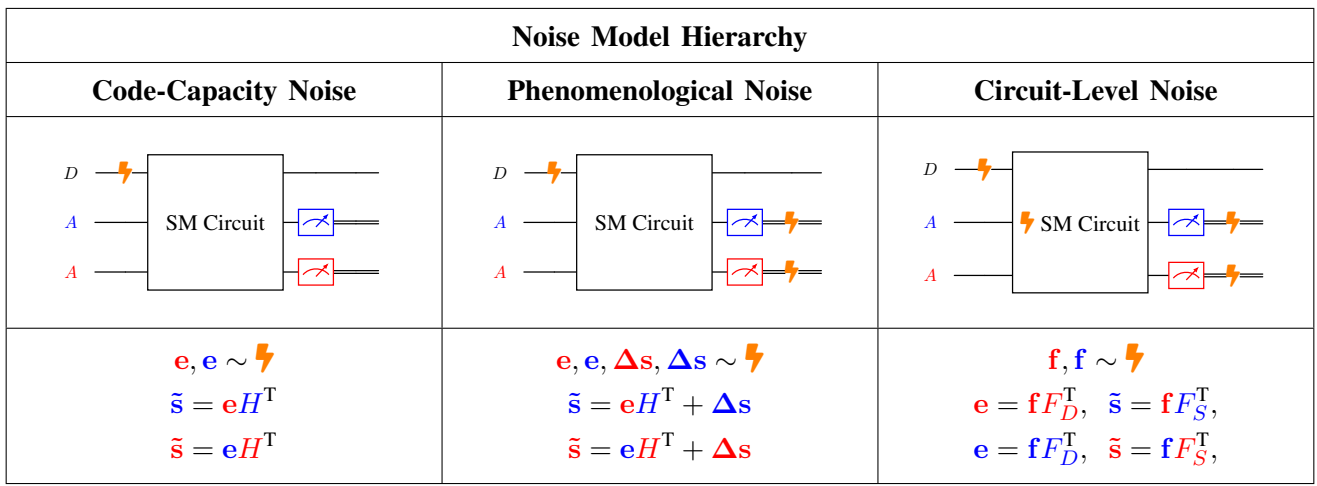

\subsection{Code-Capacity Noise Model -- A Tale of Red and Blue}
\label{sec:noise:code_capacity}
The code-capacity noise model is the simplest in our hierarchy. Let us assume that our quantum computer consists of a collection of $n$ qubits that we wish to protect from noise. In Section~\ref{sec:basics:commutativity}, we introduced the symplectic representation of Pauli operators. This notation is essentially a homomorphism from the Pauli group to a two-dimensional vector space over the field of two elements, i.e., $GF(2)^2$. This allows us to think of quantum errors as a collection of two (possibly correlated) classical bit-flip errors. Hence, we denote the noise acting on our qubits using a set of two binary-valued \emph{data error vectors}, which we denote as $\mathcolor{red}{\mathbf{e} = (\mathrm{e_1}, \mathrm{e_2}, \dots, \mathrm{e_n})}$ and $\mathcolor{blue}{\mathbf{e} = (\mathrm{e_1}, \mathrm{e_2}, \dots, \mathrm{e_n})}$, with each $(\mathcolor{red}{\mathrm{e_i}}, \mathcolor{blue}{\mathrm{e_i}})$ denoting the error on the $i^{th}$ qubit under the symplectic representation. This is essentially the code capacity model. In short, the code-capacity model can be summarized simply as a set of two different types of binary-valued errors, $\mathcolor{red}{\mathbf{e}}$ and $\mathcolor{blue}{\mathbf{e}}$, acting on our qubits. From a classical viewpoint, one may simply visualize the code-capacity model as a collection of two classical binary symmetric channels (BSCs) -- a \red{red} channel for $X$ errors and a \blue{blue} channel for $Z$ errors. Depending on the noise channel, these two channels may be correlated with one another. Indeed, these two channels do turn out to be correlated in the case of the depolarizing noise channel introduced in Section~\ref{sec:basics:errors}, which is the most popularly used noise channel for characterizing quantum error correction architectures.

In Section~\ref{sec:basics:css_codes}, we introduced the notion of CSS codes, which allows us to correct for errors in the \red{red} and \blue{blue} channels separately\footnote{This ignores any correlations between \red{red} and \blue{blue} channels. See~\cite{kuo_log-domain_2021, kuo_refined_2020} for methods to account for these correlations.} using two parity check matrices $\mathcolor{blue}{H} = H_Z$ and $\mathcolor{red}{H} = H_X$ respectively. Thus, we see that quantum error correction under the code-capacity model is not too different from classical error correction. Nevertheless, there are some important quirks introduced by the quantum nature of the systems that we are attempting to correct. Firstly, there is the measurement collapse postulate, which states that quantum systems collapse to a classical state when observed or measured. This means that the qubits that store our quantum data cannot be directly measured, leading to the requirement for syndrome decoding as discussed in Section~\ref{sec:basics:css_codes}. Secondly, there is the infamous Heisenberg uncertainty principle, which states that there exists pairs of (quantum) observables that cannot be simultaneously measured in a deterministic manner. We call such pairs \emph{complementary observables}, and as fate would have it, observables of type $\mathcolor{red}{X}$ and $\mathcolor{blue}{Z}$ form exactly such a pair. For the syndromes to be simultaneously observable, the check matrices need to satisfy the orthogonality constraint introduced in Section~\ref{sec:basics:css_codes}, i.e., $\mathcolor{red}{H} \mathcolor{blue}{H^{\text{\color{black}T}}} = 0$. This simple, innocuous constraint has been the source of much agony in the context of the QLDPC code construction problem and it has taken twenty years of effort by some of the most brilliant minds in the community~\cite{tillich_quantum_2014, panteleev_asymptotically_2021, dinur_good_2022, leverrier2022_QTannerCodes} to develop code constructions that satisfy this property while simultaneously having good rate and distance properties. We shall discuss some of these constructions in Section~\ref{sec:codes}.

\subsection{Phenomenological Noise Model -- Lies, Errors and Syndromes}
\label{sec:noise:phenomenological}
\begin{figure}
\centering
\subfigure[Phenomenological Noise Model.]{
    \begin{minipage}[b]{0.3\linewidth}
        \centering
        \scalebox{0.6}{\begin{quantikz}
    D \quad & \qwbundle{} & \noisegate{e}{} & \gate[3]{\text{SM Circuit}} & & & & \\
    \mathcolor{blue}{A} \quad & \qwbundle{} & & & \meter[style={draw=blue, fill=white}]{} & \setwiretype{c} & \deltasyndrome{red}{} & \setwiretype{c} \\
    \mathcolor{red}{A} \quad & \qwbundle{} & & & \meter[style={draw=red, fill=white}]{} & \setwiretype{c} & \deltasyndrome{blue}{} & \setwiretype{c}
\end{quantikz}}
        \vspace{15pt} 
    \end{minipage}
    \label{fig:phenom_ckt}
}
\qquad
\subfigure[Block Diagram.]{
    \includegraphics[width=0.6\linewidth]{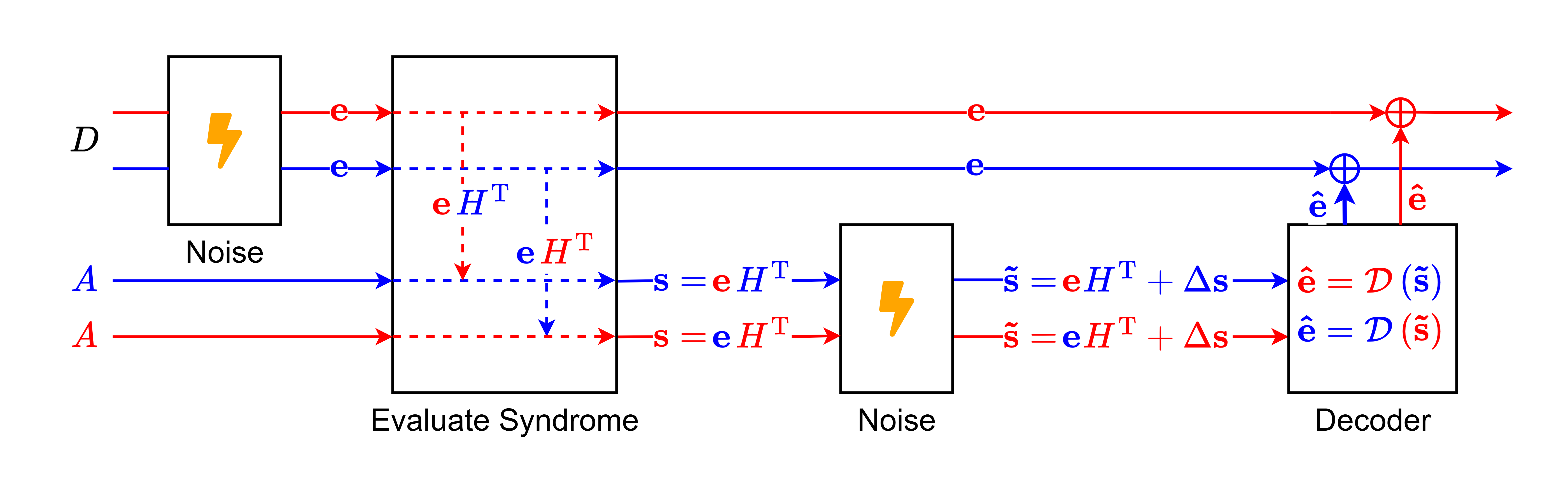}
    \label{fig:phenom_block}
}
\caption{The phenomenological noise model. Noise channels are indicated as \textcolor{orange}{\faBolt}. \subref{fig:phenom_ckt} shows the SM circuit used to acquire the syndromes, along with noise sources. \subref{fig:phenom_block} shows a block diagram of the phenomenological noise model.}
\label{fig:phenom_noise_model}
\end{figure}

In classical error correction, the syndrome vector is usually evaluated as $\mathbf{s} = \mathbf{y}H^\text{T}$. However, as discussed in the previous subsection, the measurement postulate prohibits direct access to the data qubits, requiring us to use a scheme similar to that described in Fig.~\ref{fig:measurement-circuits} for measuring out the syndrome. A simplified perspective of this approach is shown in Fig.~\ref{fig:phenom_ckt}. Specifically, we distinguish between \emph{data qubits}, which encode our (quantum) data and \emph{ancilla qubits}, which we use to measure out the syndrome. The syndrome information from the data qubits is transferred onto the ancilla qubits using a syndrome measurement (SM) circuit. This SM circuit is essentially a quantum circuit with some entangling CNOT gates between the data and ancilla qubits as described in Section~\ref{sec:basics:cnot-gate}. For now, we treat it as a black box that transfers syndrome information to the ancilla qubits, which can then be measured to read out the syndrome without affecting the data qubits. If our check matrices $(\mathcolor{blue}{H}, \mathcolor{red}{H})$ have dimensions $(\mathcolor{blue}{m} \times n, \mathcolor{red}{m} \times n)$ respectively, we require a total of $n$ data qubits and $\mathcolor{blue}{m} + \mathcolor{red}{m}$ ancilla qubits for this procedure.

The presence of the SM circuit introduces additional sources of noise in our system. Specifically, the ancilla qubits, like the data qubits, are also quantum systems, which means that they are noisy. This means, of course, that the measured syndrome is also noisy. The exact mechanism by which noise on the ancilla qubits translates to noise on the data qubits and noise on the syndrome depends on the internal structure of the SM circuit. For the phenomenological noise model however, we shall ignore this internal structure and simply assume that each ancilla qubit -- and therefore each measured syndrome bit -- is independently acted upon by an appropriate noise channel. A block diagram of the phenomenological noise model is shown in Fig.~\ref{fig:phenom_block}. Specifically, we assume that data error vectors $\mathcolor{red}{\mathbf{e}}, \mathcolor{blue}{\mathbf{e}}$ act on the data qubits, similar to the code capacity model. The actual syndromes $\mathcolor{blue}{\mathbf{s}}, \mathcolor{red}{\mathbf{s}}$ are then given by $\mathcolor{blue}{\mathbf{s}} = \mathcolor{red}{\mathbf{e}}\mathcolor{blue}{{H^{\text{\color{black}T}}}}$ and $\mathcolor{red}{\mathbf{s}} = \mathcolor{blue}{\mathbf{e}}\mathcolor{red}{{H^{\text{\color{black}T}}}}$. The measured syndromes are given by $\mathcolor{blue}{\mathbf{\tilde{s}}} = \mathcolor{blue}{\mathbf{s + \Delta s}}$ and $\mathcolor{red}{\mathbf{\tilde{s}}} = \mathcolor{red}{\mathbf{s + \Delta s}}$, where the vectors $\mathcolor{blue}{\mathbf{\Delta s}}$, $\mathcolor{red}{\mathbf{\Delta s}}$ are the \emph{syndrome error vectors}. Crucially, the phenomenological noise model assumes that the syndrome bits $\mathcolor{red}{\Delta s_i}$, $\mathcolor{red}{\Delta s_j}$ are drawn from independent probability distributions for $i \neq j$, and a similar case for $\mathcolor{blue}{\Delta s_i}$, $\mathcolor{blue}{\Delta s_j}$.\footnote{Note that $\mathcolor{red}{\Delta s_i}$, $\mathcolor{blue}{\Delta s_i}$ may still be correlated. And this is indeed the case for the commonly used depolarizing noise channel. Typically, the syndrome error vectors $\mathcolor{red}{\mathbf{\Delta s}}$, $\mathcolor{blue}{\mathbf{\Delta s}}$ are drawn from a similar noise channel as the data error vectors $\mathcolor{red}{\mathbf{e}}, \mathcolor{blue}{\mathbf{e}}$.} 

The decoder is then fed the noisy syndromes $\mathcolor{blue}{\mathbf{\tilde{s}}}, \mathcolor{red}{\mathbf{\tilde{s}}}$ and is expected to estimate $\mathcolor{red}{\mathbf{e}}, \mathcolor{blue}{\mathbf{e}}$. In any error correction decoder, syndromes provide information about the errors that occur on the data. But if the syndromes themselves are faulty, who protects the syndromes? The term \emph{syndrome miscorrection} is often used to refer to a situation in which the decoder fails due to a syndrome error. A commonly used strategy to counter this is to use several rounds of repeated measurements, so that we obtain several (noisy) copies of $\mathcolor{blue}{\mathbf{s}}, \mathcolor{red}{\mathbf{s}}$. Specifically, we assume that $R$ rounds of measurement are performed with the SM circuit being repeated each round. Each round takes a time unit of $\tau$, with decoding and recovery being performed at $T=R\tau$. This strategy essentially places the phenomenological noise model on a similar footing as the Taylor-Kuznetsov scheme, mentioned in Section~\ref{sec:intro:faulty_memory}. However, it is important to note that decoding and recovery are perfect operations in this scenario. The only source of noise is the syndrome computation step. Moreover, this noise in the syndrome computation step is assumed to be independent for each qubit. As we shall soon see, this independence assumption is not quite accurate. Indeed, correlations between errors in the syndrome computation step are the subject of the circuit-level noise model.

\subsection{Circuit-Level Noise Model -- The Correlated Conspiracy}
\label{sec:noise:circuit_level}
\begin{figure}
\centering
\subfigure[Circuit Faults.]{
    \begin{minipage}{0.45\linewidth}
        \centering{
        \scalebox{0.45}{\begin{quantikz}
    \mathrm{T} \quad \setwiretype{n} & 0 & & & 1 & & & & & 2 & & & 3 & \\
    D_1 \quad 
        & \noisegate{f}{1\hphantom{0}} \slice[style=black]{} 
        & \targ[style={draw=red}]{} 
        & & \noisegate{f}{2\hphantom{0}} 
        & & & \ctrl[style={draw=blue}]{7} 
        & & \noisegate{f}{3\hphantom{0}} 
        & & & \noisegate{f}{4\hphantom{0}} 
        & & \\
    D_2 \quad 
        & \noisegate{f}{5\hphantom{0}} 
        & & \targ[style={draw=red}]{} 
        & \noisegate{f}{6\hphantom{0}} \slice[style=black]{} 
        & & & & \ctrl[style={draw=blue}]{5} 
        & \noisegate{f}{7\hphantom{0}} 
        & & &  \noisegate{f}{8\hphantom{0}} 
        & & \\
    D_3 \quad 
        & \noisegate{f}{9\hphantom{0}} 
        & & & \noisegate{f}{10} 
        & \targ[style={draw=red}]{} 
        & & & & \noisegate{f}{11} \slice[style=black]{} 
        & \ctrl[style={draw=blue}]{5} 
        & & \noisegate{f}{12} 
        & & \\
    D_4 \quad 
        & \noisegate{f}{13} 
        & & & \noisegate{f}{14} 
        & & \targ[style={draw=red}]{} 
        & & & \noisegate{f}{15} 
        & & \ctrl[style={draw=blue}]{3} 
        & \noisegate{f}{16} \slice[style=black]{} 
        & & \\
    \mathcolor{red}{A_1} \quad 
        & \noisegate{f}{17} 
        & \ctrl[style={draw=red}]{-4} 
        & & \noisegate{f}{18} 
        & \ctrl[style={draw=red}]{-2} 
        & & & & \noisegate{f}{19} 
        & & & \noisegate{f}{20} 
        & \meter[style={draw=red, fill=white}]{} 
        & \setwiretype{c} \\
    \mathcolor{red}{A_2} \quad 
        & \noisegate{f}{21} 
        & & \ctrl[style={draw=red}]{-4} 
        & \noisegate{f}{22} 
        & & \ctrl[style={draw=red}]{-2} 
        & & & \noisegate{f}{23} 
        & & & \noisegate{f}{24} 
        & \meter[style={draw=red, fill=white}]{} 
        & \setwiretype{c} \\
    \mathcolor{blue}{A_1}  \quad 
        & \noisegate{f}{25} 
        & & & \noisegate{f}{26} 
        & & & & \targ[style={draw=blue}]{} 
        & \noisegate{f}{27} 
        & & \targ[style={draw=blue}]{} 
        & \noisegate{f}{28} 
        & \meter[style={draw=blue, fill=white}]{} 
        & \setwiretype{c} \\
    \mathcolor{blue}{A_2}  \quad 
        & \noisegate{f}{29} 
        & & & \noisegate{f}{30} 
        & & & \targ[style={draw=blue}]{} 
        & & \noisegate{f}{31} 
        & \targ[style={draw=blue}]{} 
        & & \noisegate{f}{32} 
        & \meter[style={draw=blue, fill=white}]{} 
        & \setwiretype{c}
\end{quantikz}}}
        \vspace{8pt} 
    \end{minipage}
    \label{fig:fault_ckt}
}
\qquad
\subfigure[Effective Errors.]{
    \begin{minipage}{0.45\linewidth}
        \centering{
        \scalebox{0.45}{\begin{quantikz}
    \mathrm{T} \quad \setwiretype{n} & & 1 & & & & 2 & & 3 & \\
    D_1 \quad 
        \slice[style=black]{} 
        & \targ[style={draw=red}]{} 
        & & & & \ctrl[style={draw=blue}]{7} 
        & & & & & \noisegate{e}{1} &\\
    D_2 \quad 
        & & \targ[style={draw=red}]{} 
        \slice[style=black]{} 
        & & & & \ctrl[style={draw=blue}]{5}
        & & & & \noisegate{e}{2} & \\
    D_3 \quad 
        & & & \targ[style={draw=red}]{} 
        & & & \slice[style=black]{} 
        & \ctrl[style={blue}]{5} 
        & & & \noisegate{e}{3} & \\
    D_4 \quad 
        & & & & \targ[style={draw=red}]{} 
        & & & & \ctrl[style={draw=blue}]{3} 
        \slice[style=black]{} 
        & & \noisegate{e}{4} & \\
    \mathcolor{red}{A_1} \quad 
        & \ctrl[style={draw=red}]{-4} 
        & & \ctrl[style={draw=red}]{-2} 
        & & & & & & \meter[style={draw=red, fill=white}]{} & \noisesyndrome{red}{1} \setwiretype{c} & \\
    \mathcolor{red}{A_2} \quad 
        & & \ctrl[style={draw=red}]{-4} 
        & & \ctrl[style={draw=red}]{-2} 
        & & & & & \meter[style={draw=red, fill=white}]{} & \noisesyndrome{red}{2} \setwiretype{c} & \\
    \mathcolor{blue}{A_1}  \quad 
        & & & & & & \targ[style={draw=blue}]{} 
        & & \targ[style={draw=blue}]{} 
        & \meter[style={draw=blue, fill=white}]{} & \noisesyndrome{blue}{1} \setwiretype{c} & \\
    \mathcolor{blue}{A_2}  \quad 
        & & & & & \targ[style={draw=blue}]{} 
        & & \targ[style={draw=blue}]{} 
        & & \meter[style={draw=blue, fill=white}]{} & \noisesyndrome{blue}{2} \setwiretype{c} &
\end{quantikz}}}
        \vspace{8pt} 
    \end{minipage}
    \label{fig:error_ckt}
} \\
\subfigure[Fault Maps.]{
    \begin{tcolorbox}[width=0.35\linewidth, colframe=white, colback=white]
    \tiny\[\begin{aligned}
    \mathcolor{red}{\mathrm{e_1}} & \mathcolor{red}{=}
    \mathcolor{red}{\mathrm{f_{1\hphantom{0}} + f_{2\hphantom{0}} + f_{3\hphantom{0}} + f_{4\hphantom{0}} + f_{17}}} \\[-3pt]
    \mathcolor{blue}{\mathrm{e_1}} & \mathcolor{blue}{=}
    \mathcolor{blue}{\mathrm{f_{1\hphantom{0}} + f_{2\hphantom{0}} + f_{3\hphantom{0}} + f_{4\hphantom{0}} + f_{29} + f_{30}}} \\[-3pt]
    \mathcolor{red}{\mathrm{e_2}} & \mathcolor{red}{=}
    \mathcolor{red}{\mathrm{f_{5\hphantom{0}} + f_{6\hphantom{0}} + f_{7\hphantom{0}} + f_{8\hphantom{0}} + f_{21}}} \\[-3pt]
    \mathcolor{blue}{\mathrm{e_2}} & \mathcolor{blue}{=}
    \mathcolor{blue}{\mathrm{f_{5\hphantom{0}} + f_{6\hphantom{0}} + f_{7\hphantom{0}} + f_{8\hphantom{0}} + f_{25} + f_{26}}} \\[-3pt]
    \mathcolor{red}{\mathrm{e_3}} & \mathcolor{red}{=}
    \mathcolor{red}{\mathrm{f_{9\hphantom{0}} + f_{10} + f_{11} + f_{12} + f_{17} + f_{18}}} \\[-3pt]
    \mathcolor{blue}{\mathrm{e_3}} & \mathcolor{blue}{=}
    \mathcolor{blue}{\mathrm{f_{9\hphantom{0}} + f_{10} + f_{11} + f_{12} + f_{29} + f_{30} + f_{31}}} \\[-3pt]
    \mathcolor{red}{\mathrm{e_4}} & \mathcolor{red}{=}
    \mathcolor{red}{\mathrm{f_{13} + f_{14} + f_{15} + f_{16} + f_{21} + f_{22}}} \\[-3pt]
    \mathcolor{blue}{\mathrm{e_4}} & \mathcolor{blue}{=}
    \mathcolor{blue}{\mathrm{f_{13} + f_{14} + f_{15} + f_{16} + f_{25} + f_{26} + f_{27}}} \\[-3pt]
    \mathcolor{blue}{\mathrm{\tilde{s}_1}} & \mathcolor{red}{=}
    \mathcolor{red}{\mathrm{f_{25} + f_{26} + f_{27} + f_{28} + f_{5\hphantom{0}} + f_{6\hphantom{0}} + f_{13} + f_{14} + f_{15}}} \\[-3pt]
    \mathcolor{red}{\mathrm{\tilde{s}_1}} & \mathcolor{blue}{=}
    \mathcolor{blue}{\mathrm{f_{17} + f_{18} + f_{19} + f_{20} + f_{1\hphantom{0}} + f_{9\hphantom{0}} + f_{10}}} \\[-3pt]
    \mathcolor{blue}{\mathrm{\tilde{s}_2}} & \mathcolor{red}{=}
    \mathcolor{red}{\mathrm{f_{29} + f_{30} + f_{31} + f_{32} + f_{1\hphantom{0}} + f_{2\hphantom{0}} + f_{0\hphantom{0}} + f_{10} + f_{11}}} \\[-3pt]
    \mathcolor{red}{\mathrm{\tilde{s}_2}} & \mathcolor{blue}{=}
    \mathcolor{blue}{\mathrm{f_{21} + f_{22} + f_{23} + f_{24} + f_{5\hphantom{0}} + f_{13} + f_{14}}}
    \end{aligned}\]
    \end{tcolorbox}
    \label{fig:fault_map}
}
\qquad
\subfigure[Block Diagram.]{
    \includegraphics[width=0.5\linewidth]{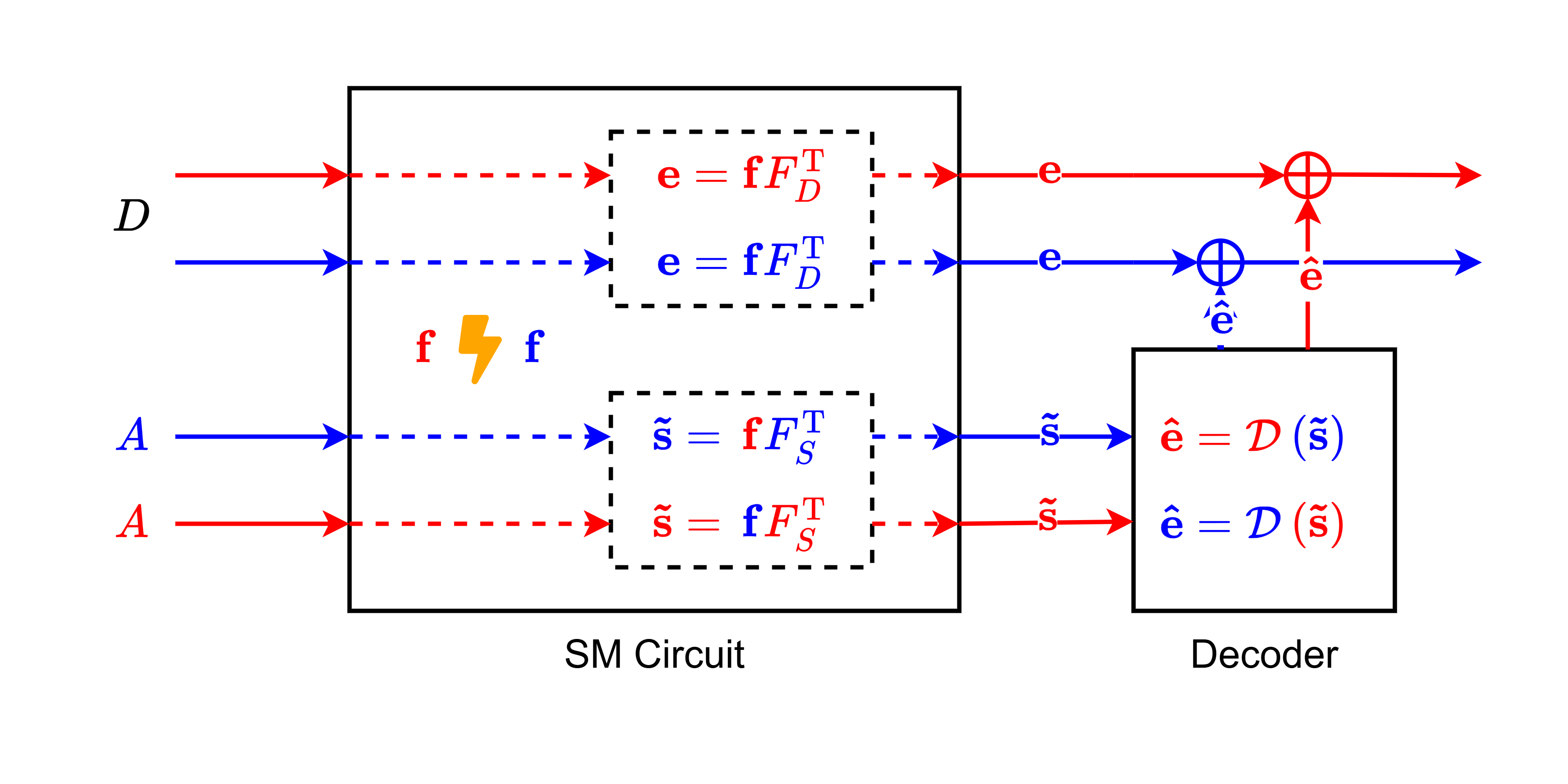}
    \label{fig:circuit_block}
}
\caption{The circuit-level noise model. Noise channels are indicated as \textcolor{orange}{\faBolt}. \subref{fig:fault_ckt} shows the full SM circuit with the locations of all noise channels. Stabilizer simulations allow us to simplify this to the form of \subref{fig:error_ckt}, where the effective errors on the data and the measured syndrome bits are related to the circuit faults via the equations shown in \subref{fig:fault_map}. A block diagram of the circuit-level noise model from a classical perspective is shown in \subref{fig:circuit_block}. It is worth noting that the noise channels marked \textcolor{orange}{\faBolt} in \subref{fig:fault_ckt} may be correlated with one another. In fact, in the commonly used circuit-level depolarizing noise model, faults occurring at the control and target wires of any given CNOT gate do turn out to be correlated.}
\label{fig:circuit_noise_model}
\end{figure}

In the circuit-level noise model, we attempt to unwrap the internal structure of the SM circuit with the goal of building a more realistic noise model. The structure of the SM circuit is best illustrated through an example. In Fig.~\ref{fig:fault_ckt}, we illustrate the SM circuit for a simple CSS code with
\begin{equation*}
    \mathcolor{red}{H} = \begin{pmatrix}
        1 & 0 & 1 & 0 \\
        0 & 1 & 0 & 1
    \end{pmatrix}, \;\;
    \mathcolor{blue}{H} = \begin{pmatrix}
        0 & 1 & 0 & 1 \\
        1 & 0 & 1 & 0
    \end{pmatrix}.
\end{equation*}
There are a total of 8 qubits, comprising 4 data qubits and 4 ancilla qubits -- two for $\mathcolor{red}{H}$ and two for $\mathcolor{blue}{H}$.\footnote{This is a dummy CSS code chosen simply for the purpose of illustration. It has a logical dimension of zero.} The data qubits contain our quantum information, and the ancilla qubits are initialized in the $\ket{0}$ or $\ket{+}$ state, depending on whether the corresponding syndrome is red or blue, respectively. The circuit itself consists of a sequence of entangling CNOT gates. Following the measurement scheme introduced in Fig.~\ref{fig:measurement-circuits}, we add a CNOT gate from the $j^{th}$ data qubit to the $i^{th}$ ancilla qubit, if $\mathcolor{blue}{H_{ij}} = 1$. Similarly, a CNOT gate in the inverse direction, namely, from the $i^{th}$ ancilla qubit to the $j^{th}$ data qubit, is added if $\mathcolor{red}{H_{ij}} = 1$. Finally, the ancilla qubits are measured in the $Z$ or $X$ basis,\footnote{Measurement in the $X$ or $Z$ basis corresponds to a projective measurement of the observables $X$ or $Z$ respectively, as described at the end of Section~\ref{sec:basics:qubits}.} again depending on syndrome color. For purposes of understanding the circuit-level noise model, the initialization state and measurement basis will not play a major role, but the CNOT gates will play a role.

To model noise in the circuit, we divide the CNOT gate sequence into a series of time-steps. At each time-step, one or more CNOT gates is applied. Two CNOT gates can be applied at the same time-step if and only if they operate on disjoint sets of qubits. In each time-step, as shown in Fig.~\ref{fig:fault_ckt}, each qubit also suffers from noise in both the \red{red} and \blue{blue} channels. We use the term \emph{circuit fault vectors} to indicate the vectors $\mathcolor{red}{\mathbf{f}}, \mathcolor{blue}{\mathbf{f}}$ comprising the faults that occur at all possible locations in the circuit. Now, circuit faults that occur at time-step $t$ are acted upon by all entangling CNOT gates at time-steps $\geq t + 1$, which induces correlated errors between qubits connected by the CNOT gates. Hence, we see that the fundamental assumption of independent errors in both the code-capacity and phenomenological noise models no longer holds in the circuit-level noise model.

To understand the nature of these correlations, we refer the reader to Fig.~\ref{fig:error_ckt}, which is an equivalent manner of modeling the scenario in Fig.~\ref{fig:fault_ckt}. Specifically, we use the CNOT propagation rules of Fig.~\ref{fig: cnot_propagation_binary} to translate the circuit faults in the SM circuit to the end of the circuit as effective data error vectors $\mathcolor{red}{\mathbf{e}}, \mathcolor{blue}{\mathbf{e}}$ and the measured syndrome vectors $\mathcolor{blue}{\mathbf{\tilde{s}}}, \mathcolor{red}{\mathbf{\tilde{s}}}$ in Fig.~\ref{fig:error_ckt}. These vectors can now be described as linear combinations of the circuit fault vectors $\mathcolor{red}{\mathbf{f}}, \mathcolor{blue}{\mathbf{f}}$ as shown in Fig.~\ref{fig:fault_map}. This linear combination relating the error vectors to the fault vectors can be described using a matrix notation as follows.
\begin{equation}
\begin{split}
    \label{eq:ckt_noise}
    \mathcolor{red}{\mathbf{e}} &= \mathcolor{red}{\mathbf{f} F_D^{\text{\color{black}T}}}, \;\; 
    \mathcolor{blue}{\mathbf{\tilde{s}}} = \mathcolor{red}{\mathbf{f}} \mathcolor{blue}{F_S^{\text{\color{black}T}}}, \\
    \mathcolor{blue}{\mathbf{e}} &= \mathcolor{blue}{\mathbf{f} F_D^{\text{\color{black}T}}}, \;\;
    \mathcolor{red}{\mathbf{\tilde{s}}} = \mathcolor{blue}{\mathbf{f}}
    \mathcolor{red}{F_S^{\text{\color{black}T}}}, \\
\end{split}    
\end{equation}
where $\mathcolor{red}{F_D}, \mathcolor{blue}{F_D}, \mathcolor{red}{F_S}, \mathcolor{blue}{F_S}$ are binary matrices, which we call \emph{fault maps}. Specifically, $\mathcolor{red}{F_D}, \mathcolor{blue}{F_D}$ are called the \emph{data fault maps} for the red and blue channels respectively and $\mathcolor{red}{F_S}, \mathcolor{blue}{F_S}$ are called the \emph{syndrome fault maps}. Note that any given choice of CSS code $(\mathcolor{blue}{H}, \mathcolor{red}{H})$ can be associated with multiple valid SM circuits, since the SM circuit depends on the CSS code as well as a choice of ordering of the ones in $\mathcolor{blue}{H}, \mathcolor{red}{H}$ into discrete time-steps. We call this choice of ordering the \emph{measurement schedule}, and the fault maps are a function of both the CSS code and the associated measurement schedule.

To summarize, let us provide a simplified overview of the circuit-level noise model from a purely classical perspective. We begin with the phenomenological noise model, where the two sources of noise are the data error vectors $\mathcolor{red}{\mathbf{e}}, \mathcolor{blue}{\mathbf{e}}$, that we wish to correct, and the syndrome error vectors $\mathcolor{blue}{\mathbf{\Delta s}}, \mathcolor{red}{\mathbf{\Delta s}}$, which cause the actual syndromes $\mathcolor{blue}{\mathbf{s}}, \mathcolor{red}{\mathbf{s}}$ to be different from the measured syndromes, $\mathcolor{blue}{\mathbf{\tilde{s}}}, \mathcolor{red}{\mathbf{\tilde{s}}}$. In the circuit noise model, instead of assuming this difference in actual and measured syndromes to be caused by independent errors, we assume that they are jointly caused by a set of hidden variables, which we call circuit faults $\mathcolor{red}{\mathbf{f}}, \mathcolor{blue}{\mathbf{f}}$. The data error vectors and the measured syndrome vectors are related to the circuit faults via the fault maps of Eq.~\ref{eq:ckt_noise}. Hence, the fault maps play the role of introducing correlations between the noise sources in the phenomenological noise model. The exact form of the fault maps can be derived from the SM circuit using the CNOT propagation rules of Fig.~\ref{fig: cnot_propagation_binary}. 

Typically, decoding for the circuit-level noise model is done by attempting to estimate the circuit fault vectors using the measured syndrome and the syndrome fault map, and then translating the circuit faults to data errors using the data fault map. However, the syndrome fault map is not usually very well suited to traditional message passing decoders such as belief propagation due to its non-sparse nature. An often-used trick to improve its sparsity is to use a variant of the fault map that does not track syndrome values, but rather changes in these values from one round to another. Also, similar to the phenomenological noise mode, the problem of syndrome miscorrections persists for the circuit-level noise model, requiring the use of multiple rounds of syndrome measurements. Thus, the parity check matrix that is ultimately used for decoding -- often referred to as the \emph{detector error matrix} -- is a variant of the syndrome fault map that involves multiple rounds of repeated measurements and tracks changes in syndrome values per round rather than the syndrome itself. In this context, the Stim software package~\cite{gidney2021stim} is an excellent resource for automating the construction of detector error matrices for the circuit-level noise model. It is important to note that the size of the detector error matrix is significantly larger than the original parity check matrices $(\mathcolor{blue}{H}, \mathcolor{red}{H})$. For instance, in the simple example of Fig.~\ref{fig:circuit_noise_model}, the check matrices have four columns since the code involves four data qubits, but the number of circuit faults is 32, which is an increase by a factor of 8. Moreover, the circuit of Fig.~\ref{fig:circuit_noise_model} simply illustrates one round of measurement, whereas repeated measurement rounds will further increase the number of circuit faults. This significantly larger size of the detector error matrix poses significant challenges to the problem of decoder design. We shall discuss the problem of decoding in more detail in Section~\ref{sec:decoder}.

Finally, we remark that an important consequence of correlations in the circuit-level noise model is that a single circuit fault may now propagate to multiple data and syndrome errors, which is severely detrimental to fault tolerance. To this end, we would like to limit the number of locations to which a single fault can propagate. It turns out that this number is directly influenced by the weight of the rows and columns of $(\mathcolor{blue}{H}, \mathcolor{red}{H})$, Specifically, a qubit participating in many parity checks increases the depth of the SM circuit (since these parity checks have to be measured sequentially), thus increasing the number of fault locations (including idle qubits), which means that we would like to limit the degree of the codes $\mathcolor{blue}{H}, \mathcolor{red}{H}$. This is the primary motivating factor for the use of LDPC codes in quantum error correction. Notably, this is different from the classical motivation for LDPC codes, which has to do with the capacity-achieving nature of classical LDPC codes.

\subsection{Other Noise Models -- Peeking into the Black Box}
\label{sec:noise:miscellaneous}
At this point, one might well wonder: can we further refine the circuit-level noise model by attempting to better understand the physics of the various gates used in the SM circuit? This brings us to the realm of hardware-aware noise models. For instance, we have simply described the CNOT gate as an idealized error propagation device in the previous subsection. Perhaps understanding how this gate is physically implemented can shed some light on the exact nature of the faults that occur before and after such a gate is applied. Indeed, this approach has been taken in several works~\cite{Noh22, hopfmueller2024bosonic}, which aim to build increasingly sophisticated noise models based on hardware specific knowledge. A prominent example is the case of bosonic codes~\cite{Gottesman01, Royer22, Grimsmo20, Xu24}, which provide soft information regarding circuit faults that can be utilized by a QLDPC decoder to significantly improve performance~\cite{Noh20, Noh22, Berent24}, \mycite{nithin_GKP_QLDPC, shantom_spacetime_gkp}. Other examples of more realistic noise channels include coherent noise models~\cite{iverson2020coherence, beale2018coherence} which model errors as a small continuous drift in the qubit state rather than a discrete Pauli operator, and non-Markovian noise, which characterizes long-range temporal correlations in a noise channel~\cite{gravier2025simulated}. Nevertheless, it is perhaps clear to the reader that this process of zooming into the finer details of any system can be repeated as many times as one desires, until one ends up encountering fundamental questions regarding the nature of quantum mechanics itself. However, that is a question which we shall leave to our friends in the Physics department. For now, we shall put on our engineering hats and concern ourselves with the construction and decoding of practical error correction systems for the three noise models described above.

\section{Main QLDPC Code Constructions}
\label{sec:codes} 
Designing QLDPC codes is significantly more challenging than designing classical LDPC codes, primarily due to the orthogonality constraint imposed by the CSS construction (Section~\ref{sec:basics:css_codes}). In the classical setting, one can generate a random LDPC code that is likely to have good properties, provided the code length is sufficiently large. Moreover, the Tanner graph can be freely manipulated by adding, removing, or permuting edges, introducing new nodes, and performing other desired modifications or optimizations, for example, to remove short cycles or to mitigate other problematic structures -- trapping sets. In contrast, QLDPC codes must preserve a rigid algebraic structure to ensure commutativity between $X$- and $Z$-type stabilizer generators (checks), which severely limits permissible changes to their Tanner graphs. Therefore, QLDPC code design places a much stronger emphasis on the algebraic aspects of the construction. While algebraically constructed classical LDPC codes do exist, this reliance on algebraic structure is far more pronounced in the quantum setting, where it quickly becomes a central feature of the design process.

In this Section, we briefly review the history of QLDPC code design, from topological codes to the recently developed and celebrated asymptotically good constructions, highlighting the key concepts and illustrating with examples the structure of such codes.

\subsection{Topological Codes}
\label{sec:codes:topological}
For historical reasons, our discussion of QLDPC code construction could hardly begin anywhere other than with topological codes. They form a subclass of the QLDPC family, characterized by stabilizer generators that act locally on qubits arranged in a low-dimensional lattice (technically, tessellations of surfaces or higher-dimensional manifolds). As the first class of QLDPC codes to be investigated -- though not coined as QLDPC codes at the time -- they demonstrated the importance of low-weight stabilizers for fault tolerance and later sparked interest in more general QLDPC constructions.

The earliest example is the Kitaev's toric code~\cite{KITAEV20032}, defined on a torus. It can be visualized as a 2D lattice with qubits on edges, $X$ checks on vertices, and $Z$ checks on faces. This lattice is placed on a torus by identifying horizontal and vertical boundaries. This construction defines an $[[n=2L^2, k=2,d=L]]$ CSS code, where $L$ denotes the size of the lattice. Interest in this construction was further boosted by its planar version, known as the surface code~\cite{PreskilMWPM_topologicalQM}, which is defined on a lattice with boundaries laid out on a plane. The surface code enables implementations on a planar layout, making it more practical for physical realizations, particularly for quantum technologies constrained by two-dimensional local (nearest-neighbor) interactions, such as superconducting qubits or spin qubits in semiconductor quantum dots. It has parameters $[[n=\Theta(L), k=1, d=L]]$, with $n=2L(L+1)$ for the conventional surface code, and $n=L^2$ for its rotated variant~\cite{PhysRevA.80.052312}. 

Color codes~\cite{PhysRevLett.97.180501} are another family of topological codes constructed on 3-valent lattices with 3-colorable faces (e.g., the honeycomb lattice) on the torus (genus $g=1$) or higher-genus closed surfaces ($g>1$), where the genus $g$ of the surface controls the number of encoded logical qubits.  They also admit a planar variant, known as triangular color codes, suitable for physical realizations within the technological limitations described earlier. 

Hyperbolic surface codes~\cite{mahmoud2025systematicapproachhyperbolicquantum}, which embed the lattice on a surface with negative curvature, or higher-dimensional generalizations, such as 3D and 4D surface codes~\cite{BRAVYI2011839}, may offer different trade-offs between the $[[n,k,d]]$ parameters and exhibit features that can be leveraged for fault-tolerant quantum computing (see also Section~\ref{sec:logical_gates}).

Despite forming a rich and versatile class, topological codes suffer from poor scaling in the number of encoded qubits and code distance, due to fundamental limits associated with encoding quantum information through spatially bounded stabilizers~\cite{bravyi2009nogo, bravyi2010tradeoffs}. For instance, for codes defined by geometrically local stabilizers on 2D lattices, it was shown that $kd^2 = O(n)$~\cite{bravyi2010tradeoffs}. This limitation has driven research toward more general constructions of QLDPC codes, which we review in the subsequent Sections. Considerable effort has been devoted to the quest for the Grail, that is, constructing families of QLDPC codes for which both $k$ and $d$ scale linearly with $n$ (i.e., constant non-vanishing rate and linear minimum distance). We will discuss these constructions in the last sub-Section~\ref{sec:codes:miscellaneous}. But let us caution the reader that such asymptotically good constructions have failed so far to yield practical codes, i.e., codes of length on the order of a few hundred to a few thousand qubits. Therefore, we begin by reviewing constructions that have established themselves as the benchmark for building good codes with practical parameters.

\subsection{Hypergraph Product Codes}
\label{sec:codes:hypergraph_product}
In 2009, Tillich and Zémor published a seminal paper~\cite{tillich_quantum_2014} where they showed that QLDPC codes can be systematically constructed from any pair of classical LDPC codes using the so-called hypergraph-product (HP) method. They showed that the well-known toric code appears as a special case of their construction, corresponding to the HP product of two classical repetition codes.  More importantly, when using asymptotically good classical LDPC codes, i.e., with constant rate and linear minimum distance, their construction yields HP codes with parameters $[[n, k=O(n), d=O(\sqrt{n})]]$, which held the state-of-the-art record for a long time.  Moreover, much of the subsequent progress in QLDPC constructions has relied on the techniques introduced in their work or adopted a similar vision, building on different graph-product variations.  

\paragraph{Hypergraph Product of Graphs}  
A fundamental tool in their construction is the \textit{hypergraph product} of graphs (more commonly known in graph theory as \emph{Cartesian product};  it appears that authors independently rediscovered it and gave it a different name).  Let $\Gamma_a = (V_a, E_a)$ and $\Gamma_b = (V_b, E_b)$ be two graphs. Their hypergraph product is a graph $\Gamma_P = (V_a \times V_b, E_P)$, where two vertices $(v_a, v_b)$ and $(u_a, u_b)$  are adjacent if and only if either  
 $v_a = u_a$ and $(v_b, u_b) \in E_b$, or  
 $v_b = u_b$ and $(v_a, u_a) \in E_a$.  

\begin{figure}[!t]
    \centering
    \subfigure[]{
        \scalebox{0.7}{
\begin{tikzpicture}[
    pathnode/.style={circle, draw=black, fill=white, minimum size=0.5cm},
    gridnode/.style={circle, draw=black, fill=white, inner sep=1pt},
    labelstyle/.style={font=\scriptsize, draw=none, fill=none}, scale=0.8, transform shape
]

\node[pathnode] (v1) at (0, 6) {$v_1$};
\node[pathnode] (v2) at (0, 4) {$v_2$};
\node[pathnode] (v3) at (0, 2) {$v_3$};

\draw (v1) -- (v2) -- (v3);

\node[pathnode] (u1) at (4, 6) {$u_1$};
\node[pathnode] (u2) at (4, 4) {$u_2$};
\node[pathnode] (u3) at (4, 2) {$u_3$};

\draw (u1) -- (u2) -- (u3);

\node[draw=none, fill=none] (times) at (2, 4) {\Huge $\times$};

\node[draw=none, fill=none] (equal) at (6, 4) {\Huge $=$};

\node[pathnode, label=above:{$(v_1,u_1)$}] (g11) at (8, 6) {};
\node[pathnode, label=45:{$(v_2,u_1)$}] (g21) at (8, 4) {};
\node[pathnode, label=below:{$(v_3,u_1)$}] (g31) at (8, 2) {};
\node[pathnode, label=above:{$(v_1,u_2)$}] (g12) at (10, 6) {};
\node[pathnode, label=45:{$(v_2,u_2)$}] (g22) at (10, 4) {};
\node[pathnode, label=below:{$(v_3,u_2)$}] (g32) at (10, 2) {};
\node[pathnode, label=above:{$(v_1,u_3)$}] (g13) at (12, 6) {};
\node[pathnode, label=45:{$(v_2,u_3)$}] (g23) at (12, 4) {};
\node[pathnode, label=below:{$(v_3,u_3)$}] (g33) at (12, 2) {};

\draw (g11) -- (g21) -- (g31);
\draw (g12) -- (g22) -- (g32);
\draw (g13) -- (g23) -- (g33);
\draw (g11) -- (g12) -- (g13);
\draw (g21) -- (g22) -- (g23);
\draw (g31) -- (g32) -- (g33);

\end{tikzpicture}
}
        \label{fig:hp-grid}
    }
    \subfigure[]{
        \includegraphics[width=0.48\linewidth]{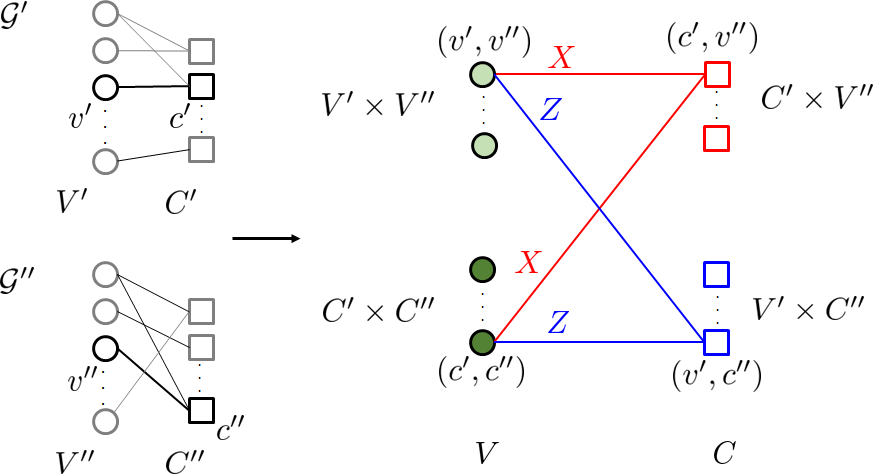}
        \label{fig:hp-graph}
    }
\caption{\subref{fig:hp-grid} Hypergraph product of two path graphs. \subref{fig:hp-graph} Hypergraph product of two Tanner graphs.}
\end{figure}
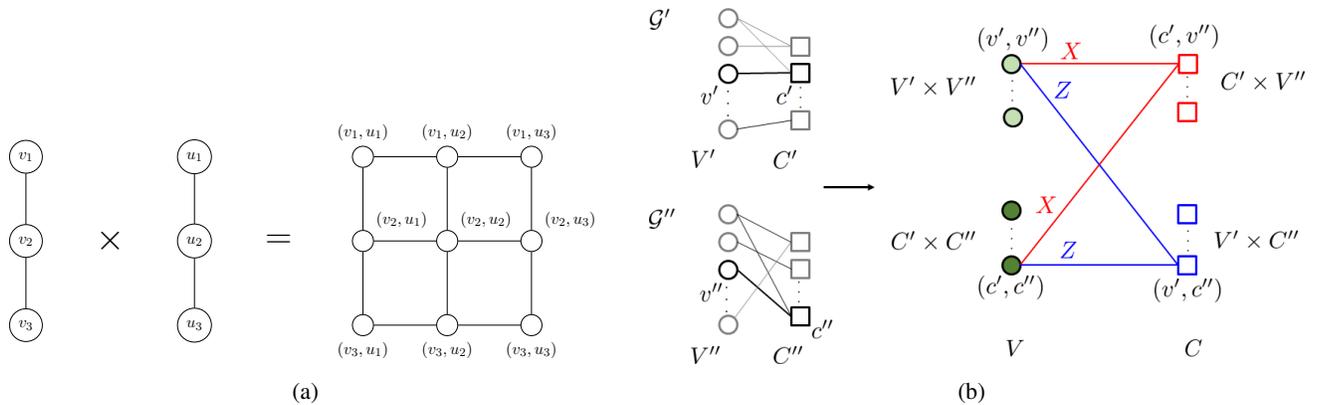

An example of a hypergraph product between two path graphs is shown in Fig.~\ref{fig:hp-grid}; the resulting product forms a grid. When the product is taken between two bipartite graphs, the result is a quadripartite graph. Let $\Gamma_a = (V_a \sqcup C_a, E_a)$ and $\Gamma_b = (V_b \sqcup C_b, E_b)$ be bipartite graphs. Their hypergraph product is a graph $\Gamma_P$ with four disjoint sets of vertices:
$$
V_a \times V_b,\quad V_a \times C_b,\quad C_a \times V_b,\quad C_a \times C_b.
$$
Each pair of edges $(v_a, c_a) \in E_a$ and $(v_b, c_b) \in E_b$ defines a square in $\Gamma_P$ with vertices
$$
\{(v_a, v_b),\ (v_a, c_b),\ (c_a, v_b),\ (c_a, c_b)\}.
$$
 
\paragraph{Construction of HP Codes}  
A hypergraph product (HP) code is defined directly from this product structure. We start with two bipartite graphs $\Gamma_a = (V_a \sqcup C_a, E_a)$ and $\Gamma_b = (V_b \sqcup C_b, E_b)$ corresponding o two classical LDPC codes. Vertices in $V_\bullet$ and $C_\bullet$ are variable- and check-nodes, corresponding to columns and rows of the parity check matrix, respectively. An edge in $E_\bullet$ connects a variable-node to a check-node if the corresponding entry in the parity check-matrix is nonzero.
The HP graph $\Gamma_P$ defines a CSS QLDPC as follows: 

\begin{itemize}
    \item Qubits correspond to vertices in $(V_a \times V_b) \cup (C_a \times C_b)$. Accordingly, qubits in $V_a \times V_b$ are often called \textit{vv-nodes}, and those in $C_a \times C_b$ are called \textit{cc-nodes}. 
    \item $X$ checks correspond to vertices in $V_a \times C_b$. Precisely, a vertex $(v_a, c_b) \in V_a \times C_b$ defines an $X$ check with support on its neighboring qubits, i.e.,  $\left(v_a \times V_b(c_b)\right) \sqcup \left(C_a(v_a) \times c_b\right)$, where $V_b(c_b)$ denotes the neighborhood of $c_b$ in $\Gamma_b$, and likewise, $C_a(v_a)$ denotes the neighborhood of $v_a$ in $\Gamma_a$.
    \item $Z$ checks correspond to vertices in $C_a \times V_b$ and they are defined in a similar manner. 
\end{itemize}

The parity-check matrices $H_X$ and $H_Z$ are the incidence matrices between these sets and admit an explicit expression in terms of the classical binary parity-check matrices 
$H_a$ and $H_b$ 
of sizes $r_a \times n_a$ and $r_b \times n_b$, respectively. 
$$
H_X = \left(H_a \otimes I_{n_b} 
\;\middle|\; I_{r_a} \otimes H_b^T \right),
$$
$$
H_Z = \left(I_{n_a} \otimes H_b 
\;\middle|\; H_a^T \otimes I_{r_b} \right).
$$

The reader can easily verify that $H_X H_Z^T = 0$. The resulting quantum code has blocklength $n = n_a n_b + r_a r_b$, dimension $ k = k_a k_b + k_a^T k_b^T$,
where $k^T$ denotes the dimension of the code with parity-check matrix $H^T$, and minimum distance
$ d = \min(d_a, d_b)$.
If both classical codes are asymptotically good, then the HP code satisfies $k = \mathcal{O}(n)$ and $d = \mathcal{O}(\sqrt{n})$. 

An interesting interpretation introduced in~\cite{grospellier:tel-03364419} relates HP codes to classical product codes: if $\mathcal{C}_a = \mathrm{ker}H_a$ and $\mathcal{C}_b = \mathrm{ker}H_b$, and $\mathcal{C}_X,\mathcal{C}_Z$ are the constituent codes of the HP code, then it can be shown that $\mathcal{C}_X^{\perp} = \mathcal{C}_a\otimes C_b^T$, and $\mathcal{C}_Z^{\perp} = \mathcal{C}_a^T\otimes C_b$, where $\otimes$ denotes the classical product code and $\mathcal{C}^T = \mathrm{ker}H^T$.  A list of finite-length examples of HP codes can be found in~\cite{panteleev2021degenerate}.

\subsection{Lifted Product Codes}
\label{sec:codes:lifted_product}
Lifted product (LP) codes can be viewed as a quantum analogue of classical \textit{protograph} LDPC codes. Here, we focus on \textit{quasi-cyclic} (QC) lifted product codes. We start with two \emph{base matrices} $B_a$ of size $r_a\times n_a$ and $B_b$ of size $r_b\times n_b$. Entries in $B_a,B_b$ are elements of $\mathbb{Z}_q\cup \{-1\} = \{-1,0,1,\ldots,q-1\}$\footnote{More generally, the entries may belong to an arbitrary group $G$ if the lift is not cyclic.}, and apply the HP construction from the previous Section to derive matrices $B_X$ and $B_Z$:
\begin{align}
    B_X &= \left(B_a \otimes I_{n_b} \;\middle|\; I_{r_a} \otimes B_b^* \right), \\
    B_Z &= \left(I_{n_a} \otimes B_b \;\middle|\; B_a^* \otimes I_{r_b} \right).
\end{align}

Here, $B^*$ denotes the \textit{transpose conjugate} of $B$, obtained by transposing the matrix and replacing each entry with its additive inverse in $\mathbb{Z}_q$, while entries equal to $-1$ remain invariant.

As in the classical case, each entry of the base matrix is replaced by a circulant binary matrix of size $q \times q$. Specifically, the entry $a \in \mathbb{Z}_q$ corresponds to the identity matrix cyclically shifted by $a$ positions: $0$ gives the (unshifted) identity matrix, $1$ gives the identity shifted by one position, and so forth; the entries equal to $-1$ are replaced by the all-zero $q \times q$ matrix. The resulting block matrices yield the parity-check matrices $H_X$ and $H_Z$. The resulting code satisfies the CSS condition and constitutes a valid QLDPC code. 

LP codes have been shown to achieve dimension  $k=\mathcal{O}(\log n)$ and distance $d=\mathcal{O}(n/\log n)$~\cite{panteleev2022quantumAlmostLinearMinD}, improving on the minimum distance scaling of the HP, but with a trade-off in the number of logical qubits. A noticeable example of finite-length LP code is obtained from the LP between the classical Tanner code of length 155 and itself~\cite{panteleev2022quantumAlmostLinearMinD,raveendran2025mindistanceLPQLDPC}.

\begin{figure}[!htbp]
  \centering

  \subfigure[Classical protograph.]{
    \includegraphics[width=0.15\textwidth]{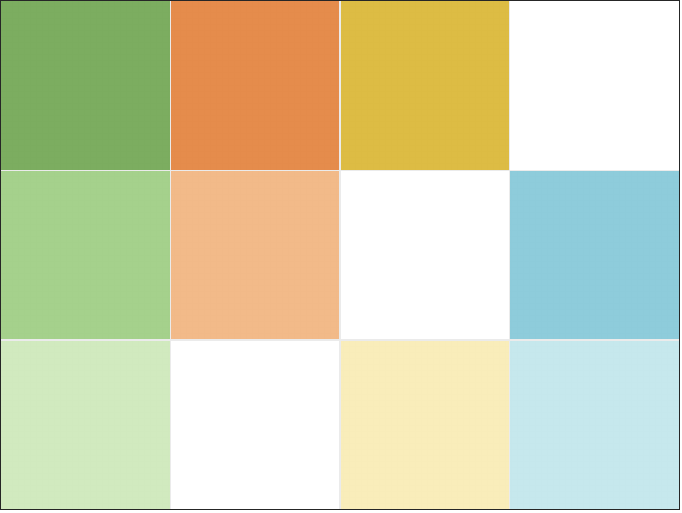}
    \label{fig:lpbase}
  }
  \hspace{12pt}
  \subfigure[$B_X$.]{
    \includegraphics[width=0.35\textwidth]{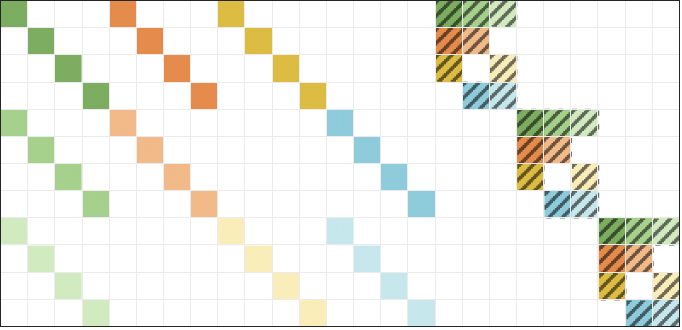}
    \label{fig:lpx}
  }
  \hfill
  \subfigure[$B_Z$.]{
    \includegraphics[width=0.35\textwidth]{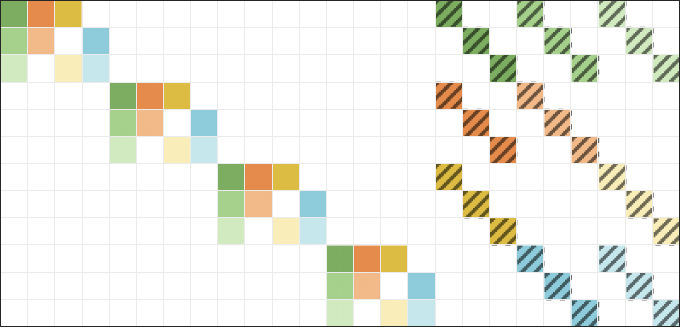}
    \label{fig:lpz}
  }
  \caption{Parity check matrix structure of {LP} codes. 
  \subref{fig:lpbase} Constituent classical base code. Different colors denote different entries in $\mathbb{Z}_q$. 
  \subref{fig:lpx} $B_X$ and \subref{fig:lpz} $B_Z$ are the two base matrices of the LP code. Striped squares denote the conjugate of the corresponding circulant.}
  \label{fig:lp}
\end{figure}

Fig.~\ref{fig:lp} illustrates the structure of the two base matrices $B_X,B_Z$ of an LP code: in Fig. \ref{fig:lpbase} the base protograph $B_a=B_b$ is depicted: each color corresponds to a different entry in $\mathbb{Z}_q$, and white blocks correspond to $-1$. In Figs. \ref{fig:lpx}, \ref{fig:lpz}, the quantum protographs are illustrated (striped squares denote the conjugate element). If we replace all the entries in $\mathbb{Z}_q$ with $1$s, we obtain a representation of the HP construction.

\subsection{Two-Block Codes}
\label{sec:codes:two_block}
Quantum two-block codes were originally constructed by MacKay \textit{et al.}~\cite{mackay_quantum} in their cyclic version called \textit{bicycle codes}. Bicycle codes are constructed from two circulant matrices 
$A$ and $B$ of sizes $\ell\times \ell$ and they are arranged in the two parity check matrices as follows to form the quantum code:
\begin{equation}
    H_X = \left(A \;\middle|\; B \right), \quad H_Z = \left(B^T\;\middle|\; A^T \right).
    \label{eq:2block}
\end{equation}

Because $A$ and $B$ commute (as all circulant matrices do), it is easy to verify that
\begin{equation*}
    H_X H_Z^T = A B + B A=0,
\end{equation*}
Asymptotically, bicycle codes have vanishing rate ($k=O(1))$ and can achieve a distance  $d = \mathcal{O}(\sqrt{n})$~\cite{Pryadko_GenBicycleCodes}.

\subsubsection{Bivariate Bicycle Codes}
More recently, a family of bicycle codes known as \emph{Bivariate Bicycle} (BB) codes~\cite{bravyi2024high} has gained attention due to their structured, geometric layout. In these codes, the circulant matrices $\mathbf{A}$ and $\mathbf{B}$ are constructed as follows:
$$
A = A_1 + A_2 + A_3, \quad B = B_1 + B_2 + B_3,
$$
where each $A_i$ and $B_j$ corresponds to a monomial in the variables $x$ or $y$, defined by:
$$
x = S_\ell \otimes I_m, \quad y = I_\ell \otimes S_m,
$$
with $S_\ell$ denoting the $\ell$-cyclic shift matrix (i.e., a one-step shift of the identity matrix).

\begin{figure}[!htbp]
  \centering
  \subfigure[$H_X$.]{
    \includegraphics[width=0.45\textwidth,trim=50pt 70pt 35pt 70pt,clip]{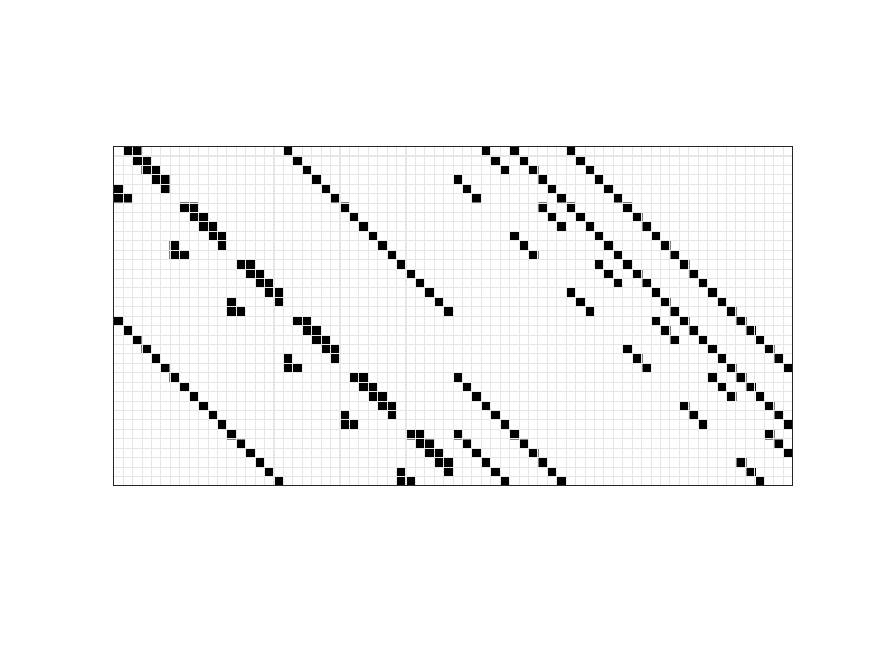}
    \label{fig:bbx}
  }
  \hfill
  \subfigure[$H_Z$.]{
    \includegraphics[width=0.45\textwidth,trim=50pt 70pt 35pt 70pt,clip]{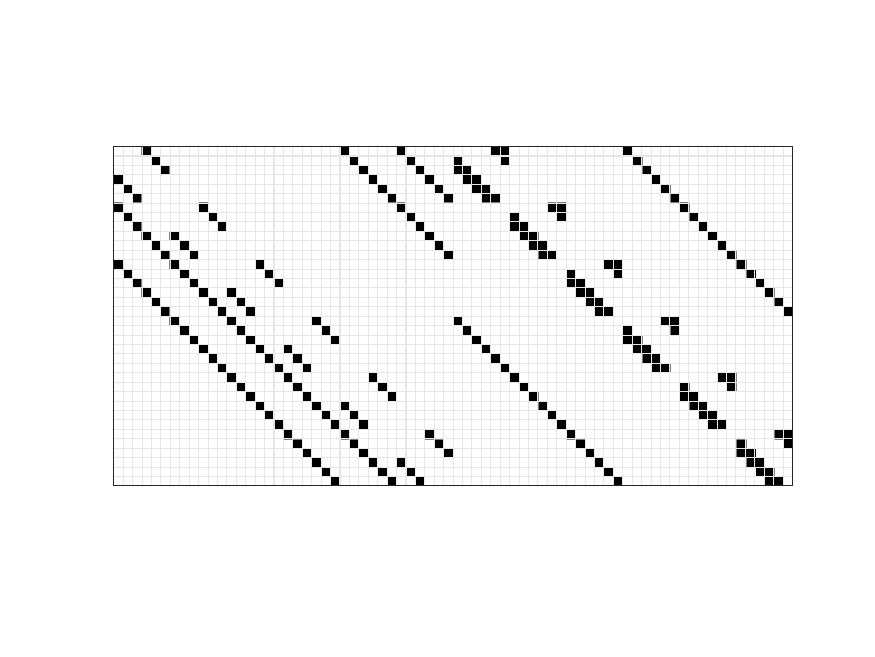}
    \label{fig:bbz}
  }
  \caption{Parity check matrix structure of the Gross code.}
  \label{fig:bb}
\end{figure}

BB codes are bicycle codes with a more elaborate structure, arising from the use of circulant matrices $A$ and $B$  constructed as described above. 
This construction ensures that the Tanner graph of the resulting QLDPC code can be embedded in two interconnected planar layers, thereby making these codes a compelling option for implementation on superconducting qubit architectures.
A prominent example of a BB code is the so-called \emph{Gross code}, with parameters $\llbracket 144,12,12 \rrbracket$, whose parity check matrices are shown in Fig.\ref{fig:bb}. This code has been identified by IBM as a leading candidate in their QEC roadmap~\cite{IBM2025roadmap}.

\subsubsection{Two-Block Group Algebra Codes} 

Two-Block Group Algebra Codes (2BGA) are a family of quantum block codes originally presented in~\cite{pryadko_2bga}. Here, we describe 2BGA codes using Cayley graphs and their adjacency matrices, providing a representation that, while differing from~\cite{pryadko_2bga}, is fully equivalent and more accessible.  
We also note that, although not immediately apparent, this family can be shown to include the BB codes discussed above.

Given a group $G$ and a subset $S\subset G$, the Cayley graph of $G$ with respect to $S$ is the graph $\mathrm{Cay}(G,S)$ with vertex set $G$, and edges $(g,gs)$ for $g\in G$ and $s\in S$, where we have considered that \emph{$S$ acts on $G$ on the right}. If we consider that \emph{$S$  acts on $G$ on the left}, the edges have the form $(g,sg)$. If $G$ is Abelian, both definitions are equivalent; otherwise, they are not.  $S$ is usually referred to as a set of generators, though the term carries a slightly unconventional meaning here: it is not a set of group generators (generating all elements of the group), but rather an arbitrary subset that ``generates'' (i.e., defines) a particular Cayley graph. 

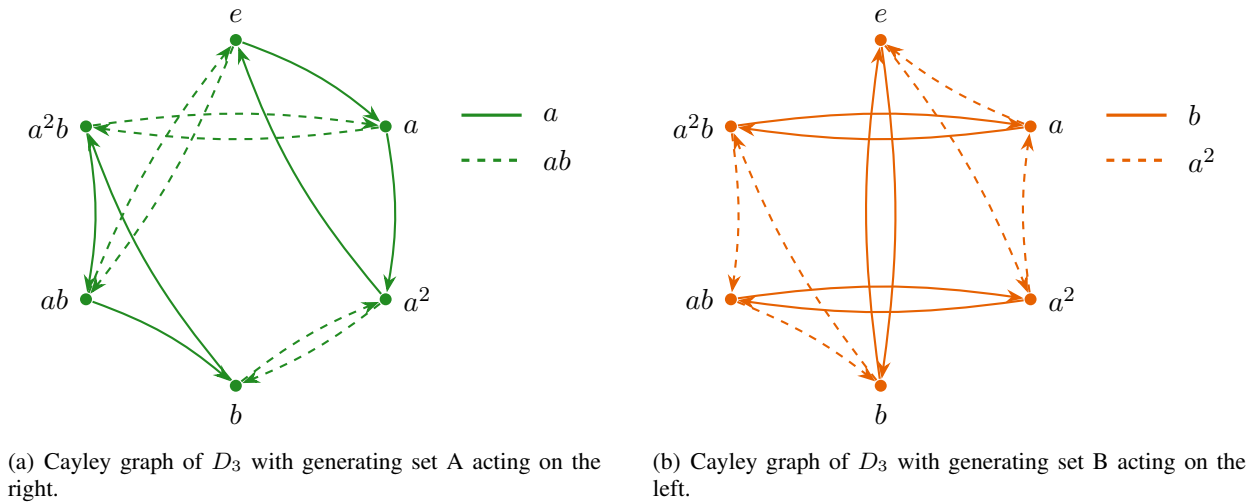
\begin{figure}[!htbp]
\centering

\subfigure[Cayley graph of $D_3$ with generating set A acting on the right.]{
    \begin{tikzpicture}[>=Stealth, thick, node distance=1.8cm]
  \definecolor{genA}{RGB}{34,139,34}   

  \node[circle, fill=genA, inner sep=1.6pt, label=above:$e$]   (e)   at (90:2.3)  {};
  \node[circle, fill=genA, inner sep=1.6pt, label=right:$a$]   (a)   at (30:2.3)  {};
  \node[circle, fill=genA, inner sep=1.6pt, label=right:$a^2$] (a2)  at (-30:2.3) {};
  \node[circle, fill=genA, inner sep=1.6pt, label=below:$b$]   (b)   at (-90:2.3) {};
  \node[circle, fill=genA, inner sep=1.6pt, label=left:$ab$]   (ab)  at (210:2.3) {};
  \node[circle, fill=genA, inner sep=1.6pt, label=left:$a^2 b$](a2b) at (150:2.3) {};

  \draw[genA,->] (e)   to[bend left=10] (a);
  \draw[genA,->] (a)   to[bend left=10] (a2);
  \draw[genA,->] (a2)  to[bend left=10] (e);
  \draw[genA,->] (b)   to[bend left=10] (a2b);
  \draw[genA,->] (ab)  to[bend left=10] (b);
  \draw[genA,->] (a2b) to[bend left=10] (ab);

  \draw[genA, dashed,->] (e)   to[bend left=8] (ab);
  \draw[genA, dashed,->] (a)   to[bend left=8] (a2b);
  \draw[genA, dashed,->] (a2)  to[bend left=8] (b);
  \draw[genA, dashed,->] (b)   to[bend left=8] (a2);
  \draw[genA, dashed,->] (ab)  to[bend left=8] (e);
  \draw[genA, dashed,->] (a2b) to[bend left=8] (a);

  \draw[genA, line width=1.2pt] (3.0,1.3) -- +(0.8,0) node[right, black]{\ \(a\)};
  \draw[genA, dashed, line width=1.2pt] (3.0,0.7) -- +(0.8,0) node[right, black]{\ \(ab\)};
\end{tikzpicture}
    \label{fig:cayley1}
}
\hspace{12pt}
\subfigure[Cayley graph of $D_3$ with generating set B acting on the left.]{
    \begin{tikzpicture}[>=Stealth, thick, node distance=1.8cm]
  \definecolor{genB}{RGB}{230,97,1}    

  \node[circle, fill=genB, inner sep=1.6pt, label=above:$e$]   (e)   at (90:2.3)  {};
  \node[circle, fill=genB, inner sep=1.6pt, label=right:$a$]   (a)   at (30:2.3)  {};
  \node[circle, fill=genB, inner sep=1.6pt, label=right:$a^2$] (a2)  at (-30:2.3) {};
  \node[circle, fill=genB, inner sep=1.6pt, label=below:$b$]   (b)   at (-90:2.3) {};
  \node[circle, fill=genB, inner sep=1.6pt, label=left:$ab$]   (ab)  at (210:2.3) {};
  \node[circle, fill=genB, inner sep=1.6pt, label=left:$a^2 b$](a2b) at (150:2.3) {};

  \draw[genB,->] (e)   to[bend left=8] (b);
  \draw[genB,->] (a)   to[bend left=8] (a2b);
  \draw[genB,->] (a2)  to[bend left=8] (ab);
  \draw[genB,->] (b)   to[bend left=8] (e);
  \draw[genB,->] (ab)  to[bend left=8] (a2);
  \draw[genB,->] (a2b) to[bend left=8] (a);

  \draw[genB, dashed,->] (e)   to[bend left=8] (a2);
  \draw[genB, dashed,->] (a)   to[bend left=8] (e);
  \draw[genB, dashed,->] (a2)  to[bend left=8] (a);
  \draw[genB, dashed,->] (b)   to[bend left=8] (a2b);
  \draw[genB, dashed,->] (ab)  to[bend left=8] (b);
  \draw[genB, dashed,->] (a2b) to[bend left=8] (ab);

  \draw[genB, line width=1.2pt]         (3.0,1.3) -- +(0.8,0) node[right, black]{\ \(b\)};
  \draw[genB, dashed, line width=1.2pt] (3.0,0.7) -- +(0.8,0) node[right, black]{\ \(a^2\)};
\end{tikzpicture}
    \label{fig:cayley2}
}

\caption{Two Cayley graph representations of $D_3$.  
\subref{fig:cayley1} First variant.  
\subref{fig:cayley2} Second variant.}
\label{fig:cayley_combined}
\end{figure}

\begin{figure}[!htbp]
  \centering
  \subfigure[$H_X$.]{
    \includegraphics[width=0.35\textwidth, trim=50pt 70pt 35pt 70pt,clip]{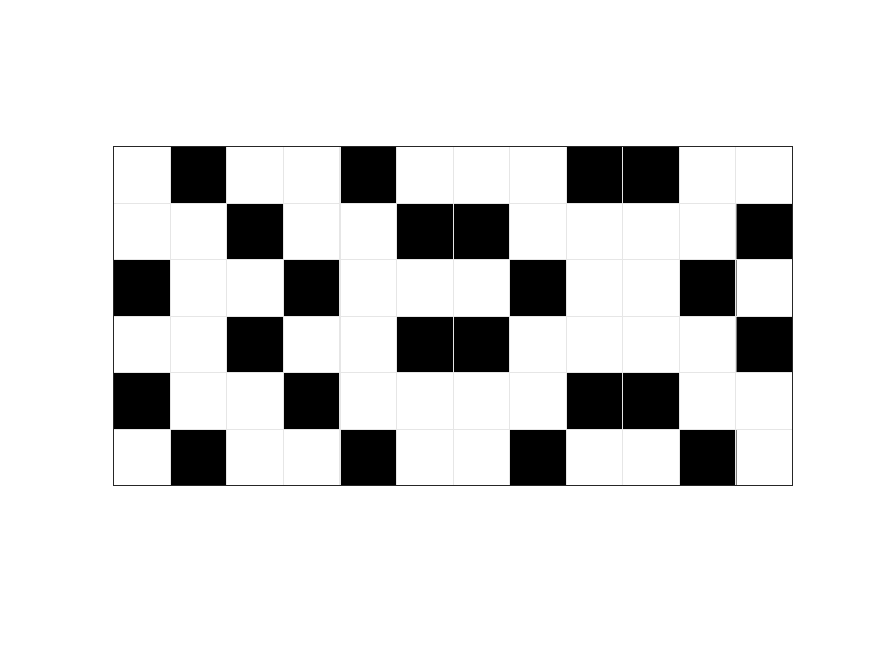}
    \label{fig:d3x}
  }
  \hspace{12pt}
  \subfigure[$H_Z$.]{
    \includegraphics[width=0.35\textwidth, trim=50pt 70pt 35pt 70pt,clip]{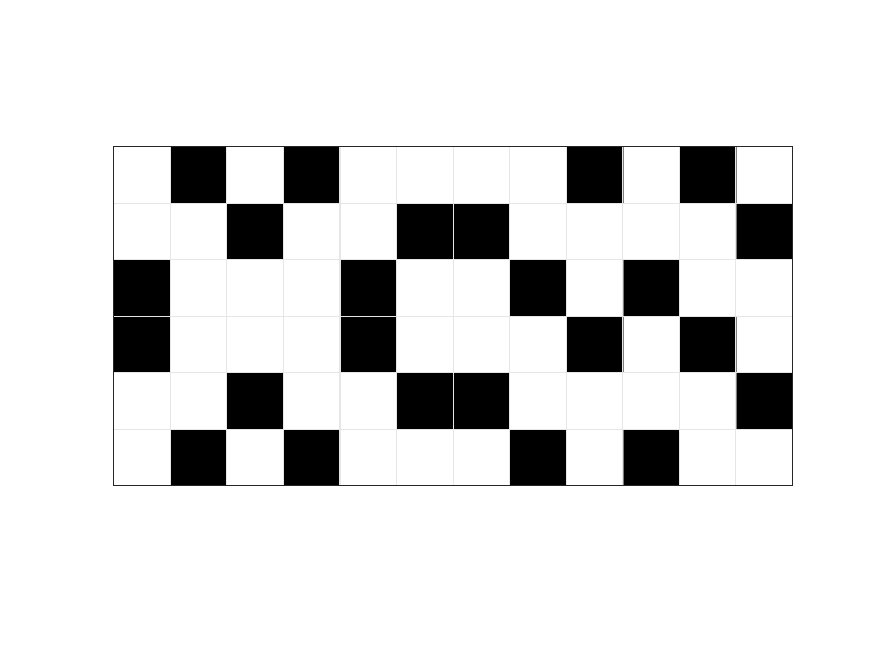}
    \label{fig:d3z}
  }
  \caption{Parity check matrices of the two-block code constructed from the two Cayley graphs of Fig. \ref{fig:cayley_combined}.}
  \label{fig:d3}
\end{figure}

For an arbitrary, possibly non-Abelian, group $G$ we consider two sets of generators $S_A, S_B\subset G$ and the corresponding Cayley graphs $\mathrm{Cay}(G,S_A)$ and $\mathrm{Cay}(G,S_B)$, with $S_A$ acting on $G$ on the right, and $S_B$ acting on the left. We then take the matrices $A$ and $B$ of our two-block quantum code to be the adjacency matrices of these graphs. 
To avoid delving into complex mathematical details, we illustrate this construction using the group of symmetries of a triangle, namely the \textit{dihedral group} 
$D_3 = \langle a, b \mid a^3 = e,\ b^2 = e \rangle$, where $e$ denotes the identity element. The group elements can be listed as:
$$
D_3 = \{e, a, a^2, b, ab, a^2 b\}.
$$
We choose two generating sets: $S_A = \{a, ab\}$, which acts on the right, and $S_B = \{b, a^2\}$, which acts on the left. The corresponding Cayley graphs are shown in Fig.~\ref{fig:cayley_combined}. Since $D_3$ is non-Abelian, the Cayley graphs are directed and their adjacency matrices are not symmetric.
The corresponding adjacency matrices are shown in Fig.~\ref{fig:d3_matrices}, with rows and columns labeled by the group elements.  
Bold entries correspond to solid edges in the Cayley graphs, while italic entries correspond to dashed edges. Finally, Fig.~\ref{fig:d3} shows the parity-check matrices of the quantum code obtained using (\ref{eq:2block}).

In~\cite{pryadko_2bga} two-block codes from Abelian and non-Abelian groups, with length up to $n=200$ have been constructed and analyzed, and empirically shown to have distance scaling as $\mathcal{O}(\sqrt{n})$. In~\mycite{quantum_margulis_allerton, pacenti2025constructiondecodingquantummargulis} a special instance of two-block codes, called \textit{quantum Margulis codes}, was constructed and analyzed. Quantum Margulis codes are the quantum counterpart of the well-known classical Margulis LDPC construction. It has been shown that quantum Margulis codes show excellent decoding performance without the aid of complex post-processing technique, while other families, such as HP, LP, or bicycle codes, do need this additional post-processing to obtain reliable performance.

\begin{figure}[h!]
\centering

\begin{minipage}[t]{0.48\linewidth}
\centering
\[A=\left(
\begin{array}{c|cccccc}
 & e & a & a^{2} & b & ab & a^{2}b \\
\hline
e      & 0 & \textcolor{genA}{\mathbf{1}} & 0 & 0 & \textcolor{genA}{\mathit{1}} & 0 \\
a      & 0 & 0 & \textcolor{genA}{\mathbf{1}} & 0 & 0 & \textcolor{genA}{\mathit{1}} \\
a^{2}  & \textcolor{genA}{\mathbf{1}} & 0 & 0 & \textcolor{genA}{\mathit{1}} & 0 & 0 \\
b      & 0 & 0 & \textcolor{genA}{\mathit{1}} & 0 & 0 & \textcolor{genA}{\mathbf{1}} \\
ab     & \textcolor{genA}{\mathit{1}} & 0 & 0 & \textcolor{genA}{\mathbf{1}} & 0 & 0 \\
a^{2}b & 0 & \textcolor{genA}{\mathit{1}} & 0 & 0 & \textcolor{genA}{\mathbf{1}} & 0
\end{array}\right)
\]
\smallskip
\end{minipage}
\hfill
\begin{minipage}[t]{0.48\linewidth}
\centering
\[B=\left(
\begin{array}{c|cccccc}
 & e & a & a^{2} & b & ab & a^{2}b \\
\hline
e      & 0 & 0 & \textcolor{genB}{\mathit{1}} & \textcolor{genB}{\mathbf{1}} & 0 & 0 \\
a      & \textcolor{genB}{\mathit{1}} & 0 & 0 & 0 & 0 & \textcolor{genB}{\mathbf{1}} \\
a^{2}  & 0 & \textcolor{genB}{\mathit{1}} & 0 & 0 & \textcolor{genB}{\mathbf{1}} & 0 \\
b      & \textcolor{genB}{\mathbf{1}}  & 0 & 0 & 0 & 0 & \textcolor{genB}{\mathit{1}} \\
ab     & 0 & 0 & \textcolor{genB}{\mathbf{1}} & \textcolor{genB}{\mathit{1}}  & 0 & 0 \\
a^{2}b & 0 & \textcolor{genB}{\mathbf{1}} & 0 & 0 & \textcolor{genB}{\mathit{1}}  & 0
\end{array}\right)
\]
\smallskip
\end{minipage}
\caption{Adjacency matrices of the Cayley graphs illustrated in Fig. \ref{fig:cayley_combined}. Bold entries correspond to solid edges, while italic entries correspond to dashed edges.}
\label{fig:d3_matrices}
\end{figure}

\subsection{Asymptotically Good QLDPC Codes}
\label{sec:codes:miscellaneous}
In this Section, we briefly review additional families of QLDPC codes which, while less commonly used in practice, offer important theoretical insights and achieve strong asymptotic parameters.

A first example is the \textit{balanced product}  codes~\cite{breuckmann2020balanced}, which originate from an HP construction between two graphs, each admitting a group action on its vertices. The resulting HP graph is then quotiented by this group action, producing a smaller graph in which each vertex represents an orbit under the action. This construction yields QLDPC whose $k$ and $d$ parameters scale sublinearly with $n$, and are therefore not asymptotically good.  However, assuming the existence of 2-sided lossless expander with free group action, balanced product codes can asymptotically achieve both positive rate and linear minimum distance~\cite{lin2022goodquantumldpccodes}. While such expander graphs were only conjectured to exist, they have actually been proposed recently in~\cite{hsieh2025explicitlosslessvertexexpanders}.

To date, three constructions of asymptotically good QLDPC codes are known, all three of which appeared within a few months of each other: the first by Panteleev and Kalachev~\cite{panteleev_asymptotically_2021}, the second by Leverrier and Zémor~\cite{leverrier2022_QTannerCodes}, and the third by Dinur \textit{et al.}~\cite{dinur_good_2022}.
All three constructions rely on similar underlying principles. The construction in~\cite{panteleev_asymptotically_2021} is based on the LP of two expander graphs, while~\cite{leverrier2022_QTannerCodes,dinur_good_2022} use the so-called \textit{left-right Cayley complex}, a quadripartite structure derived from two Cayley graphs of a group. Although a detailed discussion is beyond the scope of this paper, it can be shown that the left-right Cayley complex is essentially a balanced product of two Cayley graphs. This observation underscores the deep connections among these constructions.

A key ingredient shared across these works is the use of \textit{generalized parity checks}, where check nodes enforce constraints corresponding to local subcodes, as in classical Tanner codes (sometimes referred to as generalized LDPC codes). Leverrier and Zémor rightly named their construction  \emph{quantum Tanner codes}~\cite{leverrier2022_QTannerCodes}. Just as in the seminal work of Sipser and Spielman~\cite{96SS} for classical LDPC codes, the combination of good local code properties with the global expansion of the underlying graphs allows these constructions to achieve asymptotically good parameters, with both the rate and relative minimum distance bounded away from zero.

\section{Decoding Algorithms for QLDPC Codes}
\label{sec:decoder}
Having previously discussed both the noise models and QLDPC code construction, we will now proceed to the crucial task of decoding. 
In virtually all classical error correction systems, iterative decoders use channel log-likelihoods from the channel output vector $\mathbf{y}$ to find the closest estimated vector $\hat{\mathbf{y}}$ that has zero syndrome and thus is a valid codeword.
Instead, in quantum error correction, we simply implement an equivalent syndrome-based iterative decoder. We compute the syndrome $\mathbf{s}$ from $\mathbf{y}$, and use only that syndrome in the decoder, while all the variable nodes estimating the error pattern are initialized with the likelihood of no error. i.e., assuming $\mathbf{e}=\mathbf{0}$. 
With this initialization and only the information of $\mathbf{s}$, the iterative decoder does not try to find the estimate of $\mathbf{y}$ which matches the zero-syndrome, but the error estimate that matches the syndrome $\mathbf{s}$. 
Equivalently, the decoder produces the estimate of the error vector $\hat{\mathbf{e}}$ that matches $\mathbf{s}$. 
The obviousness of the above iterative decoder variant may offend readers familiar with classical coding theory; however, the only point we are making is that in quantum channels, the decoder has access only to the syndrome but not the channel-output vector, and hence operates just as a classical one would under the same restriction.
From here on, we only refer to these syndrome-based decoders, and in particular, focus on iterative message-passing (MP) decoders~\cite{savin2014ldpc}.

From the above, the reader may infer that all that is needed is a classical MP decoder, say \emph{Belief Propagation} (BP). Indeed, in the early stages of research on QLDPC codes, physicists attempted to use off-the-shelf decoding algorithms. The results were terrible. For the depolarizing channel, and even more for the circuit-level noise model. Now we understand well the reasons why the na\"{i}ve BP fails. However, there are still numerous open questions, and the problem of designing good decoding algorithms is far from solved - it is the element of unending immaturity. It is even harder when real-world constraints, such as decoding hardware latency and energy dissipation, are included in the picture. There has been a surge in research on this topic in the past couple of years, and a myriad of solutions of varying performance and complexity currently exist. The authors of the paper now came to a fork on the road - to give more or less comprehensive account of the most significant and creative approaches, or to give the crux of the matter, ``the marrow of the quantum decoding covenant.''

\subsection{Quantum Anointing of Classical Message Passing}
\label{sec:decoder:message_passing}
Classical LDPC codes are effectively paired with efficient iterative MP decoding algorithms~\mycite{savin2014ldpc, VNC_2014_Book}. MP decoders operate on a code’s Tanner graph by repeatedly propagating probabilistic information between variable and check nodes until they settle on a binary estimate that matches the measured syndrome. Efficiency and effectiveness of these decoders come by leveraging the random-like structure of the underlying sparse graph. 
In contrast, for QLDPC codes, firstly, the decoding graph no longer resembles a random graph due to the presence of low-weight stabilizer generators that act trivially on the code space, yet through degeneracy of error patterns, affect the convergence of iterative MP decoding. 
Secondly, the multitude of noise models that were discussed in Section III necessitate decoding over a Tanner graph that differs from the Tanner graph derived directly from the code's parity-check matrix. 
At the extreme, with the circuit-level noise model wherein errors may occur at any idling qubit in memory, ancilla qubits used for syndrome measurement, CNOT gates, or during measurement during the SM circuit, the original PCM effectively modifies into a new \emph{detector error matrix} for the corresponding SM circuit. As mentioned in Sec.~\ref{sec:noise:circuit_level}, the columns of the detector error matrix correspond to faults in the SM circuit, which can be significantly larger in number, relative to the number of data qubits. These matrices are structurally different than the original PCM because a circuit fault propagates differently through the SM circuit than a qubit error does.
Moreover, multiple distinct physical faults can yield identical syndrome patterns, producing redundant columns in the detector error matrix that must be merged along with effective fault probabilities computed according to the fault mapping described in Sec.~\ref{sec:noise:circuit_level}.
These redundancies, the large number of circuit-level fault variables, and the need to incorporate time-correlated syndrome information mean that iterative decoders face a qualitatively different problem than in classical LDPC decoding: instead of decoding the code's Tanner graph, they must operate on a larger, fault-expanded \emph{circuit-level decoding graph} whose structure reflects both the code and its measurement circuitry. 
Overall, quantum decoders, in addition to the degenerate errors which we discuss next for the code capacity model, must account for errors introduced by the entangling gates and measurement errors which can cascade into a sprawl of error propagation effects introducing correlated and the infamous \emph{hook} errors~\mycite{turbo-annihilation} - the ancilla faults during SM circuit that can evolve into multiple correlated data-qubit errors or even logical errors.

\subsection{Message Passing -- A Long and Perilous Journey}
It is probably fair to say that MP decoding is the main reason for the success of classical LDPC codes; thus, it is worth spending more time analyzing it. Both the theoretical and practical significance of MP algorithms stems from their ability to implement optimal decoding on cycle-free graphs: maximum a posteriori (MAP) decoding is implemented through BP, while maximum likelihood (ML) decoding is implemented via the min-sum (MS) algorithm~\cite{wiberg1996codes}. And even though practical (good) LDPC codes are defined by graphs with cycles, they locally resemble cycle-free graphs. Let us reconsider the previous sentence in a more precise and theoretically justified way. The detrimental effect of cycles on MP decoders manifests through correlations they induce in the exchanged messages. Specifically, the incoming messages to certain variable or check nodes become correlated, reducing the accuracy of inference in MP decoding. Such correlations occur after a number of iterations $\ell \geq g/4$, where $g$ is the girth of the graph\footnote{A message ``crosses'' two edges at each iteration; thus, after $g/4$ iterations, it may traverse two disjoint paths of $g/2$ edges each, which form together a cycle of length $g$.} (i.e., the length of the shortest cycle). However, classical LDPC codes have a favorable property: the girth and the diameter of their graph (i.e., the greatest distance between any two vertices) grow at the same rate, namely $\Theta(\log n)$, where $n$ is the code length. Such a balance between girth and diameter ensures that the information carried by the exchanged messages has been spread widely enough and that the iterative decoding process has become sufficiently strong before the first correlations begin to emerge. 

For QLDPC codes, the situation is markedly different. The orthogonality condition $\mathcolor{red}{H} \mathcolor{blue}{H}^\textbf{T}$ implies that every row of $\mathcolor{blue}{H}$ is a codeword of the code defined by $\mathcolor{red}{H}$, and vice versa. In particular, both codes have a minimum distance of $\Theta(1)$, and therefore a girth of $\Theta(1)$, while the diameter is necessarily in $\Omega(\log n)$. Such an imbalance causes the information carried by the exchanged messages to propagate more slowly and to become more easily trapped by the numerous cycles encountered within the graph. The toric code serves as a striking example, exhibiting a pronounced imbalance between its girth $g=8$ and diameter $\delta=\Theta(\sqrt{n})$.  As shown in~\mycite{ducrest2025blindness}, such a pronounced imbalance confines information propagation during MP decoding, leading to a blindness property: if the unsatisfied checks of a syndrome are separated by a distance $\geq 5$, then qubits in the neighborhood of an unsatisfied check remain unaware of any other unsatisfied checks throughout the iterative decoding process. Such blindness effects may also manifest in more general QLDPC codes; fortunately, the imbalance between girth and diameter diminishes in higher-connectivity graphs, particularly for code lengths of practical interest. While this is encouraging, small-weight codewords corresponding to the rows of the orthogonal matrix inevitably lead to degenerate errors, which will be discussed in the next section.

\subsection{Degenerate Errors -- Two Right Turns to Same Destination}
\label{sec:decoder:degeneracy}
\begin{figure}
    \centering
    \includegraphics[width=0.75\linewidth]{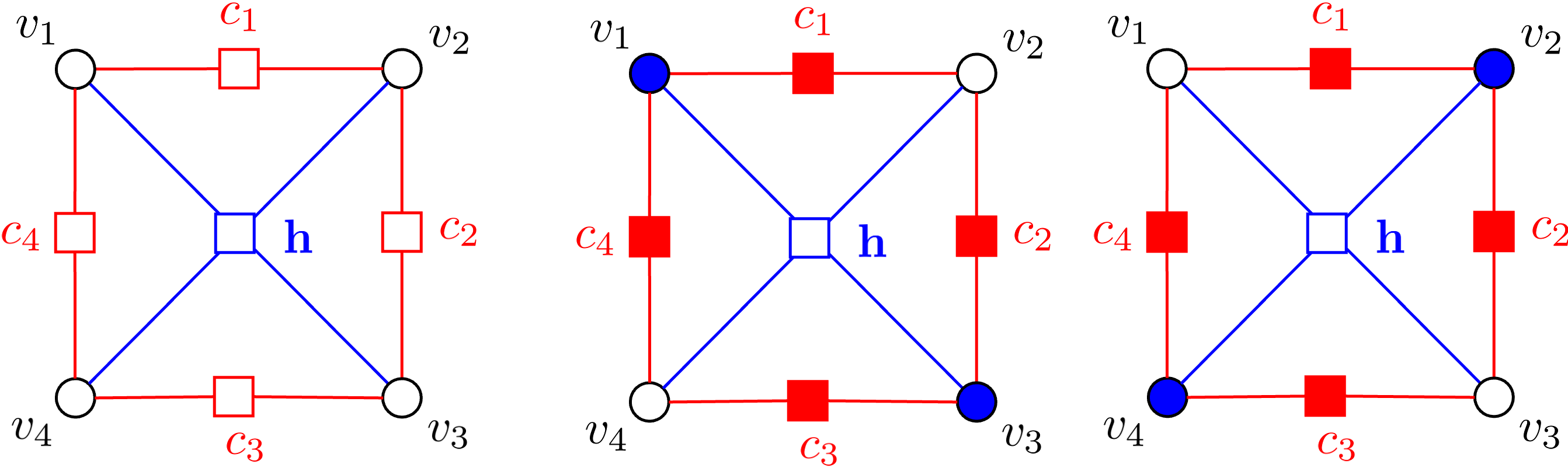}
    \caption{Illustration of degenerate errors in a QLDPC Tanner graph. Left: Stabilizer or check node $\mathbf{h} \in \mathcolor{blue}{H}$ representing a row with support on variable nodes $v_1, \ldots, v_4$ that forms a codeword. Right: Two different error configurations (\scalebox{1.5}{$\mathcolor{blue}{\bullet}$}) yield identical syndromes, as shown by the highlighted variable and check nodes, illustrating the decoding ambiguity introduced by degeneracy.}
    \label{fig:degeneracy}
\end{figure}

In QLDPC codes, the rows (and their linear combinations rowspace) of $\mathcolor{blue}{H}$ generate stabilizers — low-weight codewords of the complementary parity-check matrix $\mathcolor{red}{H}$ — meaning that error patterns with support only on these stabilizer codewords act trivially on the codespace, and vice versa. This also implies that a valid output of a quantum decoder is not just $\mathcolor{blue}{\mathbf{e_1}}$ that caused the non-zero syndrome, but also any one of the vectors $\mathcolor{blue}{\mathbf{e_2}} = \mathcolor{blue}{\mathbf{e_1}}+\mathbf{h}, \mathbf{h} \in \text{rowspace}(\mathcolor{blue}{H})$. 
We illustrate such a scenario in Fig. \ref{fig:degeneracy}. Note that both $\mathcolor{blue}{\mathbf{e_1}} = \{v_1,v_3\}$ and $\mathcolor{blue}{\mathbf{e_2}} = \{v_2,v_4\}$ have the same syndrome indicated by the shaded squares representing unsatisfied check nodes in the subgraphs on the right.  
Unfortunately, this is bad news for a symmetric iterative message passing decoder, as it fails to converge in the presence of multiple solutions. 
The decoding difficulty will manifest itself through the existence of uncorrectable low-weight error patterns sharing their supports with positions of stabilizers. 
To further complicate matters, the requirement of symplectic constraint on QLDPC parity-check matrices also inevitably introduces unavoidable length-8 cycles in the corresponding Tanner graphs. 
These cycles are intricately linked to these symmetric stabilizer codewords that the authors have collectively studied as quantum trapping sets in \mycite{NithinTrappingSetQLDPC}, along with the classical-like trapping sets. 
Formally, a \emph{symmetric stabilizer} is a stabilizer whose set of variable (qubit) nodes forms an induced subgraph with no odd-degree check nodes and can be partitioned into an even number of disjoint subsets such that: (a) The subgraphs induced by each subset of variable nodes are isomorphic, and (b) Each subset shares the same set of neighboring odd-degree check nodes. 
Hence, when the stabilizer subgraph exhibits such symmetry, with \emph{degenerate errors}\footnote{Formally, an error $\mathcolor{blue}{\mathbf{e_1}}$ is called \emph{degenerate} if it generates the same syndrome as another error $\mathcolor{blue}{\mathbf{e_1}}$ of equal probability (e.g., same weight, in the BSC model), and there exists no error with that syndrome of higher probability (e.g., smaller weight).} $\mathcolor{blue}{\mathbf{e_1}}$ and $\mathcolor{blue}{\mathbf{e_2}}$ of equal weight, decoding failure can occur as the iterative decoder attempts to converge to both patterns simultaneously, failing to satisfy the input syndrome. 

\begin{figure}
    \centering
    \includegraphics[width=0.45\linewidth]{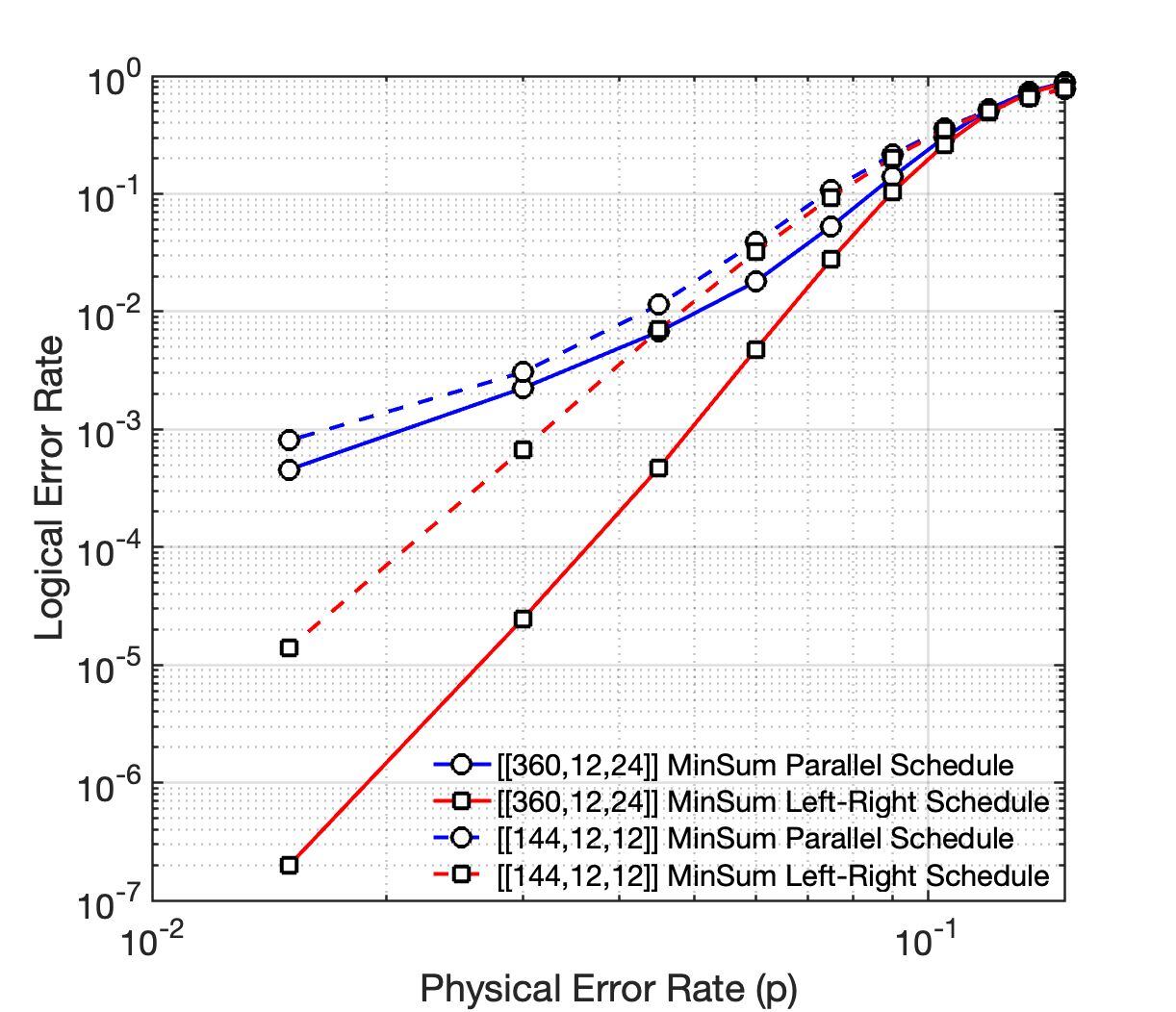}
    \caption{Performance comparison of two min-sum decoding schedules under depolarizing noise in a code-capacity model for the BB code family, like the Gross code~\cite{bravyi2024high}. Both decoders differ only in their schedules and are run for a maximum of 100 iterations with a normalization factor of 0.85. The decoder with parallel (flooding) schedule exhibits a shallow error floor due to its inability to resolve competing degenerate errors, while the layered schedule (left block $\mathbf{A}$, followed by right block $\mathbf{B}$ as in Fig. \ref{fig:bb}) that exploits the two-block structure of bivariate bicycle codes exploits code degeneracy.}
    \label{fig:bbcode_compareMS}
\end{figure}

Decoders that are unable to deal with this \emph{degeneracy} of errors suffer from significantly higher error floors. One key aspect is to introduce strategic asymmetry in the decoder to break these symmetric stabilizers. Here, we present a strategy to incorporate asymmetry in the decoder using the message-update schedule, which refers to the order in which messages are passed and beliefs are updated. Three classical schedules commonly used are flooding (all messages are updated in parallel each iteration - not able to distinguish between degenerate errors within stabilizers), serial (updates sweep the graph check by check in a fixed order - able to correct more degenerate errors with penalty of latency), and layered (variable or check nodes are partitioned into layers with disjoint support, allowing each layer to be updated in parallel - strategic layer scheduling order helps in correcting errors with nominal increase in latency). For the BB code, we demonstrate in Fig. \ref{fig:bbcode_compareMS} how significant the schedule effect can be to break symmetric stabilizer trapping sets and thereby deal with degeneracy.

However, the most common -- and unfortunately complex -- approach taken to enhancing iterative MP decoding for QLDPC codes is that after iterative decoding fails to converge, one makes use of ordered statistics decoding (OSD)~\cite{panteleev2021degenerate, roffe2020decodingQLDPC_OSD,BP_qOSD}, referred to as BP+OSD in the literature. Post-processing with OSD relies on Gaussian elimination–based inversion steps to produce the error estimate, a process that hinders efficient parallelization and also limits scalability for real-time applications, even when implemented on dedicated hardware~\mycite{ducrest2024check,francisco_quantum_min_sum_fpga}.

\subsection{Upgrading Message Passing Decoders -- A Game of Complexity and Performance}
\label{sec:decoder:complexity_performance}
The need to boost MP decoders and an overly complex OSD as a post-processing step motivated the search for newer post-processing algorithms, e.g., 
localized statistics decoding~\cite{hillmann2024localized}, ordered Tanner forest decoding~\cite{demarti2024almost}, closed branch decoding~\cite{demarti2024closed}, ambiguity clustering~\cite{wolanski2024ambiguity}, stabilizer inactivation~\mycite{ducrest2022stabilizer}, and check-agnosia~\mycite{ducrest2024check} to name a few.
These candidate decoders offer different tradeoffs between accuracy, complexity, and hardware efficiency, while building upon or refining inversion-based approaches such as OSD or clustering approaches such as the union-find decoding~\cite{delfosse2021almost}. 
Other approaches aim to improve MP accuracy after understanding their failures at the cost of added complexity, diversity, and heuristics, such as 
guided decimation guessing~\cite{gong2024toward}, collaborative check removal~\cite{bhattacharyya2025decoding}, symmetry-breaking techniques~\cite{yin2024symbreak}, refined or memory BP~\cite{kuo_refined_2020,kuo2022exploitingdegeneracyNature}, MP decoders with past influence \mycite{chytas2025enhancedMSIterdynamics}, relay BP~\cite{muller2025improvedbeliefpropagationsufficient}. 
Neural network-based decoders have also been proposed with excellent performance coming from increased complexity in learning decoder rules and noise statistics, such as neural-net BP\cite{Poulin_NN_BP,beni2025tesseract}, tensor networks~\cite{hack2024belief}, graph neural networks~\cite{ninkovic2024decoding,MP_25_ML_MP_QLDPC,Blue_25_ML_CircuitLevelBB}, and exploring decision tree decoders~\cite{ott2025decision}. 

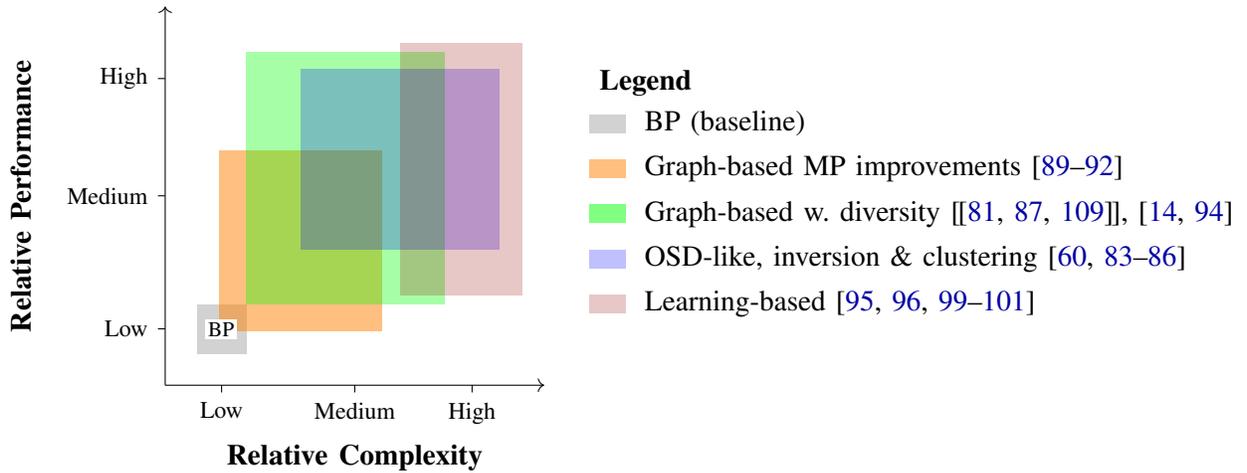
\begin{figure}
\centering
\begin{tikzpicture}[scale=1.2]

\draw[->] (0,0) -- (4.2,0);
\draw[->] (0,0) -- (0,4.2);
\node at (2.1,-0.8) {\textbf{Relative Complexity}};
\node[rotate=90] at (-1.6,2.1) {\textbf{Relative Performance}};
\foreach \x/\lbl in {0.625/{\footnotesize Low}, 2.1/{\footnotesize Medium}, 3.4/{\footnotesize High}}
  \draw (\x,0) -- (\x,-0.08) node[below] {\lbl};
\foreach \y/\lbl in {0.625/{\footnotesize Low}, 2.1/{\footnotesize Medium}, 3.4/{\footnotesize High}}
  \draw (0,\y) -- (-0.08,\y) node[left] {\lbl};

\fill[gray,opacity=0.35] (0.35,0.35) rectangle (0.9,0.9); 
\fill[orange,opacity=0.5] (0.6,0.6) rectangle (2.4,2.6); 
\fill[green,opacity=0.35] (0.9,0.9) rectangle (3.1,3.7);   
\fill[blue,opacity=0.25] (1.5,1.5) rectangle (3.7,3.5); 
\fill[red!60!black,opacity=0.22] (2.6,1) rectangle (3.95,3.8); 

\node[fill=white, inner sep=1pt] at (0.62,0.62) {\scriptsize BP};

\begin{scope}[shift={(4.7,3.1)}]
  \node[above right] at (0,0) {\textbf{Legend}};
  \fill[gray,opacity=0.35]   (0,-0.3) rectangle (0.4,-0.1);
  \node[right] at (0.5,-0.2) {BP (baseline)};
  \fill[orange,opacity=0.5] (0,-0.8) rectangle (0.4,-0.6);
  \node[right] at (0.5,-0.7) {Graph-based MP improvements~\cite{kuo2022exploitingdegeneracyNature,gong2024toward,yin2024symbreak,bhattacharyya2025decoding}};
  \fill[green,opacity=0.5]  (0,-1.3) rectangle (0.4,-1.1);
  \node[right] at (0.5,-1.2) {Graph-based w. diversity~\mycite{pradhan2025linear,ducrest2022stabilizer,ducrest2024check}, \cite{kuo_refined_2020,muller2025improvedbeliefpropagationsufficient}};
  \fill[blue,opacity=0.25]   (0,-1.8) rectangle (0.4,-1.6);
  \node[right] at (0.5,-1.7) {OSD-like, inversion \& clustering~\cite{panteleev2021degenerate,demarti2024almost,demarti2024closed,wolanski2024ambiguity,hillmann2024localized}};
  \fill[red!60!black,opacity=0.22] (0,-2.3) rectangle (0.4,-2.1);
  \node[right] at (0.5,-2.2) {Learning-based~\cite{Poulin_NN_BP,MP_25_ML_MP_QLDPC,ott2025decision,Blue_25_ML_CircuitLevelBB,beni2025tesseract}};
\end{scope}

\end{tikzpicture}
\caption{Qualitative landscape of various QLDPC decoder approaches. Each shaded region represents a class of decoders, plotted by its approximate computational complexity and relative improvement over baseline belief propagation (BP). \emph{Graph-based MP improvements} (orange) modify message‑passing to improve convergence and performance - examples include guided decimation, collaborative check removal, and symmetry‑breaking - typically achieve good accuracy at low-to-medium complexity. \emph{Graph-based w. diversity} (green) spans decoders that introduce diversity or reordering strategies, such as the linear-time algorithm of Pradhan et al., stabilizer inactivation, check-agnosia, and enhancements built on refined BP and relay‑BP; these methods deliver further gains while remaining hardware-friendly. \emph{OSD-like, inversion \& clustering} (blue) covers post-processing strategies that improve on OSD inversion (e.g., ordered Tanner forest, closed-branch, LSD) or cluster corrections (e.g., ambiguity clustering), offering higher accuracy at increased complexity. Finally, \emph{learning-based} decoders (red) employ neural networks or tensor-network methods, including neural BP, Astra, and decision-tree decoders; they can approach near-optimal decoding at the cost of the highest complexity and training requirements. Actual performance and complexity vary with specific codes, error models for benchmarking, such as code capacity and circuit-level noise models, and decoder implementation nuances; this figure illustrates qualitative trends rather than exact quantitative values.}
\label{fig: decodercomparison}
\end{figure}

In Fig. \ref{fig: decodercomparison}, we attempt to represent this tradeoff between complexity and performance qualitatively. Readers are advised that the actual behavior may vary depending on the specific code family and the particular decoder implementation. While some decoders provide reasonably good performance benefits while being fast, we can achieve very high performance with complex BP+OSD and supervised or reinforced learning approaches, pushing even higher on performance. Additionally, while depolarizing noise in code capacity setting serves as a suitable tool for benchmarking these decoders, the true performance is observed in the realistic noise models and scenarios we discuss next. It is imperative that the community endeavors to identify straightforward benchmarking metrics to facilitate comparisons among these diverse decoding algorithms.

\subsection{Syndrome Errors -- We Read the Syndrome Wrong and Say That It Deceives Us}
\label{sec:decoder:syndrome_errors}
We discussed the nuances of phenomenological and circuit-level noise models in Section \ref{sec:noise}, wherein syndrome measurements themselves can be faulty. Feeding a noisy syndrome into an iterative decoder affects its behavior: without the ``true” syndrome, stopping criteria become uncertain, and the decoder may run to its iteration limit without ever finding the correct error. Worse, treating the syndrome as perfect binary data can cause the decoder to lock onto the wrong error pattern entirely. 
The two complementary solutions typically undertaken are a) to repeat the syndrome measurement and b) to feed the \emph{soft syndrome} information to a specialized iterative decoder. 
Repeating syndrome measurements makes syndrome errors tolerable; it also introduces short cycles within the decoding graph. The iterative decoder design must be adapted to handle these cycles during the decoding process. The approach of obtaining soft-syndrome requires modifications to the syndrome readout circuitry as well as the development of suitable modified update rules to pass the additional soft information to boost iterative decoder convergence\mycite{Soft_Syndrome_QCE22}.

Several novel approaches have also emerged to make iterative decoding resilient to measurement and gate errors, including \emph{space-time decoding}~ \cite{kang2025QUITS} that operates on the circuit-level Tanner graph extended to include variables and check nodes in repeated measurements of the syndrome to jointly decode space and time dimensions, and sliding window (SW) decoding strategies to perform such decoding~\cite{skoric2023parallelWindow_RL}. 
References~\cite{Fujiwara_data_syndrome_codes,Ashikhmin_data_syndrome_codes} introduce the idea of meta-checks (redundant stabilizers) to correct syndrome errors. These approaches are intended to correct errors with minimal rounds of syndrome extraction. This idea has been extended and refined in terms of ``single-shot'' decoding in~\cite{Bombin_single_shot, campbell2019theory} 

\subsection{Matching the Tempo of Fault Tolerance}
\label{sec:decoder:fault_tolerance}
In fault-tolerant operation, the syndrome outcomes from the noisy stabilizer measurement (SM) circuits as described in Section \ref{sec:noise:circuit_level} must be decoded at least as fast as they are generated. If the generation-to-processing rate ratio $f=r_{\mathrm{gen}}/r_{\mathrm{proc}}>1$, a backlog forms and the computation can suffer an exponential slowdown with T gate-depth during feed-forwarded non-Clifford steps~\cite{DiVincenzo_2007,terhal_scalable_2019}. In other words: \emph{throughput} (rounds/s) matters, but for conditional logic, the \emph{latency} of each decode cycle - how quickly a single round produces a correction or flag - often dominates. This latency constraint is particularly acute 
on modern superconducting platforms, where full QEC cycles operate on sub-microsecond ($\mu$s) timescales. Recently, below-threshold surface-code memories demonstrated cycle times of about $1.1~\mu$s and logical error rates as low as \(1.43\times10^{-3}\) per cycle~\cite{google2024quantum}. The drive towards fault‐tolerant quantum computing requires real-time decoders that operate on microsecond timescales and are tightly integrated into control stacks~\cite{battistel2023realtime}. Latency budgets become even tighter when implementing multi-round decoding pipelines and in modular or distributed computing architectures with additional interconnect delays\cite{caune2024demonstrating,zhang2025latte,Strikis_QLDPC_ModularArch_PRXQ}. This strict timing constraint has spurred a flurry of research into fast, hardware-friendly decoders.

For instance, IBM researchers recently proposed the relay BP decoder, which dynamically adjusted memory strengths within BP to dampen oscillations and to deal with degenerate errors. It achieves accuracy comparable to minimum‑weight matching for the BB codes while remaining amenable to FPGA or ASIC implementation~\cite{muller2025improvedbeliefpropagationsufficient}. Complementing these heuristic improvements, Pradhan et al.~\cite{pradhan2025linear} provide a systematic theoretical analysis of trapping sets in HP and LP codes, designing linear‑time iterative decoders that avoid them, and thereby reducing the error floor without sacrificing complexity. 
At the hardware-algorithm interface, Garcia‑Herrero et al.~\cite{garcia2025diversityemulators} introduce an FPGA‑based emulator to rapidly test a ``diversity'' ensemble of quantized BP decoders. This ensemble achieves BP+OSD‑level accuracy while delivering 30–120\% speed‑ups and drastically reducing post‑processing latency. Together, these advances illustrate how algorithmic innovations, theoretical insights, and hardware‑aware techniques are converging to meet the stringent demands of real‑time quantum error correction.

\section{Logical Gates and Fault-Tolerance in QLDPC Codes}
\label{sec:logical_gates}

The previous sections have equipped the reader with an understanding of the key components of fault-tolerant error correction. We explained \emph{syndrome measurement}  -- a quantum procedure that retrieves classical information about the physical error affecting the system,  and \emph{syndrome decoding} -- a classical algorithm that processes the extracted syndrome to identify a classical  approximation\footnote{Here, ``classical approximation'' refers to a Pauli error. When physical errors are modeled as Pauli errors, the decoding algorithm aims to identify the actual error that occurred (up to the stabilizer group). For more general non-Pauli errors (e.g., random unitary rotations), the decoding algorithm can only identify a Pauli approximation of the physical error.} of that error.  QLDPC codes enable simple syndrome measurement, where each stabilizer generator is measured by using a single ancilla qubit that interacts with all the data qubits in its support before being measured\footnote{While this procedure may propagate errors and is not generally fault-tolerant, it remains convenient for QLDPC codes due to their low-weight generators, which limits the propagation of errors.}. 
Fault-tolerance is ensured by performing repeated syndrome measurements and jointly decoding the resulting syndromes (referred to as space-time decoding), allowing identification of errors introduced by the syndrome measurement circuit\footnote{Repeated syndrome measurements can be avoided by using codes and decoding algorithms that support single-shot error correction, see Section~\ref{sec:decoder:syndrome_errors}.}.
Finally, fault-tolerant error correction also enables the fault-tolerant preparation of logical states in the logical computational or phase  basis: this simply involves initializing all physical data qubits in the appropriate basis, followed by repeated syndrome measurements and decoding (repeated syndrome measurements are used to ensure fault tolerance in the presence of errors; otherwise, a single round of syndrome measurement would suffice). While this cannot be used to encode any arbitrary (unknown) logical state, it is sufficient for implementing any quantum algorithm, which involves initializing logical qubits in known basis states and then performing logical operations on them.

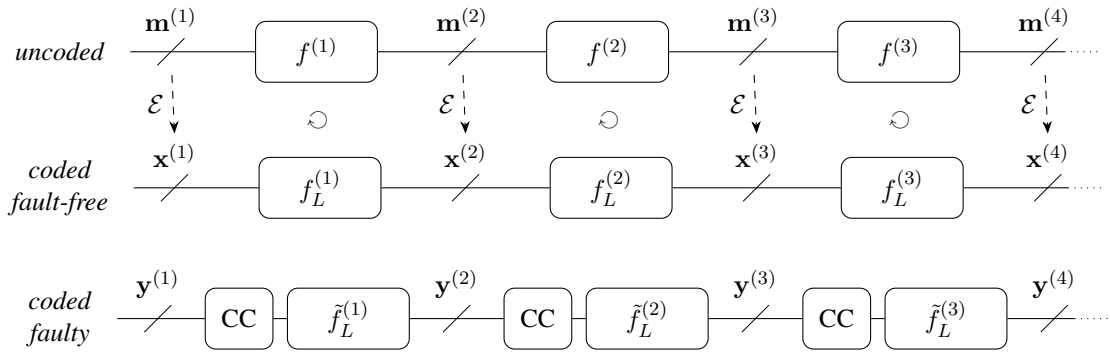
\begin{figure}
\centering
\begin{tikzpicture}[
    scale=0.9, transform shape,
    node distance=1.2cm and 2.2cm,
    box/.style={draw, rounded corners, minimum width=1.8cm, minimum height=0.9cm, align=center},
    smallbox/.style={draw, rounded corners, minimum width=1cm, minimum height=0.9cm, align=center},
    emptynode/.style={
        draw=none,
        minimum width=0pt,
        minimum height=0pt,
        inner sep=0pt,
        outer sep=0pt
    },
    >=Stealth
]

\def\sepa{0.35cm}
\def\sepb{0.2cm}
\def\sepc{0.9cm}
\newcommand{\slantnode}{\tikz{\draw (-0.2,-0.2) -- (0.2,0.2);}}

\node (u) {\emph{uncoded}};

\node[right=0.5cm of u] (m1) {\slantnode};
\node[above=-5pt of m1] {$\mathbf{m}^{(1)}$};

\node[box, right=\sepc  of m1] (f1) {$f^{(1)}$};

\node[right=\sepc of f1] (m2) {\slantnode};
\node[above=-5pt of m2] {$\mathbf{m}^{(2)}$};

\node[box, right=\sepc  of m2] (f2) {$f^{(2)}$};

\node[right=\sepc of f2] (m3) {\slantnode};
\node[above=-5pt of m3] {$\mathbf{m}^{(3)}$};

\node[box, right=\sepc  of m3] (f3) {$f^{(3)}$};

\node[right=\sepc of f3] (m4) {\slantnode};
\node[above=-5pt of m4] {$\mathbf{m}^{(4)}$};

\draw[-] ($(m1)-(0.6,0)$) -- (f1);
\draw[-] (f1) -- (f2);
\draw[-] (f2) -- (f3);
\draw[-] (f3) -- ($(m4)+(0.3,0)$);
\draw[dotted] ($(m4)+(0.3,0)$) -- ($(m4)+(0.8,0)$);

\node[below=of u, align=center] (c) {\emph{coded} \\  \emph{fault-free} };

\node[right=0.5cm of c] (x1) {\slantnode};
\node[above=-5pt of x1] {$\mathbf{x}^{(1)}$};

\node[box, right=\sepc  of x1] (fl1) {$f_L^{(1)}$};

\node[right=\sepc of fl1] (x2) {\slantnode};
\node[above=-5pt of x2] {$\mathbf{x}^{(2)}$};

\node[box, right=\sepc  of x2] (fl2) {$f_L^{(2)}$};

\node[right=\sepc of fl2] (x3) {\slantnode};
\node[above=-5pt of x3] {$\mathbf{x}^{(3)}$};

\node[box, right=\sepc  of x3] (fl3) {$f_L^{(3)}$};

\node[right=\sepc of fl3] (x4) {\slantnode};
\node[above=-5pt of x4] {$\mathbf{x}^{(4)}$};

\draw[-] ($(x1)-(0.6,0)$) -- (fl1);
\draw[-] (fl1) -- (fl2);
\draw[-] (fl2) -- (fl3);
\draw[-] (fl3) -- ($(x4)+(0.3,0)$);
\draw[dotted] ($(x4)+(0.3,0)$) -- ($(x4)+(0.8,0)$);

\draw[dashed,->] (m1) -- ($(x1)+(0,0.8)$) node[midway,left]{$\mathcal{E}$};
\draw[dashed,->] (m2) -- ($(x2)+(0,0.8)$) node[midway,left]{$\mathcal{E}$};
\draw[dashed,->] (m3) -- ($(x3)+(0,0.8)$) node[midway,left]{$\mathcal{E}$};
\draw[dashed,->] (m4) -- ($(x4)+(0,0.8)$) node[midway,left]{$\mathcal{E}$};

\node at ($(f1)!0.5!(fl1)$) {\rotatebox{90}{$\circlearrowright$}};
\node at ($(f2)!0.5!(fl2)$) {\rotatebox{90}{$\circlearrowright$}};
\node at ($(f3)!0.5!(fl3)$) {\rotatebox{90}{$\circlearrowright$}};

\node[below=0.85cm of c, align=center] (f) {\emph{coded} \\  \emph{faulty} };

\node[right=0.5cm of f] (y1) {\slantnode};
\node[above=-5pt of y1] {$\mathbf{y}^{(1)}$};

\node[smallbox, right=\sepa  of y1] (cc1) {CC};
\node[box, right=\sepb  of cc1] (yfl1) {$\tilde{f}_L^{(1)}$};

\node[right=\sepa of yfl1] (y2) {\slantnode};
\node[above=-5pt of y2] {$\mathbf{y}^{(2)}$};

\node[smallbox, right=\sepa  of y2] (cc2) {CC};
\node[box, right=\sepb  of cc2] (yfl2) {$\tilde{f}_L^{(2)}$};

\node[right=\sepa of yfl2] (y3) {\slantnode};
\node[above=-5pt of y3] {$\mathbf{y}^{(3)}$};

\node[smallbox, right=\sepa  of y3] (cc3) {CC};
\node[box, right=\sepb  of cc3] (yfl3) {$\tilde{f}_L^{(3)}$};

\node[right=\sepa of yfl3] (y4) {\slantnode};
\node[above=-5pt of y4] {$\mathbf{y}^{(4)}$};

\draw[-] ($(y1)-(0.6,0)$) -- (cc1) -- (yfl1) -- (cc2);
\draw[-] (cc2) -- (yfl2) -- (cc3);
\draw[-] (cc3) -- (yfl3) -- ($(y4)+(0.3,0)$);
\draw[dotted] ($(y4)+(0.3,0)$) -- ($(y4)+(0.8,0)$);

\end{tikzpicture}
\caption{Principle of encoded information processing. Vertical arrows from uncoded messages $\mathbf{m}^{(\ell)}$ to fault-free codewords $\mathbf{x}^{(\ell)}$ represent the encoding operation, and are included solely to illustrate the relationship between physical operations $f^{(\ell)}$ and logical operations $f_L^{(\ell)}$, namely,  $\mathcal{E}\circ f^{(\ell)} = f_L^{(\ell)} \circ \mathcal{E}$. In the faulty setting, $\mathbf{y}^{(\ell)}$ represents an error-corrupted version of $\mathbf{x}^{(\ell)}$. Correction circuits (CC) are inserted to correct errors on  $\mathbf{y}^{(\ell)}$, which may include those discussed in Section~\ref{sec:intro:quantum_constraints}. Finally, as explained in the main text, $f_L^{(\ell)}$ must be fault-tolerant,  to prevent uncontrolled spread of residual errors (after the correction circuit) as well as those arising from its faulty implementation $\tilde{f}_L^{(\ell)}$.}
    \label{fig:encoded-information-processing}
\end{figure}

However, this is not the end of the story -- rather, it marks the beginning of the journey toward fault-tolerant quantum computing. One of the main challenges we face along this journey is understanding how to operate fault-tolerantly on the error-protected logical qubits. To explain this challenge, let us step back to the classical fault-tolerance setting discussed in Section~\ref{sec:intro}, where we consider a message vector $\mathbf{m}$ composed of $k$ bits (our logical bits), encoded into an $n$-bit codeword $\mathbf{x} = \mathbf{m} G$ (our physical data bits), with $G$ denoting the generator matrix. For simplicity, we assume that the encoding is perfect, while any further operation on $\mathbf{x}$ may introduce errors. Assume that we want to perform some operation on the logical bits, which we can write as  $\mathbf{m'} = f_L(\mathbf{m})$. Reverting the encoding process and operating on $\mathbf{m}$ is not a solution, since it would leave the logical information vulnerable to errors and defeat the purpose of fault-tolerance. Hence, we need to operate directly on $\mathbf{x}$ by finding an operation $\mathbf{x'} = f(\mathbf{x})$, such that $\mathbf{x'}$ encodes $\mathbf{m'}$, i.e., $\mathbf{x'} = \mathbf{m'}G \Leftrightarrow f(\mathbf{m}G) = f_L(\mathbf{m})G$. The above equality must hold for any message (logical) vector  $\mathbf{m}$, and $f$ can be understood as a physical operation realizing the logical operation $f_L$.  Note that $f$ is not unique, and we seek a physical operation that can be implemented in a fault-tolerant manner. We will not provide a rigorous definition of fault tolerance here, yet it is easy to understand that if  $f$ is implemented through a large-depth circuit, errors could spread extensively, exceeding the error-correction capability of the code.  Hence, roughly speaking, a fault-tolerant implementation of $f$ should prevent the uncontrolled spread of errors and ensure that any induced errors remain within the error-correction capability of the code. This approach of encoded information processing is illustrated in Fig.~\ref{fig:encoded-information-processing}.

Of course, the above discussion applies directly to the fault-tolerant quantum computing setting, with the only modification being the replacement of classical encoding with quantum encoding, and classical operations with quantum operations. But before moving fully to the quantum setting, let us further illustrate our point with two examples from the classical setting. As a first example, we consider the logical operation that flips the $i$-th bit of the message vector  $\mathbf{m}$. It can be implemented at the physical level by adding the $i$-th row of $G$ to the encoded vector $\mathbf{x}$, or equivalently, by flipping the bits of $\mathbf{x}$ indexed by the support of the $i$-th row of $G$. Such a physical operation is clearly fault-tolerant, as it is implemented bitwise, and any error that occurs when flipping a bit of  $\mathbf{x}$ will not propagate to other bits. As a second example, we start with a physical operation and ask whether it implements a logical operation, and if so, which one. The physical operation we consider is the bitwise {\sc xor} (sum modulo 2) between two encoded vectors $\mathbf{x'} = \mathbf{m'} G$ and $\mathbf{x''} = \mathbf{m''} G$. This is again fault-tolerant, as it acts \emph{transversally} on the encoded vectors,  and it is easily seen to implement the logical bitwise {\sc xor}, since  $\mathbf{x'} + \mathbf{x''} = (\mathbf{m'} + \mathbf{m''}) G \mod 2$. This is an example of a two-bit gate, implemented transversally (bitwise) on all the logical qubits. However, in a practical scenario, we may want to \emph{address}  specific logical bits to {\sc xor}, i.e., the $i$-th bit of $\mathbf{m'}$ and the $j$-th bit of $\mathbf{m''}$, or the $i$-th and $j$-th bits of  $\mathbf{m'}$. This would be easy if we knew the values of the corresponding logical bits, but we only have access to the encoded vectors $\mathbf{x'}$ and $\mathbf{x''}$ (moreover, in the quantum case, directly observing the state of a qubit is prohibited by the laws of quantum mechanics). Performing addressable logical gates -- i.e., on specific logical bits -- is a more challenging task.

We move now to the quantum case, starting with a translation of what we learned from the two examples above. Transversal physical operations -- whether involving one, two, or more codeblocks -- are fault-tolerant, as errors cannot spread within any codeblock. To avoid ambiguity, let us state that a transversal operation refers to any operation consisting of the parallel application of $c$-qubit gates, where $c$ is the number of codeblocks, with each gate operating on one qubit from each codeblock, and different gates operating on disjoint sets of qubits. A fully transversal operation is a transversal operation in which every qubit in each codeblock is acted upon by some gate.
 Any logical $X$ or $Z$ operator (i.e., a logical bit-flip or phase-flip) can be implemented transversally, by applying the corresponding physical Pauli operator to the physical qubits in its support. A fully transversal CNOT gate applied between two codeblocks implements a fully transversal logical CNOT gate between the logical qubits. Addressing specific logical qubits is a more challenging task, which we will discuss later.
 
In our journey toward fault-tolerant quantum computing, we have so far encountered the concepts of \emph{transversality},  linked to the considerations of fault tolerance, and \emph{addressability}, related to the demands of quantum computation. Another concept tied to the demands of quantum computation is \emph{universality}, which expresses the need to implement a universal set of logical gates to enable the execution of any quantum algorithm. While transversal gates are naturally fault-tolerant, the Eastin-Knill theorem states that they cannot realize a universal set of logical gates~\cite{eastin2009restrictions}. Transversal operations between two or more codeblocks may also be physically infeasible in quantum systems with specific qubit layouts and connectivity constraints (e.g., 2D layouts with nearest-neighbor interactions only). These limitations motivate the use of other fault-tolerant techniques, such as lattice surgery, magic state distillation, code deformation, or code switching, to achieve universality with different families of codes.
Universality is also often discussed in conjunction with the \emph{Clifford group}, consisting of quantum operations that map Pauli operators to Pauli operators under conjugation. Although Clifford operations alone do not enable universal quantum computation -- in fact, Clifford circuits can be efficiently simulated on a classical computer~\cite{Gottesman97} -- adding any non-Clifford operation to the Clifford group suffices to achieve universality, with prominent examples being the $T$ gate or the Toffoli gate.

Having covered the key concepts on our journey toward fault-tolerant quantum computing, let us now turn our attention to the earliest and one of the most extensively studied approaches,  which is based on topological codes (Section~\ref{sec:codes:topological}).
This is already a very rich class of codes, giving rise to diverse constructions that support fault-tolerant operations in different ways. As a few examples, we may mention the 2D color codes, which support a transversal implementation of the full Clifford group~\cite{landahl2011fault}, the 3D color codes, which allow for a transversal $T$-gate~\cite{zhu2023non}, or the well-celebrated surface code, which enables universal fault-tolerant operations on 2D layouts with nearest-neighbor connectivity, through lattice surgery~\cite{horsman2012surface} and state injection and distillation techniques~\cite{bravyi2005universal,campbell2017unified,litinski2019magic}. As discussed in Sec.~\ref{sec:codes},  while topological codes
form a rich and versatile class, they encode only a constant number of logical qubits and their code distance scales poorly, limitations addressed by QLDPC constructions. However, insights from fault-tolerant operations on topological codes are informing and advancing the development of fault-tolerance techniques for more general QLDPC codes.

For QLDPC codes, having multiple logical qubits encoded within the same codeblock introduces difficulties in selectively applying fault-tolerant logical gates to individual logical qubits (such an addressability issue does not arise when each codeblock encodes a single logical qubit, e.g., the surface code). The general solution for stabilizer codes, involving teleportation of qubits or gates and measurement of logical observables with ancilla states~\cite{gottesman1999demonstrating, gottesman2013fault}, carries a prohibitive cost that could offset the low qubit-overhead benefit of QLDPC codes. Alternative low-overhead solutions are currently under active investigation, with  significant progress made recently. Concepts from topological codes, such as folding or surgery, have been adapted or generalized, providing methods to implement logical operations in suitable quantum LDPC codes~\cite{cohen2022low, breuckmann2024fold, cowtan2024css}. For bivariate bicycle codes, gauge-fixed surgery has been proposed to implement the logical Clifford group~\cite{cross2024improved}, and fold-transversal Clifford gates without overhead have been studied in~\cite{eberhardt2024logical}.
For hypergraph product codes, it has been shown that transversal operations, interleaved with error-correction, allow implementation of entangling gates between arbitrary pairs of logical qubits~\cite{quintavalle2023partitioning}. Finding fault-tolerant logical Clifford gates via automorphisms of the code's symplectic representation has been explored in~\cite{sayginel2024fault}. Very recently, a new family of QLDPC codes that enable efficient implementation of the full Clifford group via transversal operations has been proposed in~\cite{malcolm2025computing}. To conclude this section, we note that fault-tolerant protocols using quantum LDPC codes, which achieve constant-space and logarithmic or polylogarithmic time overheads, have been recently studied in~\cite{nguyen2024quantum, tamiya2024polylog}.

\section{Discussion and Perspectives}
Quantum computing is a rapidly evolving field. And within this rapidly evolving landscape, quantum error correction represents the promise of bridging the smaller-scale prototypical systems of today with the vision of a utility-scale fault-tolerant quantum system. And within the larger effort to build high performing quantum error correction architectures, QLDPC codes represent an effort to use some of the best coding theoretical techniques acquired from over fifty years of research in classical error correction systems in the quantum domain. Of course, such a transfer of technology is not facilitated easily, and the counter-intuitive nature of quantum mechanics makes this process far from straightforward. But if the last twenty years of progress in QLDPC codes is to be representative of the future, perhaps this task is not so difficult after all. Indeed, recent developments such as the construction of QLDPC codes with linear distance and constant overhead~\cite{panteleev_asymptotically_2021, dinur_good_2022, leverrier2022_QTannerCodes} and the consideration of hardware layout-friendly constructions such as the bicycle bivariate codes by IBM~\cite{bravyi2024high}, have positioned QLDPC codes as one of the leading candidates for the vision of universal fault tolerance. 

Nevertheless, the challenge is far from over. Different qubit modalities impose distinct physical constraints at the hardware level, which makes it essential to tailor QLDPC codes to the underlying platform. For example, neutral-atom quantum computers permit non-local interactions through atom rearrangement, although qubits remain confined to motion within a two-dimensional plane\cite{bluvstein2023Reconfig_Atom_Exp}. In this setting, recent work has demonstrated that quasi-cyclic QLDPC codes provide an especially favorable match to this hardware architecture~\mycite{nature_qldpc_demonstration}.
By contrast, superconducting qubits support only local two-qubit interactions, a feature that naturally aligns with 2D layouts. 
For a QLDPC code supporting a general 2D layout, the situation is rather bleak: the Bravyi-Poulin-Terhal theorem~\cite{bravyi2010tradeoffs}
establishes a maximum achievable scaling $kd^2 = \mathcal{O}(n)$, resulting in a severely limited tradeoff between the number of encoded qubits $k$, the code distance $d$, and the number of physical qubits $n$. However, as exemplified by the case of the bivariate bicycle  codes~\cite{bravyi2024high}, while a fully 2D layout cannot be achieved, the number of long-range non-local interconnections required to implement a QLDPC code can certainly be minimized. Other approaches to handling the layout issue include modular architectures based on generalized QLDPC codes~\cite{Strikis_QLDPC_ModularArch_PRXQ}, splitting long range stabilizers into smaller pieces using the idea of subsystem codes~\cite{novak2024gnarsil}, and performing long-range stabilizer checks less frequently relative to their local counterparts~\cite{berthusen2025toward}. 

Moreover, a crucial step toward fault-tolerant quantum computing is being able to perform fault-tolerant logical operations on QLDPC codes. The primary difficulty arises from the very feature that makes QLDPC codes attractive: encoding many logical qubits within the same block makes it challenging to address individual logical qubits. Existing solutions for implementing logical operations on QLDPC codes often apply to specific codes, support only restricted operations on the logical space, or entail significant overhead. Nevertheless, recent times have seen a steady progression of new ideas in this direction~\cite{yoder2025tour, he2025extractors}, with each new idea igniting a fresh debate on the way forward.

At the time of this writing, the best hardware platform to be used for the fault tolerant era is far from certain. And with each new hardware platform comes a new noise model, with the promise of significantly improved decoding performance, provided we are able to accurately model noise and are able to appropriately utilize this information in the decoding algorithm. Given the increasingly complex nature of these hardware-specific noise models, the development of coding architectures for these systems often branches out into rich and diverse problems of their own. Learning-based decoders~\cite{fitzek2020deep, andreasson2019quantum} seem to offer one way of making sense of this chaos. Indeed, in recent times, decoders based on reinforcement learning~\mycite{milad2025itw, milad2025istc, milad2025action} have seen an immense amount of success in the classical world and it is expected that these success stories will eventually find their place in the quantum domain as well.

Finally, we mention once again, the need for speed in decoding algorithms, the importance of which we have emphasized throughout this article. As the hardware continues to evolve, with greater emphasis being placed on faster, more integrated quantum systems, the demands placed on the decoding algorithm escalate accordingly. Multi-round pipelines, cryo/room-temperature interconnects, and modular links tighten the timing margins further, while power and area budgets pressure FPGA/ASIC implementations. Meeting all constraints simultaneously—accuracy, sub-\(\mu\)s reaction, low energy, and scalability—remains a formidable co-design problem spanning hardware, firmware, and algorithms. Still, with below-threshold memories running with real-time decoding, hardware decoders showing \(\mu\)s-speed feedback, and improved message-passing for QLDPC, the field is converging on decoders that keep time with fault tolerance rather than holding it. The outlook is far from bleak—steady advances across hardware, algorithms, and system integration have brought us closer to real-time, efficient, and scalable solutions. Each step forward not only sharpens our ability to keep pace with the demands of quantum error correction but also strengthens the foundation for practical, fault-tolerant quantum computation.

\section{Acknowledgment}
The University of Arizona authors acknowledge the support of the National Science Foundation under grants 
CIF-2420424, CIF-2106189, CCF-2100013, CCSS-2052751, and the generous gift from our friends and Maecenases Dora and Barry Bursey. Valentin Savin acknowledges support from the QuantERA grant EQUIP (ANR-22-QUA2-0005-01), and by the Plan France 2030 (NISQ2LSQ project, ANR-22-
PETQ-0006).

\bibliographystyle{IEEEtran}
\bibliography{./bib/references,./bib/references_qec,./bib/references_vasic,./bib/references_savin}

\begin{thebibliography}{100}
\providecommand{\url}[1]{#1}
\csname url@samestyle\endcsname
\providecommand{\newblock}{\relax}
\providecommand{\bibinfo}[2]{#2}
\providecommand{\BIBentrySTDinterwordspacing}{\spaceskip=0pt\relax}
\providecommand{\BIBentryALTinterwordstretchfactor}{4}
\providecommand{\BIBentryALTinterwordspacing}{\spaceskip=\fontdimen2\font plus
\BIBentryALTinterwordstretchfactor\fontdimen3\font minus
  \fontdimen4\font\relax}
\providecommand{\BIBforeignlanguage}[2]{{%
\expandafter\ifx\csname l@#1\endcsname\relax
\typeout{** WARNING: IEEEtran.bst: No hyphenation pattern has been}%
\typeout{** loaded for the language `#1'. Using the pattern for}%
\typeout{** the default language instead.}%
\else
\language=\csname l@#1\endcsname
\fi
#2}}
\providecommand{\BIBdecl}{\relax}
\BIBdecl

\bibitem{mackay_quantum}
D.~MacKay, G.~Mitchison, and P.~McFadden, ``Sparse-graph codes for quantum
  error correction,'' \emph{IEEE Transactions on Information Theory}, vol.~50,
  no.~10, pp. 2315--2330, Oct. 2004.

\bibitem{Gottesman97}
\BIBentryALTinterwordspacing
D.~Gottesman, ``Stabilizer codes and quantum error correction,'' Dissertation,
  California Institute of Technology, 1997. [Online]. Available:
  \url{https://resolver.caltech.edu/CaltechETD:etd-07162004-113028}
\BIBentrySTDinterwordspacing

\bibitem{gottesman2013fault}
------, ``Fault-tolerant quantum computation with constant overhead,''
  \emph{Quantum Info. Comput.}, vol.~14, no. 15–16, p. 1338–1372, Nov.
  2014.

\bibitem{taylor1}
M.~Taylor, ``Reliable information storage in memories designed from unreliable
  components,'' \emph{Bell System Technical Journal}, vol.~47, pp. 2299--2337,
  1968.

\bibitem{kuznetsov}
A.~V. Kuznetsov, ``Information storage in a memory assembled from unreliable
  components,'' \emph{Problems of Information Transmission}, vol.~9, no.~3, pp.
  254--264, 1973.

\bibitem{07_VC_J_33}
B.~Vasi\'{c} and S.~K. Chilappagari, ``An information theoretical framework for
  analysis and design of nanoscale fault-tolerant memories based on low-density
  parity-check codes,'' \emph{IEEE Transactions on Circuits and Systems I:
  Regular Papers}, vol.~54, no.~11, pp. 2438--2446, Nov. 2007.

\bibitem{07_CV_J_133}
S.~K. Chilappagari and B.~Vasi\'{c}, ``{F}ault tolerant memories based on
  expander graphs,'' in \emph{Proc. IEEE Information Theory Workshop (ITW
  '07)}, Lake Tahoe, CA, Sept. 2007, pp. 126--131.

\bibitem{06_IMCSVB_T_140}
M.~Ivkovic, S.~K. Chilappagari, and B.~Vasi\'{c}, ``{C}onstruction of memory
  circuits using unreliable components based on low-density parity-check
  codes,'' in \emph{Proceedings of IEEE Global Telecommunications Conference
  (GLOBECOM '06)}, San Francisco, CA, Nov. 2006, pp. 1--5.

\bibitem{elsa_faulty_faid_density_evolution_tcomm_2015}
E.~Dupraz, D.~Declercq, B.~Vasi\'{c}, and V.~Savin, ``Analysis and design of
  finite alphabet iterative decoders robust to faulty hardware,'' \emph{IEEE
  Transactions on Communications}, vol.~63, no.~8, pp. 2797--2809, Aug 2015.

\bibitem{ngassa2015density}
C.~K. Ngassa, V.~Savin, E.~Dupraz, and D.~Declercq, ``Density evolution and
  functional threshold for the noisy min-sum decoder,'' \emph{IEEE Transactions
  on Communications}, vol.~63, no.~5, pp. 1497--1509, 2015.

\bibitem{nielsen2010quantum}
M.~A. Nielsen and I.~L. Chuang, \emph{Quantum computation and quantum
  information}.\hskip 1em plus 0.5em minus 0.4em\relax Cambridge University
  Press, 2010.

\bibitem{calderbank1996quantum_exists}
\BIBentryALTinterwordspacing
A.~R. Calderbank and P.~W. Shor, ``Good quantum error-correcting codes exist,''
  \emph{Phys. Rev. A}, vol.~54, pp. 1098--1105, Aug. 1996. [Online]. Available:
  \url{https://link.aps.org/doi/10.1103/PhysRevA.54.1098}
\BIBentrySTDinterwordspacing

\bibitem{steane_multiple_1996}
\BIBentryALTinterwordspacing
A.~Steane, ``Multiple particle interference and quantum error correction,''
  \emph{Proc. R. Soc. Lond. A}, vol. 452, no. 1954, pp. 2551--2577, 1996.
  [Online]. Available: \url{http://arxiv.org/abs/quant-ph/9601029}
\BIBentrySTDinterwordspacing

\bibitem{kuo_refined_2020}
\BIBentryALTinterwordspacing
K.-Y. Kuo and C.-Y. Lai, ``Refined belief propagation decoding of sparse-graph
  quantum codes,'' \emph{IEEE J. on Sel. Areas in Inform. Theory}, vol.~1,
  no.~2, p. 487–498, Aug. 2020. [Online]. Available:
  \url{http://dx.doi.org/10.1109/JSAIT.2020.3011758}
\BIBentrySTDinterwordspacing

\bibitem{kuo_exploitingdegeneracy_2022}
\BIBentryALTinterwordspacing
------, ``Exploiting degeneracy in belief propagation decoding of quantum
  codes,'' \emph{npj Quantum Inform.}, vol.~8, no.~1, pp. 1--9, 2022. [Online].
  Available: \url{https://www.nature.com/articles/s41534-022-00623-2}
\BIBentrySTDinterwordspacing

\bibitem{CorrelatedBF_decodingNithin_2023}
N.~Raveendran, E.~Boutillon, and B.~Vasi{\'c}, ``{Turbo-XZ Algorithm:
  Low-Latency Decoders for Quantum LDPC Codes},'' in \emph{International
  Symposium on Topics in Coding}, Brest, France, September 2023.

\bibitem{blais2021circuit}
A.~Blais, A.~L. Grimsmo, S.~M. Girvin, and A.~Wallraff, ``Circuit quantum
  electrodynamics,'' \emph{Reviews of Modern Physics}, vol.~93, no.~2, p.
  025005, 2021.

\bibitem{krantz2019quantum}
P.~Krantz, M.~Kjaergaard, F.~Yan, T.~P. Orlando, S.~Gustavsson, and W.~D.
  Oliver, ``A quantum engineer's guide to superconducting qubits,''
  \emph{Applied physics reviews}, vol.~6, no.~2, 2019.

\bibitem{rasmussen2021superconducting}
S.~E. Rasmussen, K.~S. Christensen, S.~P. Pedersen, L.~B. Kristensen,
  T.~B{\ae}kkegaard, N.~J.~S. Loft, and N.~T. Zinner, ``Superconducting circuit
  companion — an introduction with worked examples,'' \emph{PRX Quantum},
  vol.~2, no.~4, p. 040204, 2021.

\bibitem{leibfried2003quantum}
D.~Leibfried, R.~Blatt, C.~Monroe, and D.~Wineland, ``Quantum dynamics of
  single trapped ions,'' \emph{Reviews of Modern Physics}, vol.~75, no.~1, p.
  281, 2003.

\bibitem{singer2010colloquium}
K.~Singer, U.~Poschinger, M.~Murphy, P.~Ivanov, F.~Ziesel, T.~Calarco, and
  F.~Schmidt-Kaler, ``Colloquium: Trapped ions as quantum bits: Essential
  numerical tools,'' \emph{Reviews of Modern Physics}, vol.~82, no.~3, pp.
  2609--2632, 2010.

\bibitem{castillo2023electronic}
S.~Castillo, ``The electronic control system of a trapped-ion quantum
  processor: A systematic literature review,'' \emph{IEEE Access}, vol.~11, pp.
  65\,775--65\,786, 2023.

\bibitem{luo2023recent}
W.~Luo, L.~Cao, Y.~Shi, L.~Wan, H.~Zhang, S.~Li, G.~Chen, Y.~Li, S.~Li, Y.~Wang
  \emph{et~al.}, ``Recent progress in quantum photonic chips for quantum
  communication and internet,'' \emph{Light: Science \& Applications}, vol.~12,
  no.~1, p. 175, 2023.

\bibitem{chen2021quantum}
X.~Chen, Z.~Fu, Q.~Gong, and J.~Wang, ``Quantum entanglement on photonic chips:
  a review,'' \emph{Advanced Photonics}, vol.~3, no.~6, pp. 064\,002--064\,002,
  2021.

\bibitem{wintersperger2023neutral}
K.~Wintersperger, F.~Dommert, T.~Ehmer, A.~Hoursanov, J.~Klepsch, W.~Mauerer,
  G.~Reuber, T.~Strohm, M.~Yin, and S.~Luber, ``Neutral atom quantum computing
  hardware: performance and end-user perspective,'' \emph{EPJ Quantum
  Technology}, vol.~10, no.~1, p.~32, 2023.

\bibitem{graham2022multi}
T.~Graham, Y.~Song, J.~Scott, C.~Poole, L.~Phuttitarn, K.~Jooya, P.~Eichler,
  X.~Jiang, A.~Marra, B.~Grinkemeyer \emph{et~al.}, ``Multi-qubit entanglement
  and algorithms on a neutral-atom quantum computer,'' \emph{Nature}, vol. 604,
  no. 7906, pp. 457--462, 2022.

\bibitem{de2021materials}
N.~P. De~Leon, K.~M. Itoh, D.~Kim, K.~K. Mehta, T.~E. Northup, H.~Paik,
  B.~Palmer, N.~Samarth, S.~Sangtawesin, and D.~W. Steuerman, ``Materials
  challenges and opportunities for quantum computing hardware,''
  \emph{Science}, vol. 372, no. 6539, p. eabb2823, 2021.

\bibitem{Nielsen}
M.~A. Nielsen and I.~L. Chuang, \emph{Quantum Computation and Quantum
  Information: 10th Anniversary Edition}, 10th~ed.\hskip 1em plus 0.5em minus
  0.4em\relax New York, NY, USA: Cambridge University Press, 2011.

\bibitem{Wilde_2017}
M.~M. Wilde, \emph{Quantum Information Theory}, 2nd~ed.\hskip 1em plus 0.5em
  minus 0.4em\relax Cambridge University Press, 2017.

\bibitem{geller2013efficient}
M.~R. Geller and Z.~Zhou, ``Efficient error models for fault-tolerant
  architectures and the pauli twirling approximation,'' \emph{Physical Review
  A—Atomic, Molecular, and Optical Physics}, vol.~88, no.~1, p. 012314, 2013.

\bibitem{aaronson2004improved}
\BIBentryALTinterwordspacing
S.~Aaronson and D.~Gottesman, ``Improved simulation of stabilizer circuits,''
  \emph{Physical Review A—Atomic, Molecular, and Optical Physics}, vol.~70,
  no.~5, p. 052328, 2004. [Online]. Available:
  \url{https://arxiv.org/pdf/quant-ph/0406196}
\BIBentrySTDinterwordspacing

\bibitem{kuo_log-domain_2021}
\BIBentryALTinterwordspacing
K.-Y. Kuo and C.-Y. Lai, ``Log-domain decoding of quantum {LDPC} codes over
  binary finite fields,'' \emph{{arXiv}:2104.00304 [quant-ph]}, 2021. [Online].
  Available: \url{http://arxiv.org/abs/2104.00304}
\BIBentrySTDinterwordspacing

\bibitem{tillich_quantum_2014}
\BIBentryALTinterwordspacing
J.-P. Tillich and G.~Zemor, ``Quantum {LDPC} codes with positive rate and
  minimum distance proportional to n{\textasciicircum}\{1/2\},'' \emph{{IEEE}
  Trans. Inform. Theory}, vol.~60, no.~2, pp. 1193--1202, 2014. [Online].
  Available: \url{http://arxiv.org/abs/0903.0566}
\BIBentrySTDinterwordspacing

\bibitem{panteleev_asymptotically_2021}
\BIBentryALTinterwordspacing
P.~Panteleev and G.~Kalachev, ``Asymptotically good quantum and locally
  testable classical {LDPC} codes,'' in \emph{in Proc. 54th Ann. ACM SIGACT
  Symp. Theory of Computing}, 2022, p. 375–388. [Online]. Available:
  \url{https://doi.org/10.1145/3519935.3520017}
\BIBentrySTDinterwordspacing

\bibitem{dinur_good_2022}
\BIBentryALTinterwordspacing
I.~Dinur, M.-H. Hsieh, T.-C. Lin, and T.~Vidick, ``Good quantum {LDPC} codes
  with linear time decoders,'' number: {arXiv}:2206.07750. [Online]. Available:
  \url{http://arxiv.org/abs/2206.07750}
\BIBentrySTDinterwordspacing

\bibitem{leverrier2022_QTannerCodes}
\BIBentryALTinterwordspacing
A.~Leverrier and G.~Z{\'e}mor, ``{Quantum Tanner codes},'' in \emph{IEEE 63rd
  Ann. Symp. on Foundations of Comp. Science}.\hskip 1em plus 0.5em minus
  0.4em\relax IEEE, Nov. 2022, pp. 872--883. [Online]. Available:
  \url{https://doi.ieeecomputersociety.org/10.1109/FOCS54457.2022.00117}
\BIBentrySTDinterwordspacing

\bibitem{gidney2021stim}
C.~Gidney, ``Stim: a fast stabilizer circuit simulator,'' \emph{Quantum},
  vol.~5, p. 497, 2021.

\bibitem{Noh22}
\BIBentryALTinterwordspacing
K.~Noh, C.~Chamberland, and F.~G. Brand\~ao, ``Low-overhead fault-tolerant
  quantum error correction with the surface-gkp code,'' \emph{PRX Quantum},
  vol.~3, p. 010315, Jan 2022. [Online]. Available:
  \url{https://link.aps.org/doi/10.1103/PRXQuantum.3.010315}
\BIBentrySTDinterwordspacing

\bibitem{hopfmueller2024bosonic}
F.~Hopfmueller, M.~Tremblay, P.~St-Jean, B.~Royer, and M.-A. Lemonde, ``Bosonic
  pauli+: Efficient simulation of concatenated {Gottesman-Kitaev-Preskill}
  codes,'' \emph{Quantum}, vol.~8, p. 1539, 2024.

\bibitem{Gottesman01}
\BIBentryALTinterwordspacing
D.~Gottesman, A.~Kitaev, and J.~Preskill, ``Encoding a qubit in an
  oscillator,'' \emph{Phys. Rev. A}, vol.~64, p. 012310, Jun 2001. [Online].
  Available: \url{https://link.aps.org/doi/10.1103/PhysRevA.64.012310}
\BIBentrySTDinterwordspacing

\bibitem{Royer22}
\BIBentryALTinterwordspacing
B.~Royer, S.~Singh, and S.~M. Girvin, ``Encoding qubits in multimode grid
  states,'' \emph{PRX Quantum}, vol.~3, no.~1, p. 010335, 2022. [Online].
  Available: \url{https://link.aps.org/doi/10.1103/PRXQuantum.3.010335}
\BIBentrySTDinterwordspacing

\bibitem{Grimsmo20}
\BIBentryALTinterwordspacing
A.~L. Grimsmo, J.~Combes, and B.~Q. Baragiola, ``Quantum computing with
  rotation-symmetric bosonic codes,'' \emph{Phys. Rev. X}, vol.~10, p. 011058,
  Mar 2020. [Online]. Available:
  \url{https://link.aps.org/doi/10.1103/PhysRevX.10.011058}
\BIBentrySTDinterwordspacing

\bibitem{Xu24}
\BIBentryALTinterwordspacing
Y.~Xu, Y.~Wang, and V.~V. Albert, ``Multimode rotation-symmetric bosonic codes
  from homological rotor codes,'' \emph{Phys. Rev. A}, vol. 110, p. 022402, Aug
  2024. [Online]. Available:
  \url{https://link.aps.org/doi/10.1103/PhysRevA.110.022402}
\BIBentrySTDinterwordspacing

\bibitem{Noh20}
\BIBentryALTinterwordspacing
K.~Noh and C.~Chamberland, ``Fault-tolerant bosonic quantum error correction
  with the surface--{Gottesman-Kitaev-Preskill} code,'' \emph{Phys. Rev. A},
  vol. 101, p. 012316, Jan 2020. [Online]. Available:
  \url{https://link.aps.org/doi/10.1103/PhysRevA.101.012316}
\BIBentrySTDinterwordspacing

\bibitem{Berent24}
\BIBentryALTinterwordspacing
L.~Berent, T.~Hillmann, J.~Eisert, R.~Wille, and J.~Roffe, ``Analog information
  decoding of bosonic quantum low-density parity-check codes,'' \emph{PRX
  Quantum}, vol.~5, p. 020349, May 2024. [Online]. Available:
  \url{https://link.aps.org/doi/10.1103/PRXQuantum.5.020349}
\BIBentrySTDinterwordspacing

\bibitem{nithin_GKP_QLDPC}
\BIBentryALTinterwordspacing
N.~Raveendran, N.~Rengaswamy, F.~Rozp{\k{e}}dek, A.~Raina, L.~Jiang, and
  B.~Vasi{\'c}, ``Finite rate {QLDPC-GKP} coding scheme that surpasses the
  {CSS} {Hamming} bound,'' \emph{Quantum}, vol.~6, p. 767, July 2022. [Online].
  Available: \url{https://arxiv.org/abs/2111.07029}
\BIBentrySTDinterwordspacing

\bibitem{shantom_spacetime_gkp}
S.~K. Borah, A.~K. Pradhan, N.~Raveendran, M.~Pacenti, and B.~Vasi\'c, ``Fault
  tolerant decoding of {QLDPC-GKP} codes with circuit level soft information,''
  in \emph{2025 IEEE International Conference on Quantum Computing and
  Engineering}, Sept. 2025, pp. 1--10.

\bibitem{iverson2020coherence}
J.~K. Iverson and J.~Preskill, ``Coherence in logical quantum channels,''
  \emph{New Journal of Physics}, vol.~22, no.~7, p. 073066, 2020.

\bibitem{beale2018coherence}
S.~J. Beale, J.~J. Wallman, M.~Guti{\'e}rrez, K.~R. Brown, and R.~Laflamme,
  ``Coherence in quantum error-correcting codes,'' \emph{arXiv preprint
  arXiv:1805.08802}, 2018.

\bibitem{gravier2025simulated}
O.~Gravier, T.~Ayral, B.~Vermersch, T.~Meunier, and V.~Savin, ``Simulated
  non-markovian noise resilience of silicon-based spin qubits with surface code
  error correction,'' \emph{arXiv preprint arXiv:2507.08713}, 2025.

\bibitem{KITAEV20032}
A.~Kitaev, ``Fault-tolerant quantum computation by anyons,'' \emph{Annals of
  Physics}, vol. 303, no.~1, pp. 2 -- 30, 2003.

\bibitem{PreskilMWPM_topologicalQM}
\BIBentryALTinterwordspacing
E.~Dennis, A.~Kitaev, A.~Landahl, and J.~Preskill, ``Topological quantum
  memory,'' \emph{J. of Mathematical Physics}, vol.~43, no.~9, pp. 4452--4505,
  2002. [Online]. Available: \url{https://doi.org/10.1063/1.1499754}
\BIBentrySTDinterwordspacing

\bibitem{PhysRevA.80.052312}
A.~G. Fowler, A.~M. Stephens, and P.~Groszkowski, ``High-threshold universal
  quantum computation on the surface code,'' \emph{Phys. Rev. A}, vol.~80, p.
  052312, Nov 2009.

\bibitem{PhysRevLett.97.180501}
\BIBentryALTinterwordspacing
H.~Bombin and M.~A. Martin-Delgado, ``Topological quantum distillation,''
  \emph{Phys. Rev. Lett.}, vol.~97, p. 180501, Oct 2006. [Online]. Available:
  \url{https://link.aps.org/doi/10.1103/PhysRevLett.97.180501}
\BIBentrySTDinterwordspacing

\bibitem{mahmoud2025systematicapproachhyperbolicquantum}
\BIBentryALTinterwordspacing
A.~A. Mahmoud, K.~M. Ali, and S.~Rayan, ``A systematic approach to hyperbolic
  quantum error correction codes,'' 2025. [Online]. Available:
  \url{https://arxiv.org/abs/2504.07800}
\BIBentrySTDinterwordspacing

\bibitem{BRAVYI2011839}
S.~Bravyi, B.~Leemhuis, and B.~M. Terhal, ``Topological order in an exactly
  solvable {3D} spin model,'' \emph{Annals of Physics}, vol. 326, no.~4, pp.
  839--866, 2011.

\bibitem{bravyi2009nogo}
S.~Bravyi and B.~Terhal, ``A no-go theorem for a two-dimensional
  self-correcting quantum memory based on stabilizer codes,'' \emph{New Journal
  of Physics}, vol.~11, no.~4, p. 043029, 2009.

\bibitem{bravyi2010tradeoffs}
S.~Bravyi, D.~Poulin, and B.~Terhal, ``Tradeoffs for reliable quantum
  information storage in {2D} systems,'' \emph{Physical review letters}, vol.
  104, no.~5, p. 050503, 2010.

\bibitem{grospellier:tel-03364419}
A.~Grospellier, ``{Constant time decoding of quantum expander codes and
  application to fault-tolerant quantum computation},'' Theses, {Sorbonne
  Universit{\'e}}, Nov. 2019.

\bibitem{panteleev2021degenerate}
\BIBentryALTinterwordspacing
P.~Panteleev and G.~Kalachev, ``Degenerate quantum {LDPC} codes with good
  finite length performance,'' \emph{Quantum}, vol.~5, p. 585, 2021. [Online].
  Available: \url{https://arxiv.org/abs/1904.02703}
\BIBentrySTDinterwordspacing

\bibitem{panteleev2022quantumAlmostLinearMinD}
------, ``Quantum {LDPC} codes with almost linear minimum distance,''
  \emph{IEEE Trans. Inform. Theory}, vol.~68, no.~1, pp. 213--229, 2022.

\bibitem{raveendran2025mindistanceLPQLDPC}
N.~Raveendran, D.~Declercq, and B.~Vasi{\'c}, ``{On the Minimum Distances of
  Finite-Length Lifted Product Quantum LDPC Codes},'' \emph{arXiv preprint
  arXiv:2503.07567}, 2025.

\bibitem{Pryadko_GenBicycleCodes}
\BIBentryALTinterwordspacing
A.~A. Kovalev and L.~P. Pryadko, ``Quantum kronecker sum-product low-density
  parity-check codes with finite rate,'' \emph{Phys. Rev. A}, vol.~88, p.
  012311, Jul 2013. [Online]. Available:
  \url{https://link.aps.org/doi/10.1103/PhysRevA.88.012311}
\BIBentrySTDinterwordspacing

\bibitem{bravyi2024high}
\BIBentryALTinterwordspacing
S.~Bravyi, A.~W. Cross, J.~M. Gambetta, D.~Maslov, P.~Rall, and T.~J. Yoder,
  ``High-threshold and low-overhead fault-tolerant quantum memory,''
  \emph{Nature}, vol. 627, no. 8005, pp. 778--782, 2024. [Online]. Available:
  \url{https://arxiv.org/abs/2308.07915}
\BIBentrySTDinterwordspacing

\bibitem{IBM2025roadmap}
\BIBentryALTinterwordspacing
R.~Mandelbaum, J.~Gambetta, J.~Chow, T.~Mittal, T.~J. Yoder, A.~Cross, and
  M.~Steffen, ``{How IBM will build the world’s first large-scale,
  fault-tolerant quantum computer},'' \emph{IBM Quantum Blog}, Jun. 2025.
  [Online]. Available: \url{https://www.ibm.com/quantum/blog/large-scale-ftqc}
\BIBentrySTDinterwordspacing

\bibitem{pryadko_2bga}
\BIBentryALTinterwordspacing
H.-K. Lin and L.~P. Pryadko, ``Quantum two-block group algebra codes,''
  \emph{Phys. Rev. A}, vol. 109, p. 022407, Feb 2024. [Online]. Available:
  \url{https://link.aps.org/doi/10.1103/PhysRevA.109.022407}
\BIBentrySTDinterwordspacing

\bibitem{quantum_margulis_allerton}
\BIBentryALTinterwordspacing
M.~Pacenti and B.~Vasi\'c, ``Quantum {M}argulis {C}odes,'' in \emph{60th Annual
  Allerton Conference on Communication, Control, and Computing}, Sept. 25-27
  2024, pp. 1--5. [Online]. Available: \url{https://arxiv.org/abs/2409.09830}
\BIBentrySTDinterwordspacing

\bibitem{pacenti2025constructiondecodingquantummargulis}
M.~Pacenti, D.~Chytas, and B.~Vasi\'c, ``{Construction and Decoding of Quantum
  Margulis Codes},'' \emph{arXiv preprint arXiv:2503.03936}, 2025.

\bibitem{breuckmann2020balanced}
\BIBentryALTinterwordspacing
N.~P. Breuckmann and J.~N. Eberhardt, ``Balanced product quantum codes,''
  \emph{IEEE Trans. Inform. Theory}, vol.~67, no.~10, p. 6653–6674, Oct.
  2021. [Online]. Available: \url{http://dx.doi.org/10.1109/TIT.2021.3097347}
\BIBentrySTDinterwordspacing

\bibitem{lin2022goodquantumldpccodes}
T.-C. Lin and M.-H. Hsieh, ``Good quantum {LDPC} codes with linear time decoder
  from lossless expanders,'' \emph{arXiv preprint arXiv:2203.03581}, 2022.

\bibitem{hsieh2025explicitlosslessvertexexpanders}
J.-T. Hsieh, A.~Lubotzky, S.~Mohanty, A.~Reiner, and R.~Y. Zhang, ``Explicit
  lossless vertex expanders,'' \emph{arXiv preprint arXiv:2504.15087}, 2025.

\bibitem{96SS}
M.~Sipser and D.~Spielman, ``Expander codes,'' \emph{IEEE Trans. Inform.
  Theory}, vol.~42, no.~6, pp. 1710--1722, Jun. 1996.

\bibitem{savin2014ldpc}
V.~Savin, ``{LDPC} decoders,'' in \emph{Channel coding: Theory, algorithms, and
  applications}, D.~Declercq, M.~Fossorier, and E.~Biglieri, Eds.\hskip 1em
  plus 0.5em minus 0.4em\relax Elsevier, 2014, pp. 211--260.

\bibitem{VNC_2014_Book}
B.~Vasi\'{c}, D.~Nguyen, and S.~K. Chilappagari, \emph{Chapter 6 - Failures and
  Error Floors of Iterative Decoders}.\hskip 1em plus 0.5em minus 0.4em\relax
  Oxford: Academic Press, 2014, pp. 299 -- 341.

\bibitem{turbo-annihilation}
M.~Pacenti, A.~K. Pradhan, S.~K. Borah, and B.~Vasi\'c, ``Turbo-annihilation of
  hook errors in stabilizer measurement circuits,'' in \emph{2025 IEEE
  International Conference on Quantum Computing and Engineering}, Sept. 2025,
  pp. 1--10.

\bibitem{wiberg1996codes}
N.~Wiberg, ``Codes and decoding on general graphs,'' Ph.D. dissertation,
  Likoping University, Sweden, 1996.

\bibitem{ducrest2025blindness}
J.~Du~Crest, M.~Mhalla, and V.~Savin, ``A blindness property of the min-sum
  decoding for the toric code,'' \emph{IEEE Journal on Selected Areas in
  Information Theory}, 2025.

\bibitem{NithinTrappingSetQLDPC}
\BIBentryALTinterwordspacing
N.~Raveendran and B.~Vasi{\'{c}}, ``Trapping {S}ets of {Q}uantum {LDPC}
  {C}odes,'' \emph{{Quantum}}, vol.~5, p. 562, Oct. 2021. [Online]. Available:
  \url{https://doi.org/10.22331/q-2021-10-14-562}
\BIBentrySTDinterwordspacing

\bibitem{roffe2020decodingQLDPC_OSD}
\BIBentryALTinterwordspacing
J.~Roffe, D.~R. White, S.~Burton, and E.~Campbell, ``Decoding across the
  quantum low-density parity-check code landscape,'' \emph{Phys. Rev.
  Research}, vol.~2, p. 043423, Dec. 2020. [Online]. Available:
  \url{https://link.aps.org/doi/10.1103/PhysRevResearch.2.043423}
\BIBentrySTDinterwordspacing

\bibitem{BP_qOSD}
C.-F. Kung, K.-Y. Kuo, and C.-Y. Lai, ``On belief propagation decoding of
  quantum codes with quaternary reliability statistics,'' in \emph{Proc. 12th
  Intl. Symp. Topics in Coding}, Sep. 2023.

\bibitem{ducrest2024check}
\BIBentryALTinterwordspacing
J.~du~Crest, F.~Garcia-Herrero, M.~Mhalla, V.~Savin, and J.~Valls,
  ``Check-agnosia based post-processor for message-passing decoding of quantum
  {LDPC} codes,'' \emph{Quantum}, vol.~8, p. 1334, 2024. [Online]. Available:
  \url{https://quantum-journal.org/papers/q-2024-05-02-1334/}
\BIBentrySTDinterwordspacing

\bibitem{francisco_quantum_min_sum_fpga}
J.~V. Coquillat, F.~G. Herrero, N.~Raveendran, and B.~Vasi\'c, ``Syndrome-based
  min-sum vs {OSD-0} decoders: {FPGA} implementation and analysis for quantum
  {LDPC} codes,'' \emph{{IEEE} Access}, vol.~9, pp. 138\,734--138\,743, Oct.
  2021.

\bibitem{hillmann2024localized}
\BIBentryALTinterwordspacing
T.~Hillmann, L.~Berent, A.~O. Quintavalle, J.~Eisert, R.~Wille, and J.~Roffe,
  ``Localized statistics decoding for quantum low-density parity-check codes,''
  \emph{Nature Communications}, vol.~16, no.~1, p. 8214, 2025. [Online].
  Available: \url{https://doi.org/10.1038/s41467-025-63214-7}
\BIBentrySTDinterwordspacing

\bibitem{demarti2024almost}
A.~deMarti iOlius, I.~Etxezarreta~Martinez, J.~Roffe, and
  J.~Etxezarreta~Martinez, ``An almost-linear time decoding algorithm for
  quantum ldpc codes under circuit-level noise,'' \emph{arXiv preprint
  arXiv:2409.01440}, 2024.

\bibitem{demarti2024closed}
A.~deMarti iOlius and J.~Etxezarreta~Martinez, ``The closed-branch decoder for
  quantum {LDPC} codes,'' \emph{arXiv preprint arXiv:2402.01532}, 2024.

\bibitem{wolanski2024ambiguity}
S.~Wolanski and B.~Barber, ``Ambiguity clustering: an accurate and efficient
  decoder for {qLDPC} codes,'' \emph{arXiv preprint arXiv:2406.14527}, 2024.

\bibitem{ducrest2022stabilizer}
J.~Du~Crest, M.~Mhalla, and V.~Savin, ``Stabilizer inactivation for
  message-passing decoding of quantum {LDPC} codes,'' in \emph{2022 IEEE
  Information Theory Workshop (ITW)}.\hskip 1em plus 0.5em minus 0.4em\relax
  IEEE, 2022, pp. 488--493.

\bibitem{delfosse2021almost}
N.~Delfosse and N.~H. Nickerson, ``Almost-linear time decoding algorithm for
  topological codes,'' \emph{Quantum}, vol.~5, p. 595, 2021.

\bibitem{gong2024toward}
A.~Gong, S.~Cammerer, and J.~M. Renes, ``Toward low-latency iterative decoding
  of {QLDPC} codes under circuit-level noise,'' \emph{arXiv preprint
  arXiv:2403.18901}, 2024.

\bibitem{bhattacharyya2025decoding}
M.~Bhattacharyya and A.~Raina, ``Decoding quantum {LDPC} codes using
  collaborative check node removal,'' \emph{arXiv preprint arXiv:2501.08036},
  2025.

\bibitem{yin2024symbreak}
K.~Yin, X.~Fang, J.~Ruan, H.~Zhang, D.~Tullsen, A.~Sornborger, C.~Liu, A.~Li,
  T.~Humble, and Y.~Ding, ``Symbreak: Mitigating quantum degeneracy issues in
  {QLDPC} code decoders by breaking symmetry,'' \emph{arXiv preprint
  arXiv:2412.02885}, 2024.

\bibitem{kuo2022exploitingdegeneracyNature}
K.-Y. Kuo and C.-Y. Lai, ``Exploiting degeneracy in belief propagation decoding
  of quantum codes,'' \emph{npj Quantum Information}, vol.~8, no.~1, p. 111,
  2022.

\bibitem{chytas2025enhancedMSIterdynamics}
\BIBentryALTinterwordspacing
D.~Chytas, N.~Raveendran, and B.~Vasic, ``Enhanced min-sum decoding of quantum
  codes using previous iteration dynamics,'' \emph{arXiv:2501.05021}, 2025.
  [Online]. Available: \url{https://arxiv.org/abs/2501.05021}
\BIBentrySTDinterwordspacing

\bibitem{muller2025improvedbeliefpropagationsufficient}
T.~M{\"u}ller, T.~Alexander, M.~E. Beverland, M.~B{\"u}hler, B.~R. Johnson,
  T.~Maurer, and D.~Vandeth, ``Improved belief propagation is sufficient for
  real-time decoding of quantum memory,'' \emph{arXiv preprint
  arXiv:2506.01779}, 2025.

\bibitem{Poulin_NN_BP}
\BIBentryALTinterwordspacing
Y.-H. Liu and D.~Poulin, ``Neural belief-propagation decoders for quantum
  error-correcting codes,'' \emph{Phys. Rev. Lett.}, vol. 122, p. 200501, May
  2019. [Online]. Available:
  \url{https://link.aps.org/doi/10.1103/PhysRevLett.122.200501}
\BIBentrySTDinterwordspacing

\bibitem{beni2025tesseract}
L.~A. Beni, O.~Higgott, and N.~Shutty, ``Tesseract: A search-based decoder for
  quantum error correction,'' \emph{arXiv preprint arXiv:2503.10988}, 2025.

\bibitem{hack2024belief}
P.~Hack, C.~B. Mendl, and A.~Paler, ``Belief propagation for general graphical
  models with loops,'' \emph{arXiv preprint arXiv:2411.04957}, 2024.

\bibitem{ninkovic2024decoding}
V.~Ninkovic, O.~Kundacina, D.~Vukobratovic, C.~H{\"a}ger \emph{et~al.},
  ``Decoding quantum {LDPC} codes using graph neural networks,'' \emph{arXiv
  preprint arXiv:2408.05170}, 2024.

\bibitem{MP_25_ML_MP_QLDPC}
A.~S. Maan and A.~Paler, ``Machine learning message-passing for the scalable
  decoding of qldpc codes,'' \emph{npj Quantum Information}, vol.~11, no.~1,
  p.~78, 2025.

\bibitem{Blue_25_ML_CircuitLevelBB}
J.~Blue, H.~Avlani, Z.~He, L.~Ziyin, and I.~L. Chuang, ``Machine learning
  decoding of circuit-level noise for bivariate bicycle codes,'' \emph{arXiv
  preprint arXiv:2504.13043}, 2025.

\bibitem{ott2025decision}
K.~R. Ott, B.~Het{\'e}nyi, and M.~E. Beverland, ``Decision-tree decoders for
  general quantum {LDPC} codes,'' \emph{arXiv preprint arXiv:2502.16408}, 2025.

\bibitem{Soft_Syndrome_QCE22}
\BIBentryALTinterwordspacing
N.~Raveendran, N.~Rengaswamy, A.~K.~Pradhan, and B.~Vasi{\'c}, ``Soft syndrome
  decoding of quantum {LDPC} codes to correct of data and syndrome errors,'' in
  \emph{Proc. IEEE Intl. Conf. on Quantum Computing and Engineering}, Sep.
  2022, pp. 275--281. [Online]. Available:
  \url{https://arxiv.org/abs/2205.02341}
\BIBentrySTDinterwordspacing

\bibitem{kang2025QUITS}
M.~Kang, Y.~Lin, H.~Yao, M.~Gökduman, A.~Meinking, and K.~R. Brown, ``{QUITS:
  A modular Qldpc code circUIT Simulator},'' \emph{arXiv preprint
  arXiv:2504.02673}, 2025.

\bibitem{skoric2023parallelWindow_RL}
\BIBentryALTinterwordspacing
L.~Skoric, D.~E. Browne, K.~M. Barnes \emph{et~al.}, ``Parallel window decoding
  enables scalable fault tolerant quantum computation,'' \emph{Nature
  Communications}, vol.~14, p. 7040, 2023. [Online]. Available:
  \url{https://doi.org/10.1038/s41467-023-42482-1}
\BIBentrySTDinterwordspacing

\bibitem{Fujiwara_data_syndrome_codes}
Y.~Fujiwara, ``Ability of stabilizer quantum error correction to protect itself
  from its own imperfection,'' \emph{Phys. Rev. A}, vol.~90, p. 062304, Dec
  2014.

\bibitem{Ashikhmin_data_syndrome_codes}
A.~Ashikhmin, C.-Y. Lai, and T.~A. Brun, ``Quantum data-syndrome codes,''
  \emph{IEEE Journal on Selected Areas in Communications}, vol.~38, no.~3, pp.
  449--462, 2020.

\bibitem{Bombin_single_shot}
\BIBentryALTinterwordspacing
H.~Bomb\'{\i}n, ``Single-shot fault-tolerant quantum error correction,''
  \emph{Phys. Rev. X}, vol.~5, p. 031043, Sep 2015. [Online]. Available:
  \url{https://link.aps.org/doi/10.1103/PhysRevX.5.031043}
\BIBentrySTDinterwordspacing

\bibitem{campbell2019theory}
E.~T. Campbell, ``A theory of single-shot error correction for adversarial
  noise,'' \emph{Quantum Science and Technology}, vol.~4, no.~2, p. 025006,
  2019.

\bibitem{pradhan2025linear}
A.~K. Pradhan, N.~Raveendran, N.~Rengaswamy, and B.~Vasi{\'c}, ``Linear time
  iterative decoders for hypergraph-product and lifted-product codes,''
  \emph{arXiv preprint arXiv:2504.01728}, 2025.

\bibitem{DiVincenzo_2007}
D.~P. DiVincenzo and P.~Aliferis, ``Effective fault-tolerant quantum
  computation with slow measurements,'' \emph{Phys. Rev. Lett.}, vol.~98,
  no.~2, Jan. 2007.

\bibitem{terhal_scalable_2019}
\BIBentryALTinterwordspacing
B.~Terhal, L.~Pryadko, D.~Weigand, Y.~Wang, H.~Asasi, and C.~Vuillot,
  ``Scalable quantum error correction with the bosonic {GKP} code,'' p.
  F62.004, 2019, conference Name: {APS} Meeting Abstracts. [Online]. Available:
  \url{http://adsabs.harvard.edu/abs/2019APS..MARF62004T}
\BIBentrySTDinterwordspacing

\bibitem{google2024quantum}
{Google Quantum AI and Collaborators}, ``Quantum error correction below the
  surface code threshold,'' \emph{arXiv preprint arXiv:2408.13687}, 2024.

\bibitem{battistel2023realtime}
\BIBentryALTinterwordspacing
F.~Battistel, C.~Chamberland, K.~Johar, R.~W.~J. Overwater, F.~Sebastiano,
  L.~Skoric, Y.~Ueno, and M.~Usman, ``Real-time decoding for fault-tolerant
  quantum computing: progress, challenges and outlook,'' \emph{Nano Futures},
  vol.~7, no.~3, p. 032003, Aug. 2023. [Online]. Available:
  \url{https://dx.doi.org/10.1088/2399-1984/aceba6}
\BIBentrySTDinterwordspacing

\bibitem{caune2024demonstrating}
L.~Caune, L.~Skoric, N.~S. Blunt, A.~Ruban, J.~McDaniel, J.~A. Valery, A.~D.
  Patterson, A.~V. Gramolin, J.~Majaniemi, K.~M. Barnes \emph{et~al.},
  ``Demonstrating real-time and low-latency quantum error correction with
  superconducting qubits,'' \emph{arXiv preprint arXiv:2410.05202}, 2024.

\bibitem{zhang2025latte}
K.~Zhang, J.~Xu, F.~Zhang, L.~Kong, Z.~Ji, and J.~Chen, ``{LATTE: A Decoding
  Architecture for Quantum Computing with Temporal and Spatial Scalability},''
  \emph{arXiv preprint arXiv:2509.03954}, 2025.

\bibitem{Strikis_QLDPC_ModularArch_PRXQ}
\BIBentryALTinterwordspacing
A.~Strikis and L.~Berent, ``Quantum low-density parity-check codes for modular
  architectures,'' \emph{PRX Quantum}, vol.~4, p. 020321, May 2023. [Online].
  Available: \url{https://link.aps.org/doi/10.1103/PRXQuantum.4.020321}
\BIBentrySTDinterwordspacing

\bibitem{garcia2025diversityemulators}
F.~Garcia-Herrero, J.~Valls, L.~Vergara-Picazo, and V.~Torres, ``Diversity
  methods for improving convergence and accuracy of quantum error correction
  decoders through hardware emulation,'' \emph{arXiv preprint
  arXiv:2504.01164}, 2025.

\bibitem{eastin2009restrictions}
B.~Eastin and E.~Knill, ``Restrictions on transversal encoded quantum gate
  sets,'' \emph{Physical review letters}, vol. 102, no.~11, p. 110502, 2009.

\bibitem{landahl2011fault}
A.~J. Landahl, J.~T. Anderson, and P.~R. Rice, ``Fault-tolerant quantum
  computing with color codes,'' \emph{arXiv:1108.5738}, 2011.

\bibitem{zhu2023non}
G.~Zhu, S.~Sikander, E.~Portnoy, A.~W. Cross, and B.~J. Brown, ``Non-{C}lifford
  and parallelizable fault-tolerant logical gates on constant and
  almost-constant rate homological quantum {LDPC} codes via higher
  symmetries,'' \emph{arXiv preprint arXiv:2310.16982}, 2023.

\bibitem{horsman2012surface}
D.~Horsman, A.~G. Fowler, S.~Devitt, and R.~Van~Meter, ``Surface code quantum
  computing by lattice surgery,'' \emph{New Journal of Physics}, vol.~14,
  no.~12, p. 123011, 2012.

\bibitem{bravyi2005universal}
S.~Bravyi and A.~Kitaev, ``Universal quantum computation with ideal {C}lifford
  gates and noisy ancillas,'' \emph{Physical Review A—Atomic, Molecular, and
  Optical Physics}, vol.~71, no.~2, p. 022316, 2005.

\bibitem{campbell2017unified}
E.~T. Campbell and M.~Howard, ``Unified framework for magic state distillation
  and multiqubit gate synthesis with reduced resource cost,'' \emph{Physical
  Review A}, vol.~95, no.~2, p. 022316, 2017.

\bibitem{litinski2019magic}
D.~Litinski, ``Magic state distillation: Not as costly as you think,''
  \emph{Quantum}, vol.~3, p. 205, 2019.

\bibitem{gottesman1999demonstrating}
D.~Gottesman and I.~L. Chuang, ``Demonstrating the viability of universal
  quantum computation using teleportation and single-qubit operations,''
  \emph{Nature}, vol. 402, no. 6760, pp. 390--393, 1999.

\bibitem{cohen2022low}
L.~Z. Cohen, I.~H. Kim, S.~D. Bartlett, and B.~J. Brown, ``Low-overhead
  fault-tolerant quantum computing using long-range connectivity,''
  \emph{Science Advances}, vol.~8, no.~20, p. eabn1717, 2022.

\bibitem{breuckmann2024fold}
N.~P. Breuckmann and S.~Burton, ``Fold-transversal {Clifford} gates for quantum
  codes,'' \emph{Quantum}, vol.~8, p. 1372, 2024.

\bibitem{cowtan2024css}
A.~Cowtan and S.~Burton, ``{CSS} code surgery as a universal construction,''
  \emph{Quantum}, vol.~8, p. 1344, 2024.

\bibitem{cross2024improved}
A.~Cross, Z.~He, P.~Rall, and T.~Yoder, ``Improved {QLDPC} surgery: Logical
  measurements and bridging codes,'' \emph{arXiv preprint arXiv:2407.18393},
  2024.

\bibitem{eberhardt2024logical}
J.~N. Eberhardt and V.~Steffan, ``Logical operators and fold-transversal gates
  of bivariate bicycle codes,'' \emph{IEEE Transactions on Information Theory},
  vol.~71, no.~2, pp. 1140--1152, 2025.

\bibitem{quintavalle2023partitioning}
A.~O. Quintavalle, P.~Webster, and M.~Vasmer, ``Partitioning qubits in
  hypergraph product codes to implement logical gates,'' \emph{Quantum},
  vol.~7, p. 1153, 2023.

\bibitem{sayginel2024fault}
H.~Sayginel, S.~Koutsioumpas, M.~Webster, A.~Rajput, and D.~E. Browne,
  ``{Fault-Tolerant Logical Clifford Gates from Code Automorphisms},''
  \emph{PRX Quantum}, vol.~6, p. 030343, Sep 2025.

\bibitem{malcolm2025computing}
A.~J. Malcolm, A.~N. Glaudell, P.~Fuentes, D.~Chandra, A.~Schotte, C.~DeLisle,
  R.~Haenel, A.~Ebrahimi, J.~Roffe, A.~O. Quintavalle \emph{et~al.},
  ``Computing efficiently in {QLDPC} codes,'' \emph{arXiv preprint
  arXiv:2502.07150}, 2025.

\bibitem{nguyen2024quantum}
Q.~T. Nguyen and C.~A. Pattison, ``Quantum fault tolerance with constant-space
  and logarithmic-time overheads,'' \emph{arXiv preprint arXiv:2411.03632},
  2024.

\bibitem{tamiya2024polylog}
S.~Tamiya, M.~Koashi, and H.~Yamasaki, ``Polylog-time-and
  constant-space-overhead fault-tolerant quantum computation with quantum
  low-density parity-check codes,'' \emph{arXiv preprint arXiv:2411.03683},
  2024.

\bibitem{bluvstein2023Reconfig_Atom_Exp}
\BIBentryALTinterwordspacing
D.~Bluvstein, S.~J. Evered, A.~A. Geim \emph{et~al.}, ``Logical quantum
  processor based on reconfigurable atom arrays,'' \emph{Nature}, vol. 626, no.
  7997, pp. 58--65, 2024. [Online]. Available:
  \url{https://doi.org/10.1038/s41586-023-06927-3}
\BIBentrySTDinterwordspacing

\bibitem{nature_qldpc_demonstration}
\BIBentryALTinterwordspacing
Q.~Xu, J.~P.~B. Ataides, C.~A. Pattison, N.~Raveendran, D.~Bluvstein, J.~Wurtz,
  B.~Vasi\'c, M.~D. Lukin, L.~Jiang, and H.~Zhou, ``Constant-overhead
  fault-tolerant quantum computation with reconfigurable atom arrays,''
  \emph{Nature Physics}, pp. 1--10, April 2024. [Online]. Available:
  \url{https://doi.org/10.1038/s41567-024-02479-z}
\BIBentrySTDinterwordspacing

\bibitem{novak2024gnarsil}
O.~Novak and N.~Rengaswamy, ``Gnarsil: Splitting stabilizers into gauges,'' in
  \emph{2024 IEEE International Conference on Quantum Computing and Engineering
  (QCE)}, vol.~1.\hskip 1em plus 0.5em minus 0.4em\relax IEEE, 2024, pp.
  109--116.

\bibitem{berthusen2025toward}
N.~Berthusen, D.~Devulapalli, E.~Schoute, A.~M. Childs, M.~J. Gullans, A.~V.
  Gorshkov, and D.~Gottesman, ``Toward a 2d local implementation of quantum
  low-density parity-check codes,'' \emph{PRX Quantum}, vol.~6, no.~1, p.
  010306, 2025.

\bibitem{yoder2025tour}
T.~J. Yoder, E.~Schoute, P.~Rall, E.~Pritchett, J.~M. Gambetta, A.~W. Cross,
  M.~Carroll, and M.~E. Beverland, ``Tour de gross: A modular quantum computer
  based on bivariate bicycle codes,'' \emph{arXiv preprint arXiv:2506.03094},
  2025.

\bibitem{he2025extractors}
Z.~He, A.~Cowtan, D.~J. Williamson, and T.~J. Yoder, ``Extractors: {QLDPC}
  architectures for efficient pauli-based computation,'' \emph{arXiv preprint
  arXiv:2503.10390}, 2025.

\bibitem{fitzek2020deep}
D.~Fitzek, M.~Eliasson, A.~F. Kockum, and M.~Granath, ``Deep q-learning decoder
  for depolarizing noise on the toric code,'' \emph{Physical Review Research},
  vol.~2, no.~2, p. 023230, 2020.

\bibitem{andreasson2019quantum}
P.~Andreasson, J.~Johansson, S.~Liljestrand, and M.~Granath, ``Quantum error
  correction for the toric code using deep reinforcement learning,''
  \emph{Quantum}, vol.~3, p. 183, 2019.

\bibitem{milad2025itw}
M.~Taghipour and B.~Vasi\'c, ``A decoder with reinforcement learning
  feedback,'' in \emph{2025 IEEE Information Theory Workshop (ITW)}, Sydney,
  Australia, Aug. 2025, pp. 1--5.

\bibitem{milad2025istc}
------, ``Action-list reinforcement learning decoders,'' in \emph{2025 IEEE
  International Symposium on Topics in Coding}, Los Angleles, CA, Sept. 2025,
  pp. 1 --5.

\bibitem{milad2025action}
------, ``Action-list reinforcement learning syndrome decoding for binary
  linear block codes,'' \emph{arXiv preprint arXiv:2507.17893}, 2025.

\end{thebibliography}

\newlength{\xxx} 
\setlength{\xxx}{-5mm}

\begin{IEEEbiography}
[{\includegraphics[width=1in,height=1.25in,clip,keepaspectratio]{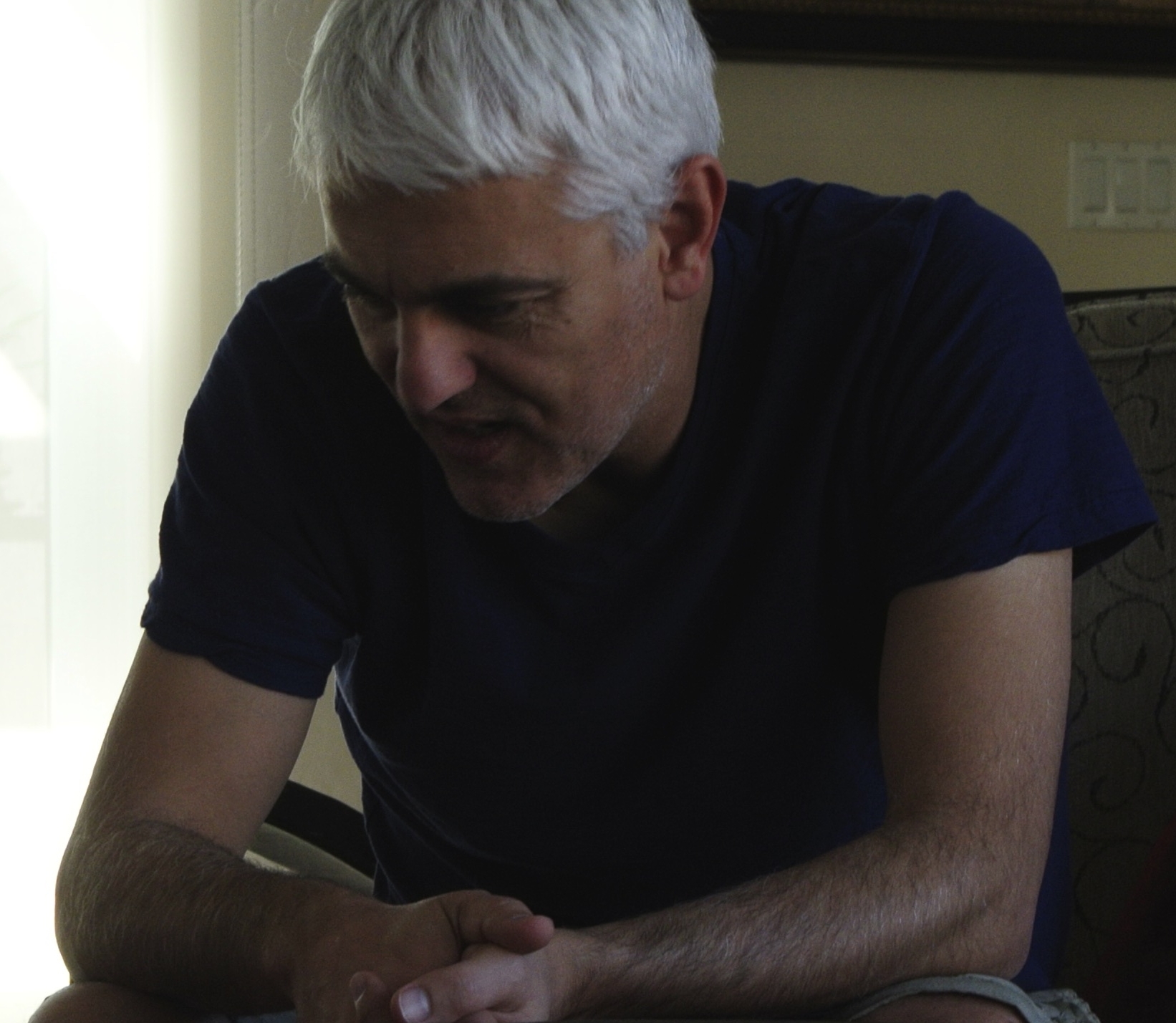}}]{Dr. Bane Vasi\'c}is a Professor of Electrical and Computer Engineering and Mathematics at the University of Arizona and a Director of the Error Correction Laboratory. He is an inventor of the soft error-event decoding algorithm for intersymbol interference channels with correlated noise, and the key architect of a detector/decoder for Bell Labs data storage read channel chips which were regarded as the best in industry. His pioneering work on structured low-density parity check (LDPC) error correcting codes based on combinatorial designs has enabled low-complexity iterative decoder implementations. Dr. Vasic leads the Quantum Error Correction Group within the Department of Energy multi-university \$115M-project led by Fermi National Laboratory to establish a Center for Superconducting Materials and Systems. He is a co-PI of the \$52M NSF Center for Quantum Network hosted at the University of Arizona.  He is also funded by NASA-Jet Propulsion Laboratory through the Strategic University Partnership Program for the development of quantum codes and error correction algorithms for NASA space missions, and is a PI on seven research grants funded by the National Science Foundation. He is an IEEE Fellow, Fulbright Scholar, da Vinci Fellow, and a past Chair of IEEE Data Storage Technical Committee.
He is a founder of Codelucida, company developing advanced error correction solutions based on LDPC codes for solid state memories for data centers since 2012. 
\vspace*{\xxx} 
\end{IEEEbiography}

\begin{IEEEbiography}
[{\includegraphics[width=1in,height=1.25in,clip,keepaspectratio]{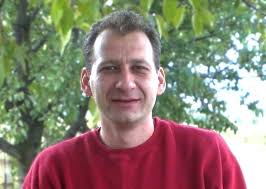}}]{Dr. Valentin Savin} is is a Senior Researcher at the Laboratory of Electronics and Information Technologies (LETI) at the French Alternative Energies and Atomic Energy Commission (CEA) and the University of Grenoble Alpes. He received his Ph.D. in Mathematics from the University Joseph Fourier in Grenoble in 2001. From 2002 to 2004, he was a Postdoctoral Researcher at the Institute of Mathematics of the Romanian Academy. Since 2005, he has been with CEA-LETI, initially as a two-year Postdoctoral Researcher, and later as a Permanent Researcher. He has authored or co-authored more than 110 papers in international journals and conference proceedings, with a focus on classical and quantum coding for data transmission, networking, and processing. He has led multiple French and European collaborative research projects in information and communications technology, facilitated several technology transfers to industry, and served as an invited expert on national and international committees and strategic seminars. Currently, he is a Fellow of the CEA, with primary research interests on quantum error correction and fault-tolerant quantum computing, with a particular emphasis on addressing the challenges of silicon spin-qubit technology under development at LETI.
\vspace*{\xxx} 
\end{IEEEbiography}

\begin{IEEEbiography}
[{\includegraphics[width=1in,height=1.25in,clip,keepaspectratio]{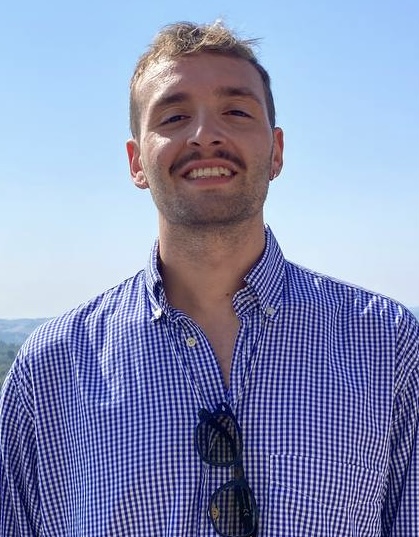}}]{Michele Pacenti} (Graduate Student Member, IEEE) received the B.S. and M.S. degrees (summa cum laude) in electronic engineering from the Polytechnic University of Marche, Italy, in 2019 and 2021, respectively. Since 2022, he has been pursuing the Ph.D. degree in electrical and computer engineering at the University of Arizona, where he is currently a doctoral candidate. As a Center for Quantum Networks (CQN) affiliated student, he participated in the Convergent Quantum Research Alliance in
Telecommunications (CoQREATE) project, spending one month in 2023 as a visiting student at University College Dublin, Ireland. His interests include QLDPC code design, decoding algorithms, noise propagation in quantum circuits, and code-based entanglement distillation.
\vspace*{\xxx} 
\end{IEEEbiography}

\begin{IEEEbiography}
[{\includegraphics[width=1in,height=1.25in,clip,keepaspectratio]{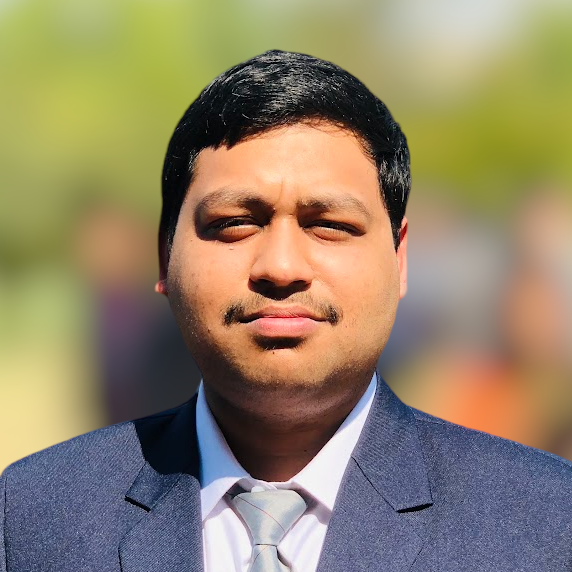}}]{Shantom Borah} is a graduate student in the Department of Electrical \& Computer Engineering at the University of Arizona. He received his B.E. in Electrical and Electronics Engineering and M.Sc. in Physics from BITS Pilani, Rajasthan, India. He is currently pursuing a Ph.D. in Electrical \& Computer Engineering with a focus on Quantum Error Correction. He is affiliated with the Superconducting Quantum Materials and Systems (SQMS) Center at the Fermi National Accelerator Laboratory, where he is involved in the design of quantum error correction architectures for quantum computers based on superconducting resonant frequency (SRF) cavities. His research interests include the development of physics-informed noise models for QLDPC error correction architectures and the interfacing of QLDPC coding schemes with continuous-variable error correction schemes such as bosonic coding.
\vspace*{\xxx} 
\end{IEEEbiography}

\begin{IEEEbiography}
[{\includegraphics[width=1in,height=1.25in,clip,keepaspectratio]{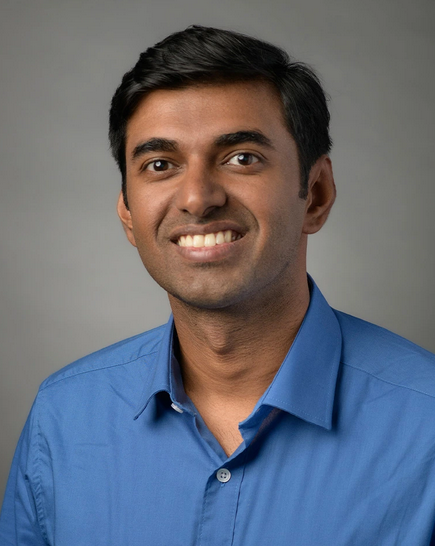}}]{Nithin Raveendran} is an Assistant Research Professor in Electrical and Computer Engineering at the University of Arizona. His research interests are in quantum computation and communication, with a focus on advancing quantum error correction through the development of improved codes and high-performance iterative decoders. He earned his Ph.D. at the University of Arizona on trapping sets in classical and quantum LDPC codes, following an M.S. from IISc Bangalore and a B.E.(Hons) from BITS Pilani. As a co-PI on a National Science Foundation–funded research grant, he is actively working on next-generation quantum error correction strategies, and his recent contributions have been published in leading journals and international conferences.
\end{IEEEbiography}

\end{document}